\documentclass[a4paper,oldversion]{aa}
\usepackage[T1]{fontenc}
\usepackage[latin9]{inputenc}
\usepackage{color}
\usepackage{amsmath}
\usepackage{graphicx}
\usepackage{amssymb}
\usepackage[authoryear]{natbib}

\providecommand{\tabularnewline}{\\}

\usepackage{xcolor}
\usepackage{ifthen}
\authorrunning{R.G. Izzard et al.}
\titlerunning{CEMP Binary Populations}
\usepackage{ifthen}
\newcommand{\modelset}[1]%
{%
\emph{%
\ifthenelse{\equal{#1}{1}}{A}{}%
\ifthenelse{\equal{#1}{2}}{Z6}{}%
\ifthenelse{\equal{#1}{3}}{Z5}{}%
\ifthenelse{\equal{#1}{4}}{CEp1}{}%
\ifthenelse{\equal{#1}{5}}{CE3}{}%
\ifthenelse{\equal{#1}{6}}{B}{}%
\ifthenelse{\equal{#1}{7}}{Ae5}{}%
\ifthenelse{\equal{#1}{8}}{8}{}%
\ifthenelse{\equal{#1}{9}}{E}{}%
\ifthenelse{\equal{#1}{10}}{10}{}%
\ifthenelse{\equal{#1}{11}}{11}{}%
\ifthenelse{\equal{#1}{12}}{12}{}%
\ifthenelse{\equal{#1}{13}}{13}{}%
\ifthenelse{\equal{#1}{14}}{14}{}%
\ifthenelse{\equal{#1}{15}}{15}{}%
\ifthenelse{\equal{#1}{16}}{C}{}%
\ifthenelse{\equal{#1}{17}}{A1}{}%
\ifthenelse{\equal{#1}{18}}{A2}{}%
\ifthenelse{\equal{#1}{19}}{D}{}%
\ifthenelse{\equal{#1}{21}}{At12}{}%
\ifthenelse{\equal{#1}{22}}{At8}{}%
\ifthenelse{\equal{#1}{26}}{F}{}%
\ifthenelse{\equal{#1}{27}}{27}{}%
\ifthenelse{\equal{#1}{28}}{28}{}%
\ifthenelse{\equal{#1}{29}}{Ap5}{}%
\ifthenelse{\equal{#1}{30}}{Ap7}{}%
\ifthenelse{\equal{#1}{31}}{31}{}%
\ifthenelse{\equal{#1}{32}}{32}{}%
\ifthenelse{\equal{#1}{33}}{33}{}%
\ifthenelse{\equal{#1}{34}}{34}{}%
\ifthenelse{\equal{#1}{}}{0}{}%
\ifthenelse{\equal{#1}{36}}{36}{}%
\ifthenelse{\equal{#1}{37}}{37}{}%
\ifthenelse{\equal{#1}{38}}{38}{}%
\ifthenelse{\equal{#1}{39}}{39}{}%
\ifthenelse{\equal{#1}{40}}{40}{}%
\ifthenelse{\equal{#1}{41}}{41}{}%
\ifthenelse{\equal{#1}{42}}{42}{}%
\ifthenelse{\equal{#1}{43}}{B2}{}%
\ifthenelse{\equal{#1}{44}}{44}{}%
\ifthenelse{\equal{#1}{45}}{45}{}%
\ifthenelse{\equal{#1}{}}{0}{}%
\ifthenelse{\equal{#1}{}}{0}{}%
\ifthenelse{\equal{#1}{47}}{47}{}%
\ifthenelse{\equal{#1}{35}}{B1}{}%
\ifthenelse{\equal{#1}{48}}{48}{}%
\ifthenelse{\equal{#1}{49}}{G}{}%
\ifthenelse{\equal{#1}{50}}{50}{}%
\ifthenelse{\equal{#1}{51}}{51}{}%
\ifthenelse{\equal{#1}{52}}{52}{}%
\ifthenelse{\equal{#1}{53}}{53}{}%
\ifthenelse{\equal{#1}{54}}{H}{}%
\ifthenelse{\equal{#1}{55}}{55}{}%
\ifthenelse{\equal{#1}{56}}{56}{}%
\ifthenelse{\equal{#1}{}}{0}{}%
\ifthenelse{\equal{#1}{57}}{57}{}%
\ifthenelse{\equal{#1}{58}}{58}{}%
\ifthenelse{\equal{#1}{59}}{59}{}%
\ifthenelse{\equal{#1}{60}}{60}{}%
\ifthenelse{\equal{#1}{61}}{61}{}%
\ifthenelse{\equal{#1}{62}}{62}{}%
\ifthenelse{\equal{#1}{}}{0}{}%
}}
\newcommand{\CEMPratio}[1]%
{%
\ifthenelse{\equal{#1}{1}}{\ensuremath{2.2}}{}%
\ifthenelse{\equal{#1}{2}}{\ensuremath{4.5}}{}%
\ifthenelse{\equal{#1}{3}}{\ensuremath{3.1}}{}%
\ifthenelse{\equal{#1}{4}}{\ensuremath{2.3}}{}%
\ifthenelse{\equal{#1}{5}}{\ensuremath{2.2}}{}%
\ifthenelse{\equal{#1}{6}}{\ensuremath{2.9}}{}%
\ifthenelse{\equal{#1}{7}}{\ensuremath{2.1}}{}%
\ifthenelse{\equal{#1}{8}}{\ensuremath{2.4}}{}%
\ifthenelse{\equal{#1}{9}}{\ensuremath{2.8}}{}%
\ifthenelse{\equal{#1}{10}}{\ensuremath{2.4}}{}%
\ifthenelse{\equal{#1}{11}}{\ensuremath{1.4}}{}%
\ifthenelse{\equal{#1}{12}}{\ensuremath{0.48}}{}%
\ifthenelse{\equal{#1}{13}}{\ensuremath{0.1}}{}%
\ifthenelse{\equal{#1}{14}}{\ensuremath{2.4}}{}%
\ifthenelse{\equal{#1}{15}}{\ensuremath{2.5}}{}%
\ifthenelse{\equal{#1}{16}}{\ensuremath{2.8}}{}%
\ifthenelse{\equal{#1}{17}}{\ensuremath{2.2}}{}%
\ifthenelse{\equal{#1}{18}}{\ensuremath{2.2}}{}%
\ifthenelse{\equal{#1}{19}}{\ensuremath{4.0}}{}%
\ifthenelse{\equal{#1}{21}}{\ensuremath{2.1}}{}%
\ifthenelse{\equal{#1}{22}}{\ensuremath{2.4}}{}%
\ifthenelse{\equal{#1}{26}}{\ensuremath{6.4}}{}%
\ifthenelse{\equal{#1}{27}}{\ensuremath{8.7}}{}%
\ifthenelse{\equal{#1}{28}}{\ensuremath{8.2}}{}%
\ifthenelse{\equal{#1}{29}}{\ensuremath{3.3}}{}%
\ifthenelse{\equal{#1}{30}}{\ensuremath{2.8}}{}%
\ifthenelse{\equal{#1}{31}}{\ensuremath{11}}{}%
\ifthenelse{\equal{#1}{32}}{\ensuremath{10}}{}%
\ifthenelse{\equal{#1}{33}}{\ensuremath{2.4}}{}%
\ifthenelse{\equal{#1}{34}}{\ensuremath{6.4}}{}%
\ifthenelse{\equal{#1}{35}}{\ensuremath{9.3}}{}%
\ifthenelse{\equal{#1}{36}}{\ensuremath{5.5}}{}%
\ifthenelse{\equal{#1}{37}}{\ensuremath{2.2}}{}%
\ifthenelse{\equal{#1}{38}}{\ensuremath{2.2}}{}%
\ifthenelse{\equal{#1}{39}}{\ensuremath{2.2}}{}%
\ifthenelse{\equal{#1}{40}}{\ensuremath{2.2}}{}%
\ifthenelse{\equal{#1}{41}}{\ensuremath{2.2}}{}%
\ifthenelse{\equal{#1}{42}}{\ensuremath{2.2}}{}%
\ifthenelse{\equal{#1}{43}}{\ensuremath{15}}{}%
\ifthenelse{\equal{#1}{44}}{\ensuremath{14}}{}%
\ifthenelse{\equal{#1}{45}}{\ensuremath{9.3}}{}%
\ifthenelse{\equal{#1}{47}}{\ensuremath{2.1}}{}%
\ifthenelse{\equal{#1}{48}}{\ensuremath{8.2}}{}%
\ifthenelse{\equal{#1}{49}}{\ensuremath{9.3}}{}%
\ifthenelse{\equal{#1}{50}}{\ensuremath{8.2}}{}%
\ifthenelse{\equal{#1}{51}}{\ensuremath{14}}{}%
\ifthenelse{\equal{#1}{52}}{\ensuremath{14}}{}%
\ifthenelse{\equal{#1}{53}}{\ensuremath{14}}{}%
\ifthenelse{\equal{#1}{54}}{\ensuremath{15}}{}%
\ifthenelse{\equal{#1}{55}}{\ensuremath{15}}{}%
\ifthenelse{\equal{#1}{56}}{\ensuremath{15}}{}%
\ifthenelse{\equal{#1}{57}}{\ensuremath{13}}{}%
\ifthenelse{\equal{#1}{58}}{\ensuremath{13}}{}%
\ifthenelse{\equal{#1}{59}}{\ensuremath{13}}{}%
}
\newcommand{\CNEMPratio}[1]%
{%
\ifthenelse{\equal{#1}{1}}{\ensuremath{0.098}}{}%
\ifthenelse{\equal{#1}{2}}{\ensuremath{0.34}}{}%
\ifthenelse{\equal{#1}{3}}{\ensuremath{0.21}}{}%
\ifthenelse{\equal{#1}{4}}{\ensuremath{0.1}}{}%
\ifthenelse{\equal{#1}{5}}{\ensuremath{0.097}}{}%
\ifthenelse{\equal{#1}{6}}{\ensuremath{0.16}}{}%
\ifthenelse{\equal{#1}{7}}{\ensuremath{0.086}}{}%
\ifthenelse{\equal{#1}{8}}{\ensuremath{0.1}}{}%
\ifthenelse{\equal{#1}{9}}{\ensuremath{0.098}}{}%
\ifthenelse{\equal{#1}{10}}{\ensuremath{0.21}}{}%
\ifthenelse{\equal{#1}{11}}{\ensuremath{0.053}}{}%
\ifthenelse{\equal{#1}{12}}{\ensuremath{0.005}}{}%
\ifthenelse{\equal{#1}{13}}{\ensuremath{0.0}}{}%
\ifthenelse{\equal{#1}{14}}{\ensuremath{0.099}}{}%
\ifthenelse{\equal{#1}{15}}{\ensuremath{0.1}}{}%
\ifthenelse{\equal{#1}{16}}{\ensuremath{0.1}}{}%
\ifthenelse{\equal{#1}{17}}{\ensuremath{0.098}}{}%
\ifthenelse{\equal{#1}{18}}{\ensuremath{0.098}}{}%
\ifthenelse{\equal{#1}{19}}{\ensuremath{0.3}}{}%
\ifthenelse{\equal{#1}{21}}{\ensuremath{0.093}}{}%
\ifthenelse{\equal{#1}{22}}{\ensuremath{0.11}}{}%
\ifthenelse{\equal{#1}{26}}{\ensuremath{0.1}}{}%
\ifthenelse{\equal{#1}{27}}{\ensuremath{0.17}}{}%
\ifthenelse{\equal{#1}{28}}{\ensuremath{0.1}}{}%
\ifthenelse{\equal{#1}{29}}{\ensuremath{0.2}}{}%
\ifthenelse{\equal{#1}{30}}{\ensuremath{0.16}}{}%
\ifthenelse{\equal{#1}{31}}{\ensuremath{0.26}}{}%
\ifthenelse{\equal{#1}{32}}{\ensuremath{0.22}}{}%
\ifthenelse{\equal{#1}{33}}{\ensuremath{0.12}}{}%
\ifthenelse{\equal{#1}{34}}{\ensuremath{0.1}}{}%
\ifthenelse{\equal{#1}{35}}{\ensuremath{0.1}}{}%
\ifthenelse{\equal{#1}{36}}{\ensuremath{0.1}}{}%
\ifthenelse{\equal{#1}{37}}{\ensuremath{0.096}}{}%
\ifthenelse{\equal{#1}{38}}{\ensuremath{0.096}}{}%
\ifthenelse{\equal{#1}{39}}{\ensuremath{0.097}}{}%
\ifthenelse{\equal{#1}{40}}{\ensuremath{0.098}}{}%
\ifthenelse{\equal{#1}{41}}{\ensuremath{0.097}}{}%
\ifthenelse{\equal{#1}{42}}{\ensuremath{0.097}}{}%
\ifthenelse{\equal{#1}{43}}{\ensuremath{0.3}}{}%
\ifthenelse{\equal{#1}{44}}{\ensuremath{0.29}}{}%
\ifthenelse{\equal{#1}{45}}{\ensuremath{0.1}}{}%
\ifthenelse{\equal{#1}{47}}{\ensuremath{0.0}}{}%
\ifthenelse{\equal{#1}{48}}{\ensuremath{0.1}}{}%
\ifthenelse{\equal{#1}{49}}{\ensuremath{0.1}}{}%
\ifthenelse{\equal{#1}{50}}{\ensuremath{0.1}}{}%
\ifthenelse{\equal{#1}{51}}{\ensuremath{0.3}}{}%
\ifthenelse{\equal{#1}{52}}{\ensuremath{0.3}}{}%
\ifthenelse{\equal{#1}{53}}{\ensuremath{0.3}}{}%
\ifthenelse{\equal{#1}{54}}{\ensuremath{0.3}}{}%
\ifthenelse{\equal{#1}{55}}{\ensuremath{0.3}}{}%
\ifthenelse{\equal{#1}{56}}{\ensuremath{0.3}}{}%
\ifthenelse{\equal{#1}{57}}{\ensuremath{0.28}}{}%
\ifthenelse{\equal{#1}{58}}{\ensuremath{0.28}}{}%
\ifthenelse{\equal{#1}{59}}{\ensuremath{0.28}}{}%
}
\newcommand{\NEMPratio}[1]%
{%
\ifthenelse{\equal{#1}{1}}{\ensuremath{0.25}}{}%
\ifthenelse{\equal{#1}{2}}{\ensuremath{0.19}}{}%
\ifthenelse{\equal{#1}{3}}{\ensuremath{0.19}}{}%
\ifthenelse{\equal{#1}{4}}{\ensuremath{0.25}}{}%
\ifthenelse{\equal{#1}{5}}{\ensuremath{0.25}}{}%
\ifthenelse{\equal{#1}{6}}{\ensuremath{0.24}}{}%
\ifthenelse{\equal{#1}{7}}{\ensuremath{0.24}}{}%
\ifthenelse{\equal{#1}{8}}{\ensuremath{0.24}}{}%
\ifthenelse{\equal{#1}{9}}{\ensuremath{0.25}}{}%
\ifthenelse{\equal{#1}{10}}{\ensuremath{0.12}}{}%
\ifthenelse{\equal{#1}{11}}{\ensuremath{0.052}}{}%
\ifthenelse{\equal{#1}{12}}{\ensuremath{0.014}}{}%
\ifthenelse{\equal{#1}{13}}{\ensuremath{0.0}}{}%
\ifthenelse{\equal{#1}{14}}{\ensuremath{0.28}}{}%
\ifthenelse{\equal{#1}{15}}{\ensuremath{0.29}}{}%
\ifthenelse{\equal{#1}{16}}{\ensuremath{0.29}}{}%
\ifthenelse{\equal{#1}{17}}{\ensuremath{0.25}}{}%
\ifthenelse{\equal{#1}{18}}{\ensuremath{0.25}}{}%
\ifthenelse{\equal{#1}{19}}{\ensuremath{0.17}}{}%
\ifthenelse{\equal{#1}{21}}{\ensuremath{0.24}}{}%
\ifthenelse{\equal{#1}{22}}{\ensuremath{0.24}}{}%
\ifthenelse{\equal{#1}{26}}{\ensuremath{0.24}}{}%
\ifthenelse{\equal{#1}{27}}{\ensuremath{0.13}}{}%
\ifthenelse{\equal{#1}{28}}{\ensuremath{0.24}}{}%
\ifthenelse{\equal{#1}{29}}{\ensuremath{0.14}}{}%
\ifthenelse{\equal{#1}{30}}{\ensuremath{0.19}}{}%
\ifthenelse{\equal{#1}{31}}{\ensuremath{0.048}}{}%
\ifthenelse{\equal{#1}{32}}{\ensuremath{0.083}}{}%
\ifthenelse{\equal{#1}{33}}{\ensuremath{0.32}}{}%
\ifthenelse{\equal{#1}{34}}{\ensuremath{0.24}}{}%
\ifthenelse{\equal{#1}{35}}{\ensuremath{0.24}}{}%
\ifthenelse{\equal{#1}{36}}{\ensuremath{0.24}}{}%
\ifthenelse{\equal{#1}{37}}{\ensuremath{0.24}}{}%
\ifthenelse{\equal{#1}{38}}{\ensuremath{0.24}}{}%
\ifthenelse{\equal{#1}{39}}{\ensuremath{0.24}}{}%
\ifthenelse{\equal{#1}{40}}{\ensuremath{0.24}}{}%
\ifthenelse{\equal{#1}{41}}{\ensuremath{0.24}}{}%
\ifthenelse{\equal{#1}{42}}{\ensuremath{0.24}}{}%
\ifthenelse{\equal{#1}{43}}{\ensuremath{0.17}}{}%
\ifthenelse{\equal{#1}{44}}{\ensuremath{0.23}}{}%
\ifthenelse{\equal{#1}{45}}{\ensuremath{0.24}}{}%
\ifthenelse{\equal{#1}{47}}{\ensuremath{0.018}}{}%
\ifthenelse{\equal{#1}{48}}{\ensuremath{0.24}}{}%
\ifthenelse{\equal{#1}{49}}{\ensuremath{0.24}}{}%
\ifthenelse{\equal{#1}{50}}{\ensuremath{0.24}}{}%
\ifthenelse{\equal{#1}{51}}{\ensuremath{0.2}}{}%
\ifthenelse{\equal{#1}{52}}{\ensuremath{0.2}}{}%
\ifthenelse{\equal{#1}{53}}{\ensuremath{0.2}}{}%
\ifthenelse{\equal{#1}{54}}{\ensuremath{0.2}}{}%
\ifthenelse{\equal{#1}{55}}{\ensuremath{0.2}}{}%
\ifthenelse{\equal{#1}{56}}{\ensuremath{0.2}}{}%
\ifthenelse{\equal{#1}{57}}{\ensuremath{0.16}}{}%
\ifthenelse{\equal{#1}{58}}{\ensuremath{0.16}}{}%
\ifthenelse{\equal{#1}{59}}{\ensuremath{0.16}}{}%
}
\newcommand{\CEMPsratio}[1]%
{%
\ifthenelse{\equal{#1}{1}}{\ensuremath{29}}{}%
\ifthenelse{\equal{#1}{2}}{\ensuremath{75}}{}%
\ifthenelse{\equal{#1}{3}}{\ensuremath{65}}{}%
\ifthenelse{\equal{#1}{4}}{\ensuremath{29}}{}%
\ifthenelse{\equal{#1}{5}}{\ensuremath{29}}{}%
\ifthenelse{\equal{#1}{6}}{\ensuremath{47}}{}%
\ifthenelse{\equal{#1}{7}}{\ensuremath{28}}{}%
\ifthenelse{\equal{#1}{8}}{\ensuremath{20}}{}%
\ifthenelse{\equal{#1}{9}}{\ensuremath{44}}{}%
\ifthenelse{\equal{#1}{10}}{\ensuremath{35}}{}%
\ifthenelse{\equal{#1}{11}}{\ensuremath{20}}{}%
\ifthenelse{\equal{#1}{12}}{\ensuremath{0.8}}{}%
\ifthenelse{\equal{#1}{13}}{\ensuremath{0.015}}{}%
\ifthenelse{\equal{#1}{14}}{\ensuremath{28}}{}%
\ifthenelse{\equal{#1}{15}}{\ensuremath{27}}{}%
\ifthenelse{\equal{#1}{16}}{\ensuremath{29}}{}%
\ifthenelse{\equal{#1}{17}}{\ensuremath{86}}{}%
\ifthenelse{\equal{#1}{18}}{\ensuremath{95}}{}%
\ifthenelse{\equal{#1}{19}}{\ensuremath{65}}{}%
\ifthenelse{\equal{#1}{21}}{\ensuremath{28}}{}%
\ifthenelse{\equal{#1}{22}}{\ensuremath{30}}{}%
\ifthenelse{\equal{#1}{26}}{\ensuremath{22}}{}%
\ifthenelse{\equal{#1}{27}}{\ensuremath{95}}{}%
\ifthenelse{\equal{#1}{28}}{\ensuremath{99}}{}%
\ifthenelse{\equal{#1}{29}}{\ensuremath{19}}{}%
\ifthenelse{\equal{#1}{30}}{\ensuremath{22}}{}%
\ifthenelse{\equal{#1}{31}}{\ensuremath{73}}{}%
\ifthenelse{\equal{#1}{32}}{\ensuremath{81}}{}%
\ifthenelse{\equal{#1}{33}}{\ensuremath{34}}{}%
\ifthenelse{\equal{#1}{34}}{\ensuremath{97}}{}%
\ifthenelse{\equal{#1}{35}}{\ensuremath{99}}{}%
\ifthenelse{\equal{#1}{36}}{\ensuremath{98}}{}%
\ifthenelse{\equal{#1}{37}}{\ensuremath{28}}{}%
\ifthenelse{\equal{#1}{38}}{\ensuremath{28}}{}%
\ifthenelse{\equal{#1}{39}}{\ensuremath{28}}{}%
\ifthenelse{\equal{#1}{40}}{\ensuremath{29}}{}%
\ifthenelse{\equal{#1}{41}}{\ensuremath{28}}{}%
\ifthenelse{\equal{#1}{42}}{\ensuremath{28}}{}%
\ifthenelse{\equal{#1}{43}}{\ensuremath{73}}{}%
\ifthenelse{\equal{#1}{44}}{\ensuremath{74}}{}%
\ifthenelse{\equal{#1}{45}}{\ensuremath{94}}{}%
\ifthenelse{\equal{#1}{47}}{\ensuremath{24}}{}%
\ifthenelse{\equal{#1}{48}}{\ensuremath{99}}{}%
\ifthenelse{\equal{#1}{49}}{\ensuremath{94}}{}%
\ifthenelse{\equal{#1}{50}}{\ensuremath{99}}{}%
\ifthenelse{\equal{#1}{51}}{\ensuremath{96}}{}%
\ifthenelse{\equal{#1}{52}}{\ensuremath{99}}{}%
\ifthenelse{\equal{#1}{53}}{\ensuremath{99}}{}%
\ifthenelse{\equal{#1}{54}}{\ensuremath{97}}{}%
\ifthenelse{\equal{#1}{55}}{\ensuremath{99}}{}%
\ifthenelse{\equal{#1}{56}}{\ensuremath{99}}{}%
\ifthenelse{\equal{#1}{57}}{\ensuremath{96}}{}%
\ifthenelse{\equal{#1}{58}}{\ensuremath{99}}{}%
\ifthenelse{\equal{#1}{59}}{\ensuremath{99}}{}%
}

 \date{}

\newcommand{\Change}[1]{#1}
\newcommand{\ChangeA}[1]{#1}

\newcommand{\LabelG}[1]{\fontfamily{ptm}{{\normalsize\textsf{\textbf{#1)}}}}}

\begin{document}

\title{Population Synthesis of Binary\\
Carbon-enhanced Metal-poor Stars }

\author{Robert G. Izzard$^{1,2}$, Evert Glebbeek$^{1,3}$, Richard J. Stancliffe$^{4,5}$
and Onno Pols$^{1}$}

\institute{1 Sterrenkundig Instituut, Universiteit Utrecht, P.O. Box 80000,
NL-3508 TA Utrecht, The Netherlands.\\
2 Institut d'Astronomie et d'Astrophysique, Université Libre de
Bruxelles, Boulevard du Triomphe, B-1050 Brussels, Belgium\thanks{Present address. Email Robert.Izzard@ulb.ac.be}.\\
3 Department of Physics and Astronomy, McMaster University, Hamilton,
Ontario, L8S 4M1, Canada.\\
4 Institute of Astronomy, University of Cambridge, Madingley Road,
Cambridge, CB3 0HA, United Kingdom.\\
5 School of Mathematical Sciences, PO Box 28M, Monash University,
Victoria 3800, Australia.}

\abstract{The carbon-enhanced metal-poor (CEMP) stars constitute approximately
one fifth of the metal-poor ($[\mathrm{Fe}/\mathrm{H}]\lesssim-2$)
population but their origin is not well understood. The most widely
accepted formation scenario, at least for the majority of CEMP stars
which are also enriched in $s$-process elements, invokes mass-transfer
of carbon-rich material from a thermally-pulsing asymptotic giant
branch (TPAGB) primary star to a less massive main-sequence companion
which is seen today. Recent studies explore the possibility that an
initial mass function biased toward intermediate-mass stars is required
to reproduce the observed CEMP fraction in stars with metallicity
$[\mathrm{Fe}/\mathrm{H}]<-2.5$. These models also implicitly predict
a large number of nitrogen-enhanced metal-poor (NEMP) stars which
is not seen. In this paper we investigate whether the observed CEMP
and NEMP to extremely metal-poor (EMP) ratios can be explained \emph{without}
invoking a change in the initial mass function. We construct binary-star
populations in an attempt to reproduce the observed number and chemical
abundance patterns of CEMP stars at a metallicity $[\mathrm{Fe}/\mathrm{H}]\sim-2.3$.
Our binary-population models include synthetic nucleosynthesis in
TPAGB stars and account for mass transfer and other forms of binary
interaction. This approach allows us to explore uncertainties in the
CEMP-star formation scenario by parameterization of uncertain input
physics. In particular, we consider the uncertainty in the physics
of third dredge up in the TPAGB primary, binary mass transfer and
mixing in the secondary star. We confirm earlier findings that with
current detailed TPAGB models, in which third dredge up is limited
to stars more massive than about $1.25\mathrm{\, M_{\odot}}$, the
large observed CEMP fraction cannot be accounted for. We find that
efficient third dredge up in low-mass (less than $1.25\mathrm{\, M_{\odot}}$),
low-metallicity stars may offer at least a partial explanation to
the large observed CEMP fraction while remaining consistent with the
small observed NEMP fraction. }

\maketitle

\section{Introduction}

\label{sec:Introduction}One of the most interesting problems in modern
stellar astronomy is to explain the existence of a population of carbon-enhanced
metal-poor (CEMP%
\footnote{$[\mathrm{C}/\mathrm{Fe}]\geq1$, $\mathrm{\left[\mathrm{Fe}/\mathrm{H}\right]}\lesssim-2$,
where the logarithmic abundance ratio $[X/Y]=\log_{10}(X/Y)-\log_{10}(X_{\odot}/Y_{\odot})$.%
}) stars in the Galactic halo. The HK \citep{1992Beers++} and Hamburg/ESO
\citep{2001A&A...375..366C} surveys find a large number of CEMP stars
among the metal-poor (EMP, $[\mathrm{Fe}/\mathrm{H}]\lesssim-2$)
population, at a fraction around 20\% (e.g. $9\pm2\%$ \citealp{2006ApJ...652.1585F};
$21\pm2\%$ \citealp{2006ApJ...652L..37L}; up to $30\%$ from the
SAGA database of \citealp{2008PASJ...60.1159S} -- see below for details).

The CEMP stars are subdivided into four groups depending on the presence
or absence of the heavy elements barium and europium (see e.g. \citealp{2005ARA&A..43..531B}).
The most populous group consists of the $s$-process rich CEMP stars,
the so-called CEMP-$s$ stars (e.g. \citealp{2007ApJ...655..492A})
which display barium enhancements of $[\mathrm{Ba}/\mathrm{Fe}]>+0.5$.
These account for about 80 per cent of all CEMP stars. There are also
CEMP stars with $r$-process enhancements (the CEMP-$r$ class) and
some with both $r$- and $s$-process enhancements (CEMP-$r$+$s$,
e.g. \citealp{2006A&A...451..651J}). Finally, there is a class of
CEMP stars which show no enhancement of neutron-capture elements.
These are called the CEMP-no stars \citep{2002Aoki++_no_ncapt}. A
detailed review of the various CEMP subgroups can be found in \citet{2009arXiv0901.4737M}.

A quantitative understanding of the origin of CEMP stars touches on
many branches of stellar astronomy. The most likely formation mechanism
for the $s$-process rich CEMP stars involves mass transfer in binary
systems. Carbon-rich material from the TPAGB primary star pollutes
the lower-mass main sequence secondary such that it becomes enriched
in carbon and $s$-process elements. We observe only the secondary
today; the primary is an unseen white dwarf. Surveys of radial velocity
shifts find that the binary fraction of CEMP-$s$ stars is consistent
with them all being binaries \citep{2004MmSAI..75..772T,2005ApJ...625..825L}.
This binary mass transfer scenario is the same as that which is invoked
to explain the Ba and CH stars \citep{1983ARA&A..21..271I,1984ApJ...280L..31M,1990mcclure_woodsworth,1997mcclure}.
However, only about $1\%$ of population I/II stars are Ba/CH stars,
respectively \citep{1989A&A...219L..15T,1991ApJS...77..515L}. The
CH stars are also carbon-rich but not as metal-poor as the CEMP stars,
with $[\mathrm{Fe}/\mathrm{H}]\sim-1$. The carbon-rich fraction of
1 per cent at higher metallicity is in stark contrast to the observed
CEMP fraction of around 20 per cent. 

The mass-transfer scenario involves many processes that are not well
understood. There are uncertainties associated with stellar evolution,
particularly with respect to nucleosynthesis in TPAGB stars. Carbon
and $s$-process enhancements are thought to occur via third dredge
up, but other processes may also play an important role in low-metallicity
nucleosynthesis. These include hot-bottom burning (e.g. \citealp{1975ApJ...196..525I,1993ApJ...416..762B,2004Herwig}),
dual core flashes (also known as helium flash driven deep mixing)
and dual shell flashes (helium flash driven deep mixing during a thermal
pulse, \citealt*{1990ApJ...349..580F,2002Schlattl++,2007ApJ...667..489C,2008A&A...490..769C}),
{}``extra mixing'' on the first giant branch and perhaps on the
TPAGB (e.g. \citealt*{2000A&A...356..181W,2003ApJ...582.1036N,2008ApJ...677..581E})
and convective overshooting \citep{2000A&A...360..952H}. 

The CEMP-formation scenario also requires knowledge of the physics
of stellar interaction in binary systems. The primary star must transfer
material to the secondary star which in turn might dilute and burn
it. A wind mass-transfer scenario in wide binaries, e.g. by a mechanism
similar to that of \citet{1944MNRAS.104..273B}, likely plays a role
in CEMP formation. Closer binaries which undergo Roche-lobe overflow
(RLOF) from a TPAGB star on to a less massive main-sequence star are
expected to enter a common-envelope phase. Such stars would undergo
few thermal pulses with little accretion on to the secondary (although
see \citealp{2008ApJ...672L..41R} for details of accretion in a common
envelope and associated uncertainties).

The fate of the accreted material is also uncertain. The molecular
weight of accreted material is certainly greater than that of the
secondary star, as carbon-enhanced TPAGB stars should also be helium
rich. Accreted material should thus sink by the thermohaline instability
\citep{2007A&A...464L..57S} but this may be inhibited by gravitational
settling \citep{2008MNRAS.389.1828S,2008ApJ...677..556T}. Furthermore,
radiative levitation of some chemical species may be important \citep{2002ApJ...580.1100R,2002ApJ...568..979R}.
When the secondary ascends the first giant branch, its convection
zone mixes any accreted material which may remain in the surface layers
with material from deep inside the star. The surface abundance distribution
depends on whether material has mixed deep into the star or not, because
if it has it may have undergone some nuclear burning, the ashes of
which are mixed to the surface.

Two studies have considered population models in an attempt to reproduce
the observed CEMP fraction of about $20\%$. The models of \citet{2005ApJ...625..833L}
and \citet{2007ApJ...658..367K} both concluded that in order to make
enough CEMP stars the initial mass function (IMF) at low metallicity
must be significantly different to that observed in the solar neighbourhood.
In particular, they enhanced the number of intermediate-mass stars
relative to low-mass stars -- this has the effect of increasing the
number of $\sim2\mathrm{\, M_{\odot}}$ stars which are responsible
for the production of most of the carbon. 

The \citet{2007ApJ...658..367K} model differentiates between two
metallicity regimes. In their model, stars with masses greater than
$1.5\mathrm{\, M_{\odot}}$ undergo third dredge up irrespective of
metallicity. For stars with $[\mathrm{Fe}/\mathrm{H}]\leq-2.5$ and
mass less than $1.5\mathrm{\, M_{\odot}}$ they invoke proton ingestion
at the helium flash (the dual core flash) or at the first thermal
pulse (the dual shell flash) as the source of carbon. This implies
that the CEMP fraction should be smaller for $[\mathrm{Fe}/\mathrm{H}]>-2.5$,
but the Stellar Abundances for Galactic Archeology (SAGA) database
\citep{2008PASJ...60.1159S} shows that the CEMP fraction is approximately
constant as a function of metallicity up to $[\mathrm{Fe}/\mathrm{H}]\approx-2$.

A related problem is that of the nitrogen-enhanced metal-poor (NEMP)
stars which have $[\mathrm{N}/\mathrm{Fe}]>0.5$ and $[\mathrm{C}/\mathrm{N}]<-0.5$
as defined by \citet{2007ApJ...658.1203J}. Such stars are expected
to result from mass transfer in binaries with TPAGB primaries more
massive than about $3\, M_{\odot}$ in which hot bottom burning has
converted most of the dredged-up carbon into nitrogen. The observed
NEMP to EMP ratio is small, less than one in twenty-one according
to \citet{2007ApJ...658.1203J} or less than $7\%$ in the metallicity
range $[\mathrm{Fe}/\mathrm{H]=-2.3\pm0.5}$ according to the SAGA
database (see Table~\ref{tab:Observations-table}). The \citet{2007ApJ...658..367K}
models, with an enhanced number of intermediate-mass relative to low-mass
stars, should make many more NEMP stars than are observed.

The aim of this paper is to investigate which physical scenarios are
able to reproduce the CEMP and NEMP to EMP ratios, at metallicity
$[\mathrm{Fe}/\mathrm{H}]\sim-2.3$, \emph{without} altering the initial
mass function. We combine a synthetic nucleosynthesis model with a
binary population synthesis code to simulate populations of low-metallicity
binaries (see Section~\ref{sec:Model}). The power of the population
synthesis approach is that it can efficiently explore the available
parameter space. Much of the input physics is uncertain (as we have
described above) but population synthesis allows us to explore the
consequences of these uncertainties by varying the model free parameters
within reasonable bounds. We try to reproduce the observed CEMP and
NEMP to EMP ratios, surface chemistry distributions, binary period
distributions and chemical abundance correlations. It is thus a powerful
tool to apply to this problem. 

In order to compare our models to observations we choose a subset
of the SAGA database which corresponds to giants and turn-off stars
as described in Section~\ref{sec:Observational-database}. The results
of our simulations and comparison with the sample of observations
are given in Section~\ref{sec:Results}. The implications of our
results and outstanding problems are discussed in Section~\ref{sec:Discussion}
while Section~\ref{sec:Conclusions} concludes.

\section{Models}

\label{sec:Model}In this section we describe our binary population
synthesis model (Sections \ref{sub:Physics}-\ref{sub:Population-Synthesis}),
initial distributions (Section~\ref{sub:Stellar-distributions}),
the choices of parameters for the various model sets (Section~\ref{sub:modelsets})
and our criteria for selecting CEMP and NEMP stars (Section~\ref{sub:model-Selection-criteria}).

\subsection{Input physics}

\label{sub:Physics}Our binary population synthesis model is based
on the synthetic nucleosynthesis models of \citet{Izzard_et_al_2003b_AGBs}
and \citet{2006A&A...460..565I}. Binary stellar evolution is followed
according to the rapid binary stellar evolution (BSE) prescription
of \citet{2002MNRAS_329_897H} in which detailed stellar evolution
model results are approximated by fitting functions. Coupled with
a binary evolution algorithm which includes mass transfer due to both
RLOF and winds, tidal circularisation and common envelope evolution,
this approach allows the simulation of millions of binary stars in
less than a day on a modern computer. 

Stellar evolution is augmented by a nucleosynthesis algorithm which
follows the evolution of stars through the first, second and third
dredge ups, altering surface abundances as necessary. This is mostly
based on the \citet{Parameterising_3DUP_Karakas_Lattanzio_Pols} and
\citet[hereafter K02 and K07 respectively]{2007PASA...24..103K} detailed
models. 

We include a prescription for hot-bottom burning (HBB) in sufficiently
massive AGB stars, $M\gtrsim2.75\mathrm{\, M_{\odot}}$ at $Z=10^{-4}$.
\Change{The mass at which HBB switches on may be greater than our
models suggest, e.g. $>5\mathrm{\, M_{\odot}}$ in \citet[see their table B.4 for $Z=0.0005$]{2009arXiv0903.2155W},
$3-4\mathrm{\, M_{\odot}}$ \citet{2009MNRAS.396.1046L} or $3\mathrm{\, M_{\odot}}$
in the latest models by \citet[MNRAS submitted]{Karakas2009update}.
These models use different input physics and/or nuclear reaction rates
to the K02/K07 models on which our synthetic model is based. The impact
of such changes on the number of CEMP stars is discussed in Section~\ref{sub:default-initial-parameter-space}.
Proton-capture reaction-rate uncertainties affect mainly hot-bottom
burning stars, i.e. NEMP progenitors, rather than CEMP stars (see
e.g. \citealp{2007A&A...466..641I} for a discussion of the effect
of proton-capture reaction rate uncertainties which affect Ne-Al in
massive AGB stars).}

We model binary mass transfer by both stellar winds according to the
Bondi-Hoyle prescription \citep{1944MNRAS.104..273B} and Roche-lobe
overflow. Common-envelope evolution follows the prescription of \citet{2002MNRAS_329_897H}.
We have updated some of the physical prescriptions in our model which
are relevant to CEMP star formation. We describe below the most important
changes to our binary code since \citet{2006A&A...460..565I}.

\Change{Our binary population synthesis model has been applied to
a number of problems including the higher-metallicity equivalents
of CEMP stars, the barium stars \citep{2003ASPC..303..290} and CH
stars \citep{Binary_Origin_low_L_C_Stars}. Our model approximately
matches the observed Ba star to G/K giant ratio of ${\sim1\%}$ \citep{1991ApJS...77..515L}.
An extended version of our model was used by \citet{2008A&A...480..797B}
to successfully model the eccentricities of the barium stars.}

\subsubsection{Metallicity}

The \citet{2002MNRAS_329_897H} fitting formulae are limited to metallicities
above and including $Z=10^{-4}$ and hence our stellar evolution model
is not valid below this metallicity\Change{. Similarly, the K02/07
models extend down to $Z=10^{-4}$ or, equivalently, $[\mathrm{Fe}/\mathrm{H}]=-2.3$.
As a consequence we compare our models only to observations with $[\mathrm{Fe}/\mathrm{H}]\sim-2.3$
(Section~\ref{sec:Observational-database}). We cannot compare our
models to stars of significantly lower metallicity because our model
lacks algorithms to describe phenomena such as proton ingestion at
the first thermal pulse (see Section~\ref{sub:Missing-physics}).
}

\subsubsection{First dredge up}

\label{sub:First-Dredge-Up}Abundance changes at first dredge up are
interpolated from a grid of detailed stellar evolution models made
with the {\sc stars} code. The {\sc{stars}} code was originally
written by \citet{1971MNRAS.151..351E} and has been updated by many
authors e.g. \citet{1995MNRAS.274..964P} and \citet{2009MNRAS.396.1699S}.
The version used here employs the nucleosynthesis routines of \citet{2005MNRAS.360..375S},
which follow forty isotopes from $\mathrm{D}$ to $^{32}\mathrm{S}$
and important iron group elements. Model sequences are evolved from
the pre-main sequence to the tip of the red giant branch using 499
mesh points. Convective overshooting is employed via the prescription
of \citet{1997MNRAS.285..696S} with an overshooting parameter of
$\delta_{\mathrm{ov}}=0.12$. Thermohaline mixing on the RGB is included
via the prescription of \citet{1980A+A-91-175K}. The diffusion coefficient
is multiplied by a factor of 100, following the work of \citet{2007A&A...467L..15C}.

In single stars at low metallicity first dredge up has a small effect
on surface abundances. However, in secondary stars which have been
polluted by a companion first dredge up may either dilute accreted
material which is sitting on the stellar surface or mix material from
inside the star which has been been burned, depending on the efficiency
of thermohaline mixing of the accreted material (Section~\ref{sub:Thermohaline-mixing};
see also \citealp{2007A&A...464L..57S,2007A&A...467L..15C}). We take
such processes into account. A detailed description of our algorithm
is given in Appendix \ref{sub:First-DUP-appendix}.

\subsubsection{Third dredge up}

\label{sub:Third-Dredge-Up}Third dredge up is the primary mechanism
by which carbon made by helium burning is brought to the stellar surface
in AGB stars. As stars evolve up the TPAGB their core mass $M_{\mathrm{c}}$
increases and every $\tau_{\mathrm{IP}}$ years a thermal pulse occurs.
Once $M_{\mathrm{c}}$ exceeds a threshold mass $M_{\mathrm{c},\mathrm{min}}$
third dredge up occurs with efficiency $\lambda$, the ratio of the
mass dredged up to the core growth during the previous interpulse
phase. The values of $\lambda$ and $M_{\mathrm{c},\mathrm{min}}$
are fitted as a function of mass and metallicity to the detailed models
of K02/K07. Without modification of this prescription single stars
with initial mass greater than $1.25\mathrm{\, M_{\odot}}$ become
carbon stars at a metallicity of $Z=10^{-4}$.

The correction factors $\Delta M_{\mathrm{c},\mathrm{min}}$ and $\lambda_{\mathrm{min}}$
were introduced by \citet{Izzard_et_al_2003b_AGBs} to enhance dredge
up in low-mass stars relative to the detailed models%
\footnote{These parameters modify the fits to the detailed models such that
$M_{\mathrm{c},\mathrm{min}}\rightarrow M_{\mathrm{c},\mathrm{min}}+\Delta M_{\mathrm{c},\mathrm{min}}$
and $\lambda\rightarrow\max(\lambda,\lambda_{\mathrm{min}})$. A negative
$\Delta M_{\mathrm{c},\mathrm{min}}$ allows dredge up in lower initial-mass
stars and a positive $\lambda_{\mathrm{min}}$ increases the amount
of material dredged up once dredge up begins. %
}. \Change{They found that dredge up should occur earlier on the TPAGB
and with greater efficiency than predicted by the K02 models. With
the parameter choices $\Delta M_{\mathrm{c},\mathrm{min}}\sim-0.07\mathrm{\, M_{\odot}}$
and $\lambda_{\mathrm{min}}\approx0.8-37.5Z$ the carbon-star luminosity
functions in the Magellanic clouds are approximately fitted by the
model.} However, these parameters are poorly constrained, especially
at metallicities less than that of the Small Magellanic Cloud,\Change{
so $\Delta M_{\mathrm{c},\mathrm{min}}$ and $\lambda_{\mathrm{min}}$
should be considered free parameters at $Z=10^{-4}$.}

We introduce a parameter, $M_{\mathrm{env},\mathrm{min}}$, the minimum
\emph{envelope mass} for third dredge up. This is $0.5\mathrm{\, M_{\odot}}$
by default, following solar-metallicity models \citep{1997ApJ...478..332S},
but we treat it as a free parameter. Recent detailed models calculated
with the {\sc{stars}} code (\citealp{2008MNRAS.389.1828S}) find
third dredge up in an initially $0.9\mathrm{\, M_{\odot}}$, $Z=10^{-4}$
model which, when it reaches the AGB, has an envelope mass of only
$0.31\mathrm{\, M_{\odot}}$. Also, \citet{2007MNRAS.375.1280S} find
third dredge up continues to occur even when the envelope mass drops
below $0.5\mathrm{\, M_{\odot}}$. In the latest models of \citet{Karakas2009update}
dredge up is also found for small envelope mass (down to $0.1\mathrm{\, M_{\odot}}$)
at $Z=10^{-4}$. We use this as justification of our decision to reduce
$M_{\mathrm{env,min}}$ below $0.5\mathrm{\, M_{\odot}}$ but note
that the value of $M_{\mathrm{env,min}}$ is not accurately known.

A choice of $\Delta M_{\mathrm{c},\mathrm{min}}=-0.07$, $\lambda_{\mathrm{min}}=0.5$
and $M_{\mathrm{env,min}}=0$ leads to dredge up in all stars which
reach the TPAGB in the age of the galaxy, i.e. a minimum initial mass
for third dredge up of $\sim0.8\mathrm{\, M_{\odot}}$.

In low-metallicity TPAGB stars dredge up of the hydrogen-burning shell
is an important source of $^{13}\mathrm{C}$ and $^{14}\mathrm{N}$.
We include an approximate prescription which well fits the K02/07
models (see Appendix \ref{sub:Third-DUP-appendix}).

\subsubsection{Thermohaline mixing}

\label{sub:Thermohaline-mixing}Most of our model sets assume thermohaline
mixing of accreted material according to the prescription of \citet{2006A&A...460..565I}
in which accreted material sinks and mixes instantaneously with the
stellar envelope. The calculations of \citet{2007A&A...464L..57S}
suggest this is reasonable in some cases. However, gravitational settling
prior to accretion may prevent thermohaline mixing \citep{2008ApJ...677..556T,2008AIPC..990..330B}.
In order to account for both possibilities, either efficient and instantaneous
or highly inefficient thermohaline mixing, we have also run models
in which the accreted material remains on the stellar surface (regardless
of its molecular weight) until mixed in by convection. Recent calculations
show that the situation is somewhat more complicated than either of
the extremes we test here \citep{2008MNRAS.389.1828S}.

\subsubsection{Parameter choices}

\label{sub:Default-physical-model}Our binary nucleosynthesis model
has many free parameters, some of which have been constrained by previous
studies, some which have not. We list here the most important parameters
and our default choices.
\begin{itemize}
\item Abundances are solar-scaled with a mixture according to \citet{1989GeCoA..53..197A}.
We do not include an $\alpha$-element enhancement. Most of our models
have a metallicity of $Z=10^{-4}$ (equivalent to $[\mathrm{Fe}/\mathrm{H}]=-2.3$).
\item Wind mass-loss rates are parameterised according to the Reimers formula
with $\eta=0.5$ on the first giant branch and \citet{1993ApJ...413..641V}
on the AGB (as in K02). We modulate the AGB mass-loss rate with a
factor $f_{\mathrm{VW}}$ which is one by default. We apply a correction
$\Delta P_{\mathrm{VW}}$, zero by default, to the Mira period relation
used in the \citeauthor{1993ApJ...413..641V} prescription to simulate
a delayed superwind on the AGB. We also consider both Reimers and
Van Loon mass-loss rates on the AGB \citep{1975psae.book..229R,2005A&A...438..273V}.
Appendix \ref{sec:Mass-loss-prescriptions} describes the mass-loss
formulae in detail.
\item The Bondi-Hoyle accretion efficiency factor $\alpha_{\mathrm{BH}}=3/2$.
We also consider $\alpha_{\mathrm{BH}}=5$, however unphysical this
may be, to simulate enhanced stellar-wind mass transfer. 
\item Third dredge up parameters $\Delta M_{\mathrm{c},\mathrm{min}}=0\mathrm{\, M_{\odot}}$,
$\lambda_{\mathrm{min}}=0$ and $M_{\mathrm{env},\mathrm{min}}=0.5\mathrm{\, M_{\odot}}$
which correspond to the detailed TPAGB models of K02/07. Enhanced
third dredge up is simulated in some model sets by choosing a negative
$\Delta M_{\mathrm{c},\mathrm{min}}$, positive $\lambda_{\mathrm{min}}$
and zero $M_{\mathrm{env},\mathrm{min}}$.
\item Common envelope efficiency $\alpha=1$ according to the prescription
of \citet{2002MNRAS_329_897H}. The common-envelope structure parameter
$\lambda_{\mathrm{CE}}$ is fitted to the detailed models of \citet{2000A&A...360.1043D}.
We do not include accretion on to the secondary star during the common
envelope phase by default but allow up to $0.05\mathrm{\, M_{\odot}}$
to be accreted in some model sets. We also consider the alternative
common-envelope prescription of \citet{2005MNRAS.356..753N}.
\item The $^{13}\mathrm{C}$ pocket efficiency $\xi_{13}$ is set to $1$
by default as defined by Eq.~A.10 of \citet{2006A&A...460..565I}.
\item The efficiency of the Companion Reinforced Attrition Process (CRAP,
\citealp{1988MNRAS.231..823T}) $B$ is set to zero by default.
\end{itemize}

\subsubsection{Missing physics}

\label{sub:Missing-physics}Our synthetic models do not include any
extra mixing which may be responsible for the conversion of $^{12}\mathrm{C}$
to $^{13}\mathrm{C}$ and $^{14}\mathrm{N}$ in low-mass stars (see
e.g. \citealp*{2003ApJ...582.1036N}; \citealp{2007ApJ...671..802B};
\citealt*{2008ApJ...677..581E} and references therein). Also, we
do not include any prescription which describes mixing events induced
by proton ingestion at the helium flash (the \emph{dual core flash})
or during thermal pulses (the \emph{dual shell flashes}, both of which
are also known as helium-flash-driven deep mixing; \citealp{1990ApJ...349..580F,2004A&A...422..217W,2007ApJ...667..489C,2008A&A...490..769C}).
Current stellar models suggest these events occur only if $[\mathrm{Fe}/\mathrm{H}]\lesssim-3$
while our models have $[\mathrm{Fe}/\mathrm{H}]=-2.3$ and our observational
sample includes stars with $[\mathrm{Fe}/\mathrm{H}]=-2.3\pm0.5$.
The latest results of the Teramo group show that at the lowest masses,
around $0.8\mathrm{\, M_{\odot}}$ for $[\mathrm{Fe}/\mathrm{H}]=-2.3$
(with $Z_{\odot}\sim0.01$), a dual shell flash may occur at the beginning
of the TPAGB with significant C, N and $s$-process production (Cristallo,
private communication) -- we leave the analysis of such a case to
future work.

\subsection{\Change{An example CEMP system}}

\begin{figure*}
\includegraphics[angle=270,scale=0.65]{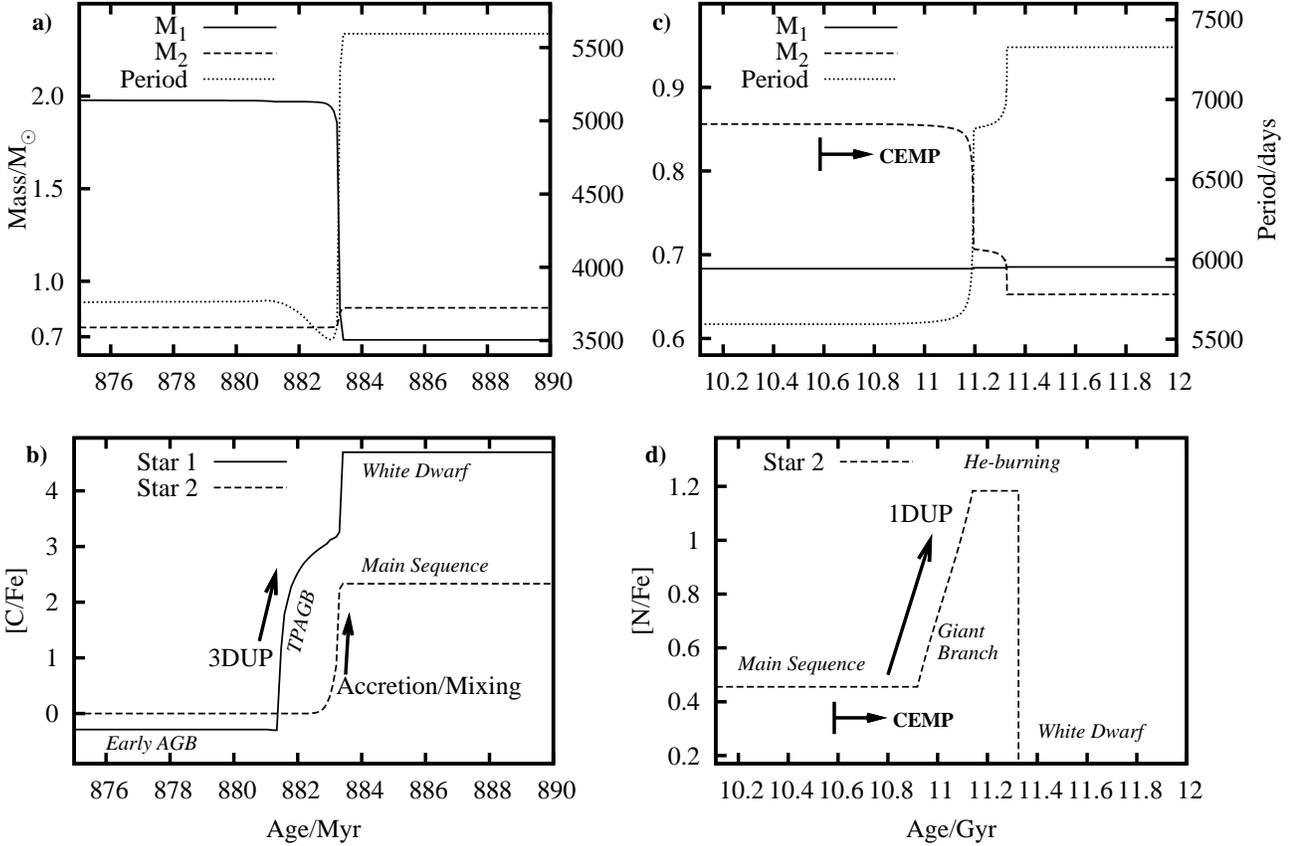}

\caption{\label{fig:example-CEMP}\Change{An example CEMP system with our
default parameter choices as in model set \protect \modelset{1}.
Initially $M_{1}=2\mathrm{\, M_{\odot}}$, $M_{2}=0.75\mathrm{\, M_{\odot}}$
and $P=3700\,\mathrm{days}$. Panels \emph{a }and \emph{b} show the
evolution of the primary up the AGB and accretion onto the secondary.
Panels \emph{c }and \emph{d }show the evolution of the secondary through
the CEMP phase. In panels \emph{a }and \emph{c }the left ordinate
denotes the mass of the stars in $\mathrm{M_{\odot}}$ (solid and
dashed lines for the primary and secondary respectively), the right
ordinate gives the orbital period in days (dotted lines). In panel
\emph{b} the left ordinate shows $[\mathrm{C}/\mathrm{Fe}]$ while
in panel \emph{d }the left axis shows $[\mathrm{N}/\mathrm{Fe}]$.
In all panels the abscissa\emph{ }shows the age of the system. Labels
in italic show the evolution phases, labels in Roman type show the
dredge up, mixing and accretion phases, and arrows in panels \emph{c
}and \emph{d }indicate when the star is tagged as a CEMP star according
to our selection criteria (older than $10\,\mbox{Gyr}$, $\log g<4$
and $[\mbox{C}/\mbox{Fe}]>1$, see Section~\ref{sub:model-Selection-criteria}).}}

\end{figure*}

\Change{We preempt the results of Section~\ref{sec:Results} with
an example of our synthetic stellar evolution algorithm for a CEMP
system as shown in Fig.~\ref{fig:example-CEMP}. Initially it has
$M_{1}=2\mathrm{\, M_{\odot}}$, $M_{2}=0.75\mathrm{\, M_{\odot}}$,
$P=3700\,\mathrm{days}$ ($a=1400\mathrm{\, R_{\odot}}$), $Z=10^{-4}$,
 $e=0$ and otherwise has the default physics as described above.
The primary mass is chosen to illustrate the case of maximum carbon
accretion. After $880\,\mathrm{Myr}$ the primary evolves onto the
TPAGB. Third dredge up increases its surface carbon abundance to $\mathrm{[\mathrm{C}/\mathrm{Fe}]}=+3.2$
(Fig.~\ref{fig:example-CEMP}\emph{b}). Of its wind $0.11\mathrm{\, M_{\odot}}$
is accreted by and mixed into the main-sequence secondary such that
its mass increases to $0.86\mathrm{\, M_{\odot}}$ (Fig.~\ref{fig:example-CEMP}\emph{a}),
it has a carbon abundance $[\mathrm{C}/\mathrm{Fe}]=+2.3$ and nitrogen
abundance $[\mathrm{N}/\mathrm{Fe}]=+0.35$ (Fig.~\ref{fig:example-CEMP}\emph{b}).
The binary orbit expands to a period of $5500\,\mathrm{days}$ (Fig.~\ref{fig:example-CEMP}\emph{a}).
At this moment the secondary becomes a CEMP star, although in our
population synthesis we do not count it as such until it has evolved
to $\log g<4$ (see Section~\ref{sub:model-Selection-criteria}),
which occurs after $10.59\,\mbox{Gyr}$.}

\Change{Subsequently the secondary also ascends the giant branch
twice (Fig.~\ref{fig:example-CEMP}\emph{c }and \emph{d}). During
the first ascent, dredge up reduces the surface abundance of carbon
only slightly but increases $[\mathrm{N}/\mathrm{Fe}]$ to $+1.2$
when burnt material is mixed to the surface. The orbit expands when
the secondary loses $0.15\mathrm{\, M_{\odot}}$ through its stellar
wind at the tip of the giant branch (Fig.~\ref{fig:example-CEMP}\emph{c}).
At $11.3\,\mathrm{Gyr}$ it too ascends the AGB with a mass of $0.7\mathrm{\, M_{\odot}}$
and ceases to be a CEMP star when its envelope is lost in a wind and
it becomes a white dwarf. The system is then a pair of carbon-oxygen
white dwarfs with a period of about $20\,\mathrm{years}$.} 

\Change{Further examples of our stellar evolution and nucleosynthesis
algorithms and comparisons to detailed stellar evolution models are
given in \citet{Izzard_et_al_2003b_AGBs,2006A&A...460..565I}.}

\subsection{Population synthesis}

\label{sub:Population-Synthesis}Each of our population synthesis
simulations consists of $N^{3}$ stars in $\ln M_{1}$-$\ln M_{2}$-$\ln a$
parameter space, where $M_{1,2}$ are the initial masses of the primary
and secondary, $a$ is the initial separation and $N=128$. We then
count the number of stars of a particular type according to the sum\begin{eqnarray}
n_{\mathrm{type}}\!\! & =\! & S\hspace{-2mm}\sum_{M_{\mathrm{1,min}}}^{M_{\mathrm{1,max}}}\!\sum_{M_{2,\mathrm{min}}}^{M_{2,\mathrm{max}}}\!\sum_{a_{\mathrm{min}}}^{a_{\mathrm{max}}}\!\sum_{t_{\mathrm{min}}}^{t_{\mathrm{max}}}\delta(\mathrm{type})\,\Psi\,\delta M_{1}\,\delta M_{2}\,\delta a\,\delta t\,,\label{eq:popsyn-n-sums}\end{eqnarray}
where $S$ is the star formation rate, $\Psi$ is the initial distribution
function and $\delta(\mathrm{type})=1$ when a star is of the required
type and zero otherwise. The grid cell size is given by $\delta M_{1}\cdot\delta M_{2}\cdot\delta a$
while the timestep is $\delta t$. We further assume that $\Psi$
is separable, \begin{eqnarray}
\Psi & = & \psi(M_{1})\,\phi(M_{2})\,\chi(a)\,,\label{eq:Psi}\end{eqnarray}
 where the functions $\psi(M_{1})$, $\phi(M_{2})$ and $\chi(a)$
are the initial distributions of $M_{1}$, $M_{2}$ and $a$ respectively
(see Section~\ref{sub:Stellar-distributions}). The star formation
rate $S$ is a function of time or, given an age-metallicity relation,
metallicity. Because we calculate ratios of numbers of stars with
the same metallicity, i.e. the same age and same $S$, the star formation
rate simply cancels out.

We set the limits of our population synthesis grid as follows:
\begin{itemize}
\item $M_{1,\mathrm{min}}=0.7\mathrm{\, M_{\odot}}$ -- stars less massive
than this do not evolve off the main sequence within the lifetime
of the Universe. $M_{\mathrm{1},\mathrm{max}}=8\mathrm{\, M_{\odot}}$
-- we do not consider massive stars which end their lives as supernovae.
\item $M_{2,\mathrm{min}}=0.1\mathrm{\, M_{\odot}}$ -- There is no simple
limit on $M_{2,\mathrm{min}}$ as a secondary star may accrete an
arbitrary fraction of the mass lost from its companion. However, with
the Bondi-Hoyle accretion formalism the accreted mass is typically
less than $0.2\mathrm{\, M_{\odot}}$, so we catch all possible CEMP
stars (which must have masses around $0.8\mathrm{\, M_{\odot}}$).
$M_{2,\mathrm{max}}=0.9\mathrm{\, M_{\odot}}$ -- stars more massive
than this have already evolved to white dwarfs after $\sim10\,\mathrm{Gyr}$
and so cannot be CEMP stars. 
\item $a_{\mathrm{min}}=3\mathrm{\, R_{\odot}}$, typically $a_{\mathrm{max}}=10^{5}\mathrm{\, R_{\odot}}$.
The upper limit is chosen to include all CEMP stars made in our models.
Our assumption is that \emph{all} stars are made in this range of
separations, i.e. a binary fraction of $100\%$. In reality some systems
are wider and/or single. This can easily be accommodated by lowering
the binary fraction.
\item $t_{\mathrm{min}}=10\,\mathrm{Gyr}$ -- Our stars must be old halo
stars, so a $10\,\mathrm{Gyr}$ limit is reasonable. Our results are
not sensitive to changes of $\pm2\,\mathrm{Gyr}$ in this limit. The
age of the Universe gives the upper limit $t_{\mathrm{max}}=13.7\,\mathrm{Gyr}$.
\end{itemize}
Selection criteria for $\delta(\mathrm{type})$ are given in Section~\ref{sub:model-Selection-criteria}.

\subsection{Stellar distributions}

\label{sub:Stellar-distributions}The choice of distributions of initial
primary mass $M_{1}$, initial secondary mass $M_{2}$ (or, alternatively,
$q=M_{2}/M_{1}$) and initial separation $a$ (or initial period $P$)
affects the final number counts and distribution of CEMP parameters.
Unfortunately, in the Galactic halo \emph{all} the relevant initial
distributions are unknown. We are forced to assume solar neighbourhood
distributions:
\begin{itemize}
\item The primary mass distribution $\psi(M_{1})$ is the initial mass function
of \citet*[KTG93]{KTG1993MNRAS-262-545K}.
\item The secondary mass distribution $\phi(M_{2})$ is flat in $q=M_{2}/M_{1}$,
i.e. any mass ratio is equally likely. 
\item The separation distribution $\chi(a)$ is flat in $\ln a$ between
$a_{\mathrm{min}}$ and $a_{\mathrm{max}}$ (i.e. $\chi(a)\sim1/a$). 
\item We start all binaries in circular orbits, i.e. eccentricity $e=0$,
and assume a binary fraction of $100\%$.
\end{itemize}

\subsection{Model sets}

\label{sub:modelsets}As described in Section~\ref{sub:Default-physical-model},
many physical parameters associated with binary evolution are uncertain.
Table~\ref{tab:modelsets} lists the most important parameter sets
we consider and how they differ from our default model set (set \protect \modelset{1},
with parameter values defined in Section~\ref{sub:Default-physical-model},
see also Table~\ref{tab:The-full-list-of-population-parameters}
for a list of all the model sets considered, including those not discussed
in detail). In some model sets, e.g. \modelset{29} and \modelset{30},
we altered the CEMP selection criteria rather than the physical parameters.
All our model sets use the initial distributions outlined in Section~\ref{sub:Stellar-distributions}
(except model set \modelset{7} with initial eccentricity $e=0.5$).
\begin{table}
\begin{centering}
\begin{tabular}{|c|c|}
\hline 
Model set & %
\begin{minipage}[t][1.1\totalheight]{6cm}%
Physical parameters \\
(differences from model set \protect \modelset{1})%
\end{minipage}\tabularnewline
\hline
\modelset{1} & -\tabularnewline
\modelset{29}, \modelset{30} & $\left[\mathrm{C}/\mathrm{Fe}\right]_{\mathrm{min}}=0.5$ and $0.7$
respectively\tabularnewline
\modelset{17}, \modelset{18} & $\xi_{13}=0.1$ and $0.01$ respectively\tabularnewline
\modelset{6} & Bondi-Hoyle efficiency $\alpha_{\mathrm{BH}}=5$\tabularnewline
\modelset{16} & common envelope accretion $0.05\mathrm{\, M_{\odot}}$\tabularnewline
\modelset{19} & no thermohaline mixing\tabularnewline
\modelset{9} & $\Delta M_{\mathrm{c,min}}=-0.07\mathrm{\, M_{\odot}}$, $\lambda_{\mathrm{min}}=0.8$\tabularnewline
\modelset{26} & $M_{\mathrm{env},\mathrm{min}}=0\mathrm{\, M_{\odot}}$\tabularnewline
 & \vspace{-3mm}\tabularnewline
 \modelset{49} & %
\begin{minipage}[t][1.1\totalheight]{6cm}%
\begin{center}
$\Delta M_{\mathrm{c,min}}=-0.1\mathrm{\, M_{\odot}}$, $\lambda_{\mathrm{min}}=0.8$,\\
$M_{\mathrm{env},\mathrm{min}}=0\mathrm{\, M_{\odot}}$, $\xi_{\mathrm{13}}=0.1$
\par\end{center}%
\end{minipage}\tabularnewline
 & \vspace{-3mm}\tabularnewline
\modelset{54} & %
\begin{minipage}[t][1.1\totalheight]{6cm}%
\begin{center}
as model \modelset{49}, with: \\
common envelope accretion $0.05\mathrm{\, M_{\odot}}$\\
no thermohaline mixing
\par\end{center}%
\end{minipage}\tabularnewline
\hline
\end{tabular}
\par\end{centering}

\caption{\label{tab:modelsets}Physical parameters corresponding to the most
important of our binary population models. A full list of all our
model sets can be found in Table~\ref{tab:The-full-list-of-population-parameters}.}

\end{table}

\subsection{Model selection criteria}

\label{sub:model-Selection-criteria}Stars are selected from our model
population as metal poor (EMP) giants if their surface gravity $\log_{10}g\leq4$
and they are older than $10\,\mathrm{Gyr}$ corresponding to the approximate
age of the Galactic halo. We then define the following subtypes:
\begin{description}
\item [{$\mathrm{\mathbf{CEMP}}$}] Surface carbon abundance $[\mathrm{C}/\mathrm{Fe}]\geq[\mathrm{C}/\mathrm{Fe}]_{\mathrm{min}}$,
where $[\mathrm{C}/\mathrm{Fe}]_{\mathrm{min}}=1.0$ by default. If
in addition $[\mathrm{Ba}/\mathrm{Fe}]\geq0.5$ the star is classified
as CEMP-$s$.
\item [{$\mathrm{\mathbf{NEMP}}$}] $[\mathrm{N}/\mathrm{Fe}]\geq1.0$
and $[\mathrm{C}/\mathrm{N}]<-0.5$. The former is somewhat more restrictive
than the criterion $[\mathrm{N}/\mathrm{Fe}]>0.5$ applied by \citet{2007ApJ...658.1203J}
and more consistent with the CEMP criterion. We find that with the
\citet{2007ApJ...658.1203J} criterion normal giants that have undergone
strong CN cycling, presumably due to extra mixing on the giant branch
which is not included in our models, enter our observational selection
(see Section~\ref{sec:Observational-database}). Therefore we define
the NEMP class to contain only stars affected by third dredge-up and
HBB and subsequent mass transfer, i.e. the nitrogen-rich equivalents
of CEMP stars.
\item [{$\mathrm{\mathbf{FEMP}}$}] Surface fluorine abundance $\mathrm{\left[\mathrm{F}/\mathrm{Fe}\right]}\geq1.0$
\citep{2008A&A...484L..27L}.
\end{description}
These three subtypes are not mutually exclusive. It turns out that
in our models the FEMP and CEMP subtypes nearly coincide (see Section~\ref{sub:default-C-and-F}).
The CEMP and NEMP classes also partially overlap, these are designated
as CNEMP.

\section{Observational database and selection criteria}

\label{sec:Observational-database}%
\begin{figure*}
\begin{centering}
\begin{tabular}{cc}
\includegraphics[scale=0.33,angle=270]{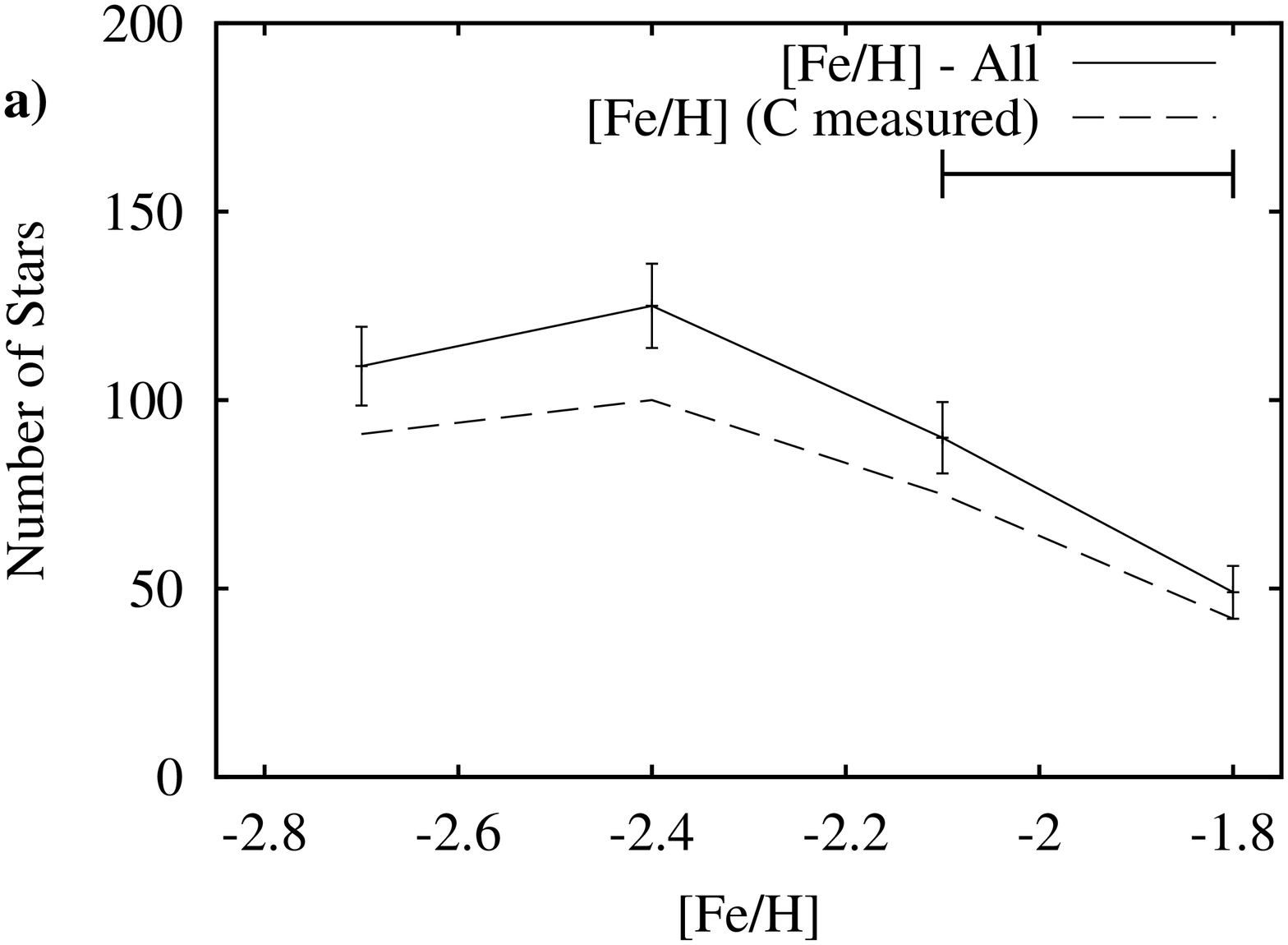} & \includegraphics[scale=0.33,angle=270]{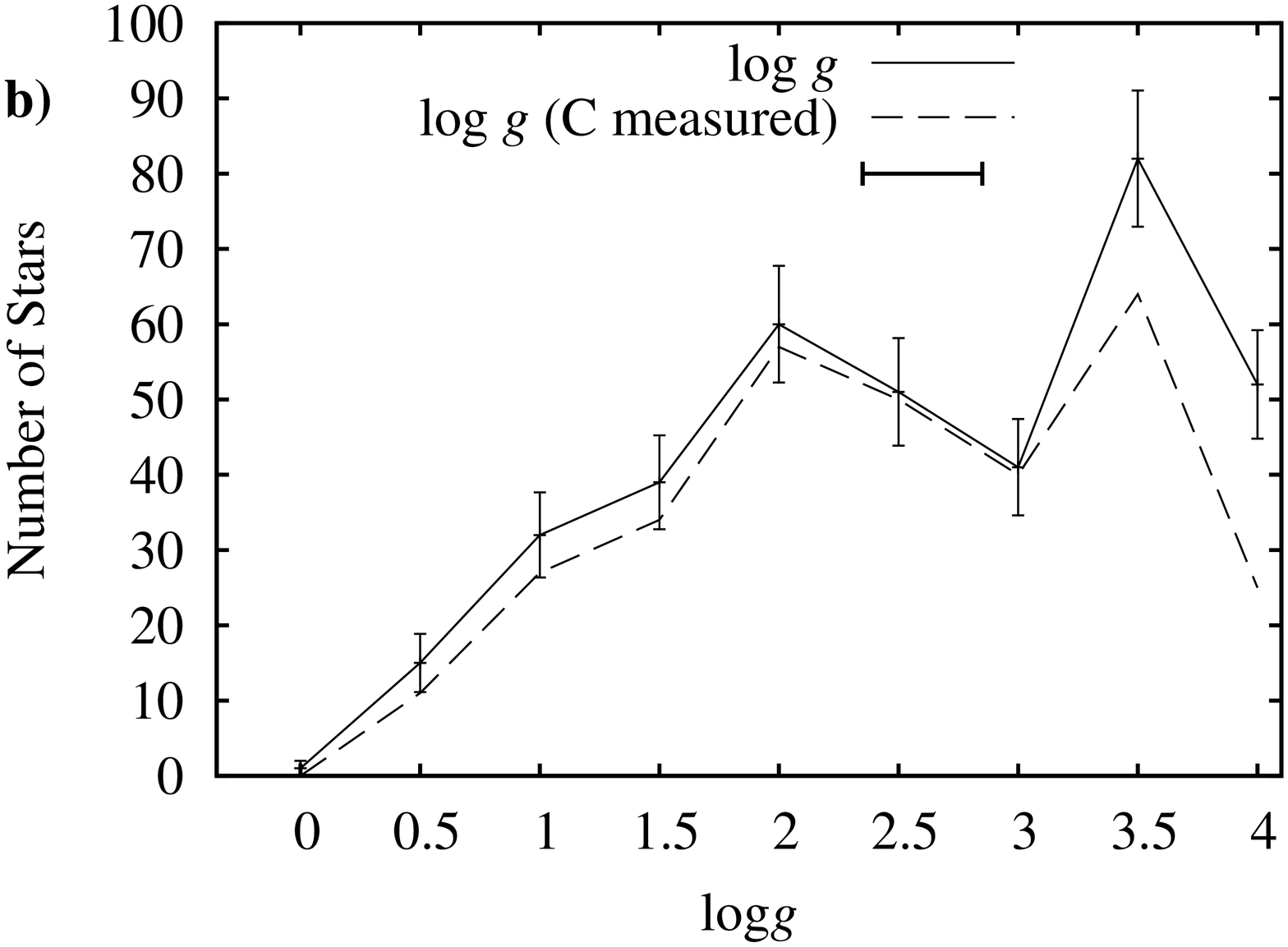}\tabularnewline
\end{tabular}
\par\end{centering}

\caption{\label{fig:SAGA-distributions}Selected distributions from our sample
of the SAGA database. All stars in our selection have $\log g\leq4$
and $[\mathrm{Fe}/\mathrm{H}]=-2.3\pm0.5$. \ChangeA{ panels \textbf{a}
and \textbf{b} show} the distributions of $[\mathrm{Fe}/\mathrm{H}]$
(the metallicity distribution function) and $\log g$ \ChangeA{respectively}
for the sets of 1) all stars in our selection and 2) those stars in
our selection with a carbon measurement. \Change{The horizontal error
bars, and hence the bin widths, are typical $1\sigma$ errors for
a single observation, $\pm0.15\,\mathrm{dex}$ in $[\mathrm{Fe}/\mathrm{H}]$
and $\pm0.25\,\mathrm{dex}$ in $\log g$.} }

\end{figure*}

Our database of observed EMP stars is based on 2376 observations of
1300 stars in the SAGA database%
\footnote{With the addition of $\log g$ values from \citet[Beers, private communication]{2006ApJ...652L..37L}.%
} \citep{2008PASJ...60.1159S}. \Change{When observed values of a
parameter are available from multiple sources for one star, we simply
use the arithmetic mean of these values in our database} (see Appendix~\ref{sec:appendix-Observation-Database}
for details). We ignore observations which provide only an upper or
lower limit. We select stars which correspond to our model EMP giants
and turn-off stars as follows:
\begin{enumerate}
\item The observed star must have metallicity in the range $[\mathrm{Fe}/\mathrm{H}]=-2.3\pm0.5\,\mathrm{dex}$.
Fig.~\ref{fig:SAGA-distributions}a shows that the number distribution
of stars in this range varies by a factor of two with no clear trend
in number as a function of $[\mathrm{Fe}/\mathrm{H}]$.
\item The star must be a giant or sub-giant, i.e. $\log g\leq4.0$. Fig.~\ref{fig:SAGA-distributions}b
shows the distribution of the number of stars as a function of $\log g$.
\end{enumerate}
This selection leaves us with $373$ stars, of which $308$ have a
measured carbon abundance and $96$ have $[\mathrm{C}/\mathrm{Fe}]\geq1$
-- these are our CEMP stars%
\footnote{When $[\mathrm{C}/\mathrm{Fe}]$ is not available but $[\mathrm{CH}/\mathrm{Fe}]$
is, we use $[\mathrm{CH}/\mathrm{Fe}]$ as a proxy for $[\mathrm{C}/\mathrm{Fe}]$.%
}. The CEMP to EMP fraction, for stars \emph{with measured carbon},
is thus $31\%$. Fig.~\ref{fig:SAGA-distributions}a shows that whether
a star has measured carbon, or not, does not depend on metallicity
but is sensitive to $\log g$. In particular, $90\%$ of stars with
$\log g\leq3.5$ have carbon measurements, whereas about half of the
highest-gravity stars ($\log g>3.5$) do not. Presumably this is because
of observational difficulties, these stars being relatively dim and
hot.

If we assume that stars which have no carbon measurement actually
have no carbon enhancement, i.e. $[\mathrm{C}/\mathrm{Fe}]<1$, the
CEMP/EMP ratio drops to $26\%$. The latter statistic is best for
comparison with our models, but it rests on a possibly dubious assumption:
are stars with \emph{no} carbon measurement \emph{really} not enhanced
in carbon? Certainly the CEMP fraction depends on how it is counted,
as shown in Table~\ref{tab:Observations-table}. The CEMP fraction
also varies depending on the survey under consideration: \citet{2006ApJ...652.1585F},
\citet{2005ApJ...633L.109C} and \citet{2006ApJ...652L..37L} find
$9$, $14$ and $21\%$ respectively.

Application of the NEMP criteria%
\footnote{We use $[\mathrm{CN}/\mathrm{Fe}]$ or $[\mathrm{NH}/\mathrm{Fe}]$
as proxies for $[\mathrm{N}/\mathrm{Fe}]$ when $[\mathrm{N}/\mathrm{Fe}]$
is not available in the SAGA database.%
}, $[\mathrm{N}/\mathrm{Fe}]\geq1$ and $[\mathrm{C}/\mathrm{N}]<-0.5$,
leaves us with zero NEMP stars in the $[\mathrm{Fe}/\mathrm{H}]$
range considered. This is to some extent a statistical fluke, because
NEMP stars are found both at higher (in the case of HE0400-2040) and
especially at lower metallicity. However, even when counting all stars
regardless of $[\mathrm{Fe}/\mathrm{H}]$ the number of NEMP stars
is small compared to the total number of EMP or CEMP stars (see Table~\ref{tab:Observations-table}).
Use of the less restrictive $[\mathrm{N}/\mathrm{Fe}]\geq0.5$ criterion
would introduce seven NEMP stars into the sample, but these are mostly
carbon-depleted giants that have presumably undergone strong CN-cycling
during their RGB evolution, as discussed in Section~\ref{sub:model-Selection-criteria}.

\begin{figure*}
\includegraphics[bb=50bp 50bp 554bp 770bp,scale=0.34,angle=270]{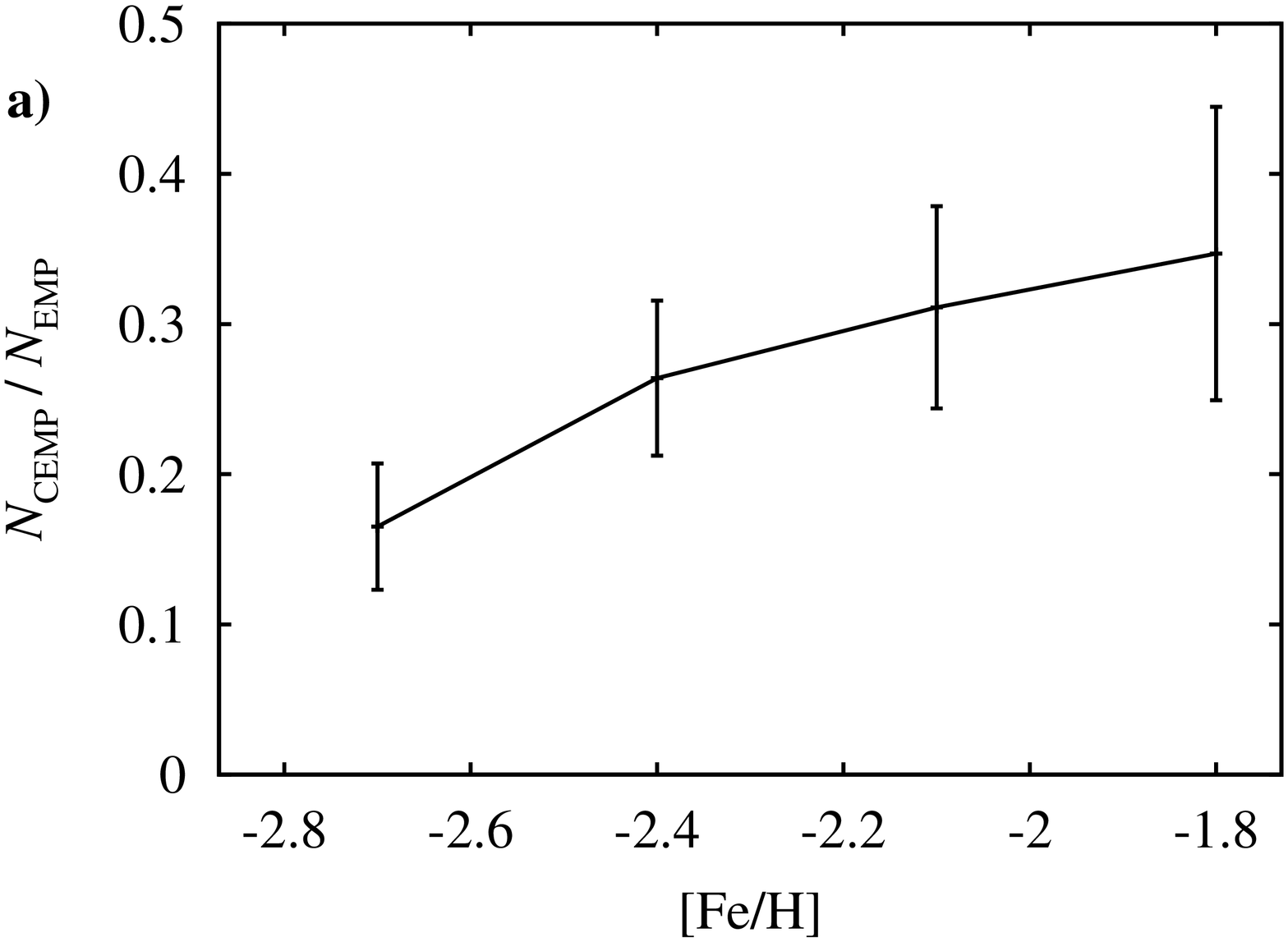}\includegraphics[bb=50bp -50bp 554bp 770bp,scale=0.34,angle=270]{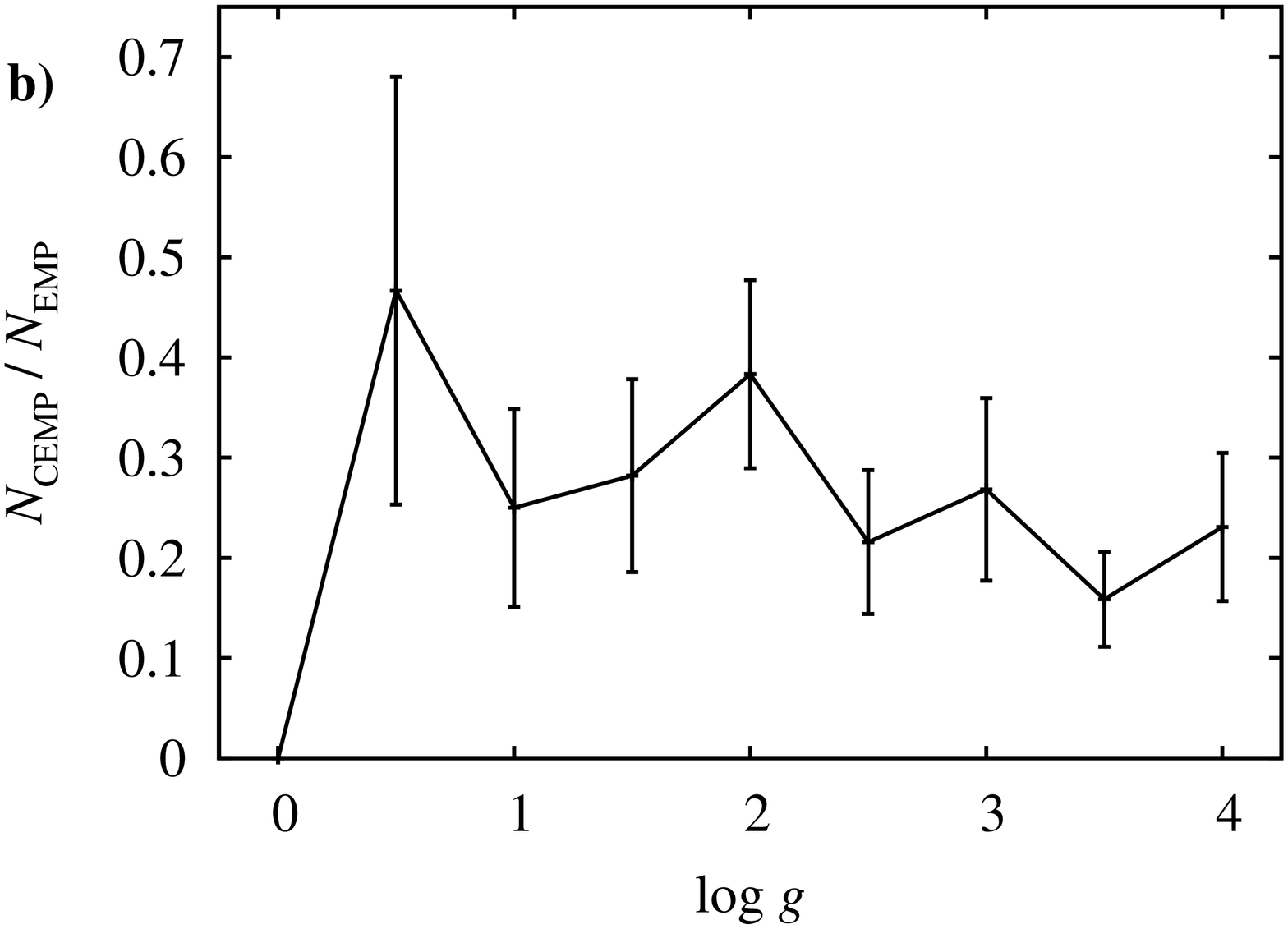}

\caption{\label{fig:CEMP-frac-as-f-logg-and-FeH}The CEMP/EMP ratio for our
selection from the SAGA database ($[\mathrm{Fe}/\mathrm{H}]=-2.3\pm0.5$
and $\log g<4$) as\textbf{ }\protect \\
\textbf{a)} a function of metallicity $[\mathrm{Fe}/\mathrm{H}]$
and \textbf{}\protect \\
\textbf{b)} a function of gravity $\log g$. \protect \\
Error bars are based on Poisson ($\sqrt{N}$) statistics\Change{
and bin widths are as in Fig.~\ref{fig:SAGA-distributions}}.}

\end{figure*}

\begin{table*}
\begin{tabular}{|c|c|c|c|c|c|c|c|l|}
\hline 
{\scriptsize Metallicity} & {\scriptsize $\log g$} & {\scriptsize EMP} & {\scriptsize CEMP} & {\scriptsize CNEMP} & {\scriptsize NEMP} &  & {\scriptsize CEMP/EMP} & {\scriptsize Additional Selection Criteria}\tabularnewline
{\scriptsize $[\mathrm{Fe}/\mathrm{H}]$} &  &  &  &  & {\scriptsize $[\mathrm{N}/\mathrm{Fe}]>1$} & {\scriptsize $[\mathrm{N}/\mathrm{Fe}]>0.5$} & {\scriptsize fraction} & \tabularnewline
\hline
$-2.3\pm0.5$ & $\leq4$ & 308 & 96 &  &  &  & 31\% & {\scriptsize only stars with C measured}\tabularnewline
$-2.3\pm0.5$ & $\leq4$ & 373 & 96 &  &  &  & 26\% & {\scriptsize all EMP stars}\tabularnewline
$-2.3\pm0.5$ & $\leq4$ & 104 & 88 & 0 & 0 & 7 &  & {\scriptsize only stars with C and N measured}\tabularnewline
$-2.3\pm0.5$ & $\leq4$ & 373 & 96 & 0 & 0 & 7 & 26\% & {\scriptsize $\left\{ \begin{array}{c}
\mathrm{C\, and\, N\, measured\, for\,(C)NEMPs,}\\
\mathrm{C\, measured\, for\, CEMPs},\end{array}\right.$}\tabularnewline
all & $\leq4$ & 779 & 144 & 4 & 14 & 22 & 18\% & $\mathrm{"}$\tabularnewline
$-2.3\pm0.5$ & all & 479 & 115 & 0 & 0 & 7 & 24\% & $\mathrm{"}$\tabularnewline
all & all & 1366 & 177 & 6 & 17 & 23 & 13\% & $\mathrm{"}$\tabularnewline
\hline
$\leq2.0$ & giants & 132 & 11 &  &  &  & $9\pm2\%$ & \citet{2006ApJ...652.1585F}\tabularnewline
$\leq2.0$ &  & 270 & 58 &  &  &  & $>21\pm2\%$ & \citet{2006ApJ...652L..37L}\tabularnewline
$\leq2.0$ & giants &  &  &  &  &  & $14\pm4\%$ & \citet{2005ApJ...633L.109C}\tabularnewline
\hline
\end{tabular}

\caption{\label{tab:Observations-table}Observed numbers of EMP, CEMP, CNEMP
and NEMP stars from the SAGA database for our various selection criteria.
The first line of data counts only stars which have a $[\mathrm{C}/\mathrm{Fe}]$
measurement. The second line of data additionally counts stars \emph{without}
measured $[\mathrm{C}/\mathrm{Fe}]$ as EMPs, i.e. it is assumed that
$[\mathrm{C}/\mathrm{Fe}]<1$ for these stars. In the next four lines
stars with both measured C and N are used to count (C)NEMP stars while
stars with C measured count the CEMP stars. \protect \\
Also shown are estimates of the CEMP/EMP ratio from \citet{2005ApJ...633L.109C},
\citet{2006ApJ...652.1585F} and \citet{2006ApJ...652L..37L}. }

\end{table*}
\Change{Our selected sample may be biased, as shown in Fig. \ref{fig:CEMP-frac-as-f-logg-and-FeH}b.}
 \Change{The CEMP/EMP ratio increases slightly as $\log g$ decreases
but a constant value is consistent with the error bars. The CEMP
fraction may increase slightly with metallicity (Fig.~\ref{fig:CEMP-frac-as-f-logg-and-FeH}a)
at least over the narrow range we consider ($[\mbox{Fe}/\mbox{H}]=-2.3\pm0.5$).
If dual-core and/or dual-shell flashes occur exclusively for $[\mathrm{Fe}/\mathrm{H}]\lesssim-2.5$
and are responsible for the formation of most CEMP stars we expect
the CEMP fraction to drop as $[\mathrm{Fe}/\mathrm{H}]$ increases
beyond $-2.5$. This is not seen. We note that the SAGA database was
not designed to be complete in any statistical sense. We await a more
complete census of metal-poor stars before definite conclusions on
the CEMP fraction and its dependence on metallicity and evolutionary
stage, can be drawn.}

\section{Results and comparison with observations}

\label{sec:Results}The ratios of the number of CEMP and NEMP stars
to EMP stars as selected from our model sets of Table~\ref{tab:modelsets}
 are shown in Table~\ref{tab:percentages}. %
\begin{table*}
\begin{centering}
\begin{tabular}{|c|c|c|c|c|}
\hline
Model Set & CEMP/EMP $\%$ & NEMP/EMP $\%$ & FEMP/EMP $\%$ & CEMP-$s$/CEMP $\%$ \tabularnewline
\hline
$\modelset{1}$ & $2.30 \pm 0.04$ & $0.267 \pm 0.007$ & $2.03$ & $28.3$ \tabularnewline
$\modelset{30}$ & $2.99 \pm 0.04$ & $0.267 \pm 0.005$ & $2.03$ & $21.8$ \tabularnewline
$\modelset{29}$ & $3.46 \pm 0.04$ & $0.267 \pm 0.005$ & $2.03$ & $18.8$ \tabularnewline
$\modelset{17}$ & $2.30 \pm 0.04$ & $0.267 \pm 0.007$ & $2.03$ & $86.3$ \tabularnewline
$\modelset{18}$ & $2.30 \pm 0.04$ & $0.267 \pm 0.007$ & $2.03$ & $94.1$ \tabularnewline
$\modelset{6}$ & $3.05 \pm 0.04$ & $0.323 \pm 0.006$ & $2.74$ & $47.2$ \tabularnewline
$\modelset{16}$ & $2.94 \pm 0.04$ & $0.311 \pm 0.007$ & $2.36$ & $29.1$ \tabularnewline
$\modelset{19}$ & $4.21 \pm 0.04$ & $0.409 \pm 0.004$ & $3.74$ & $63.1$ \tabularnewline
$\modelset{9}$ & $2.90 \pm 0.04$ & $0.267 \pm 0.007$ & $2.43$ & $43.5$ \tabularnewline
$\modelset{26}$ & $6.47 \pm 0.03$ & $0.267 \pm 0.006$ & $5.16$ & $22.3$ \tabularnewline
$\modelset{49}$ & $9.43 \pm 0.04$ & $0.266 \pm 0.006$ & $7.81$ & $94.1$ \tabularnewline
$\modelset{54}$ & $15.52 \pm 0.07$ & $0.426 \pm 0.004$ & $13.5$ & $97.2$ \tabularnewline
\hline
\end{tabular}

\par\end{centering}

\caption{\label{tab:percentages}Percentage of CEMP, NEMP and FEMP (sub-)giants
relative to total EMP giants in our model binary populations of Table~\ref{tab:modelsets}
(see Section~\ref{sub:model-Selection-criteria} for selection criteria
and Table~\ref{tab:percentages-full} for the full set of results).
The errors convey Poisson statistics only. The final column gives
the number ratio of CEMP-$s$ to CEMP stars.}

\end{table*}
Our results fall into two categories:
\begin{enumerate}
\item Most of our model sets have a CEMP to EMP ratio of \Change{$2-4\%$}.
This is not much larger than the $1\%$ of more metal-rich giants
which are CH stars. We describe below one of these, our default model
set (model set~\modelset{1}), in detail.%
\begin{figure*}
\begin{tabular}{cc}
\put(5,-15){\LabelG{a}}\includegraphics[bb=80bp 50bp 600bp 770bp,scale=0.34,angle=270]{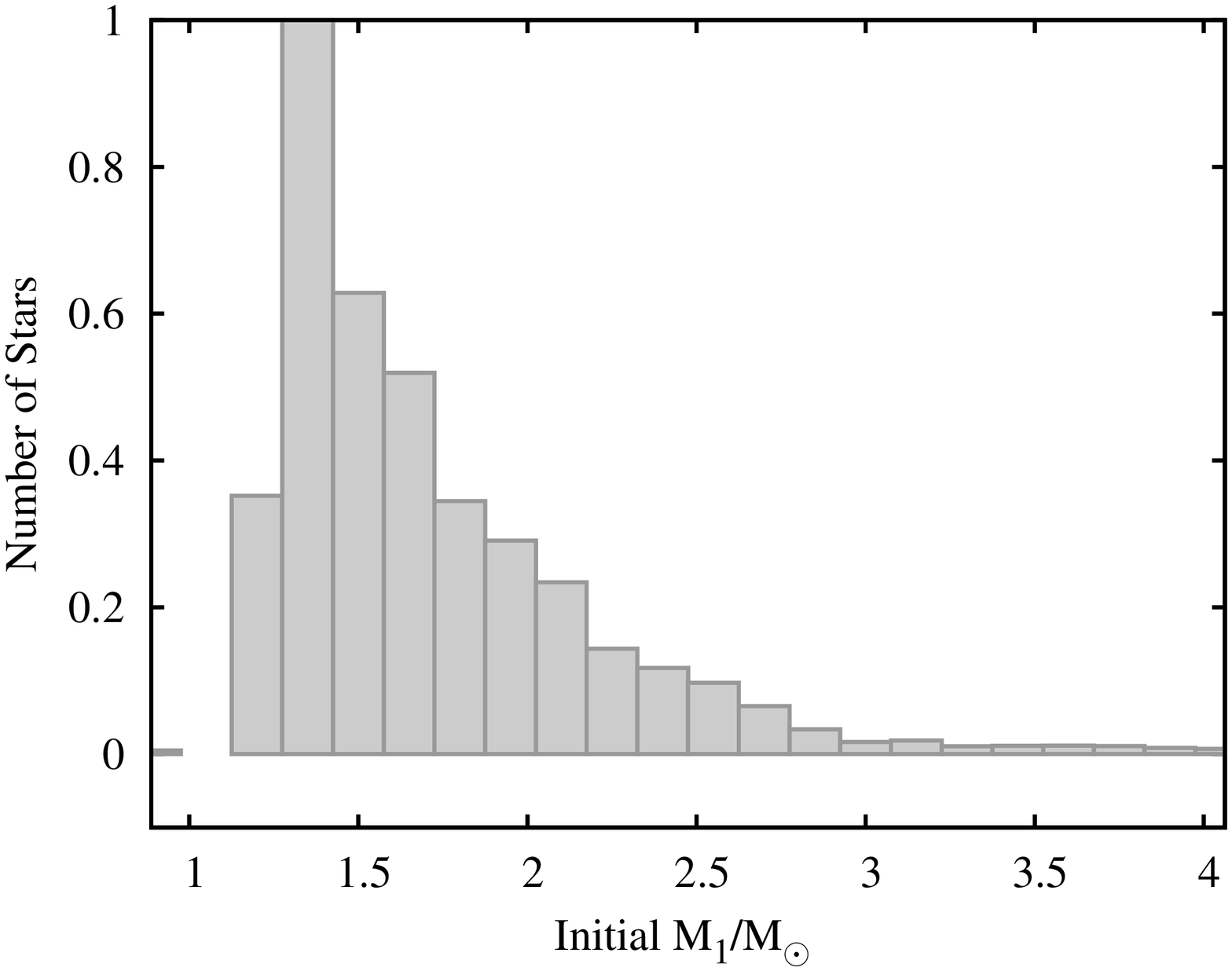} & \put(5,-15){\LabelG{b}}\includegraphics[bb=80bp 50bp 600bp 770bp,scale=0.34,angle=270]{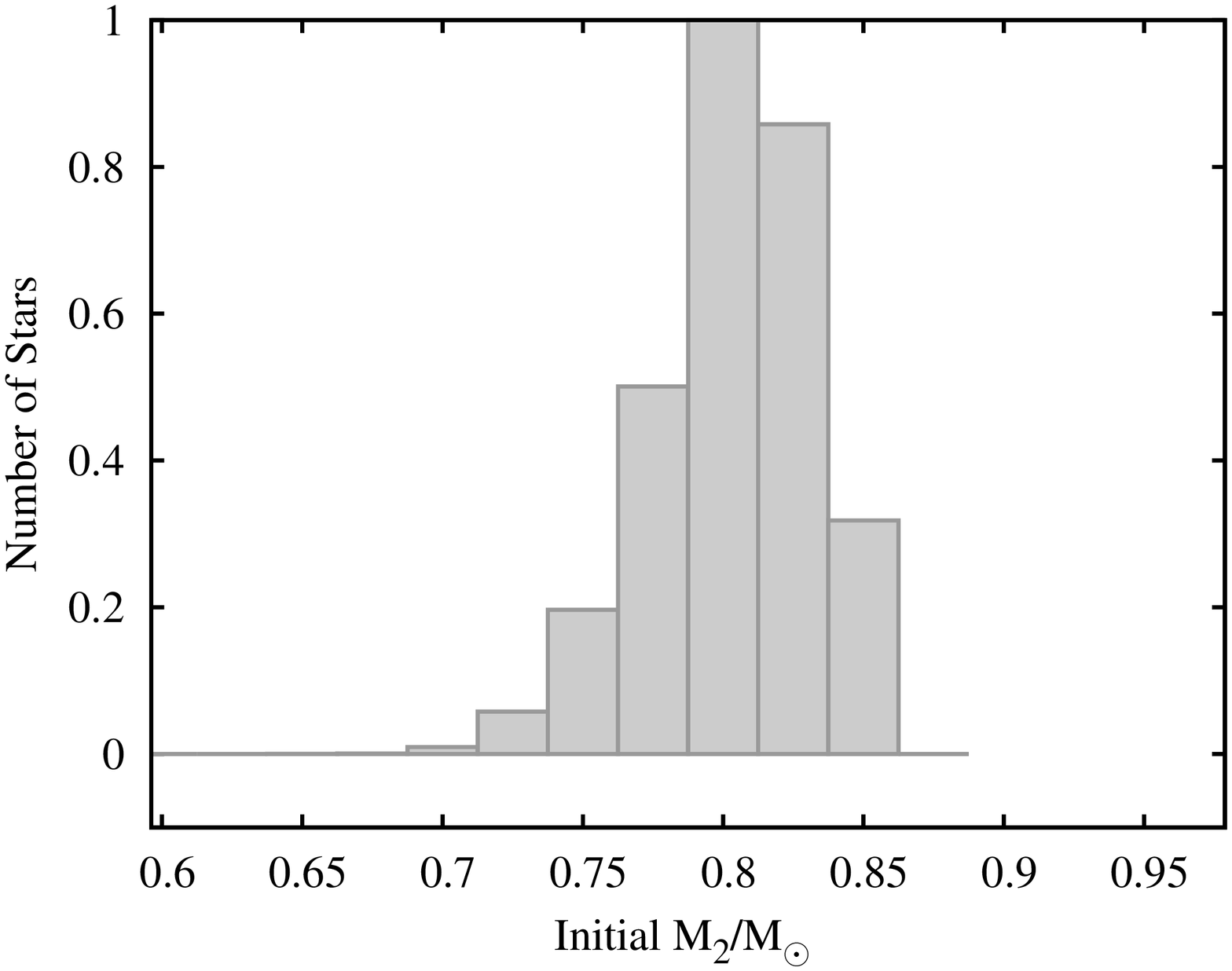}\tabularnewline
\put(5,-3){\LabelG{c}}\includegraphics[bb=120bp 50bp 554bp 770bp,scale=0.34,angle=270]{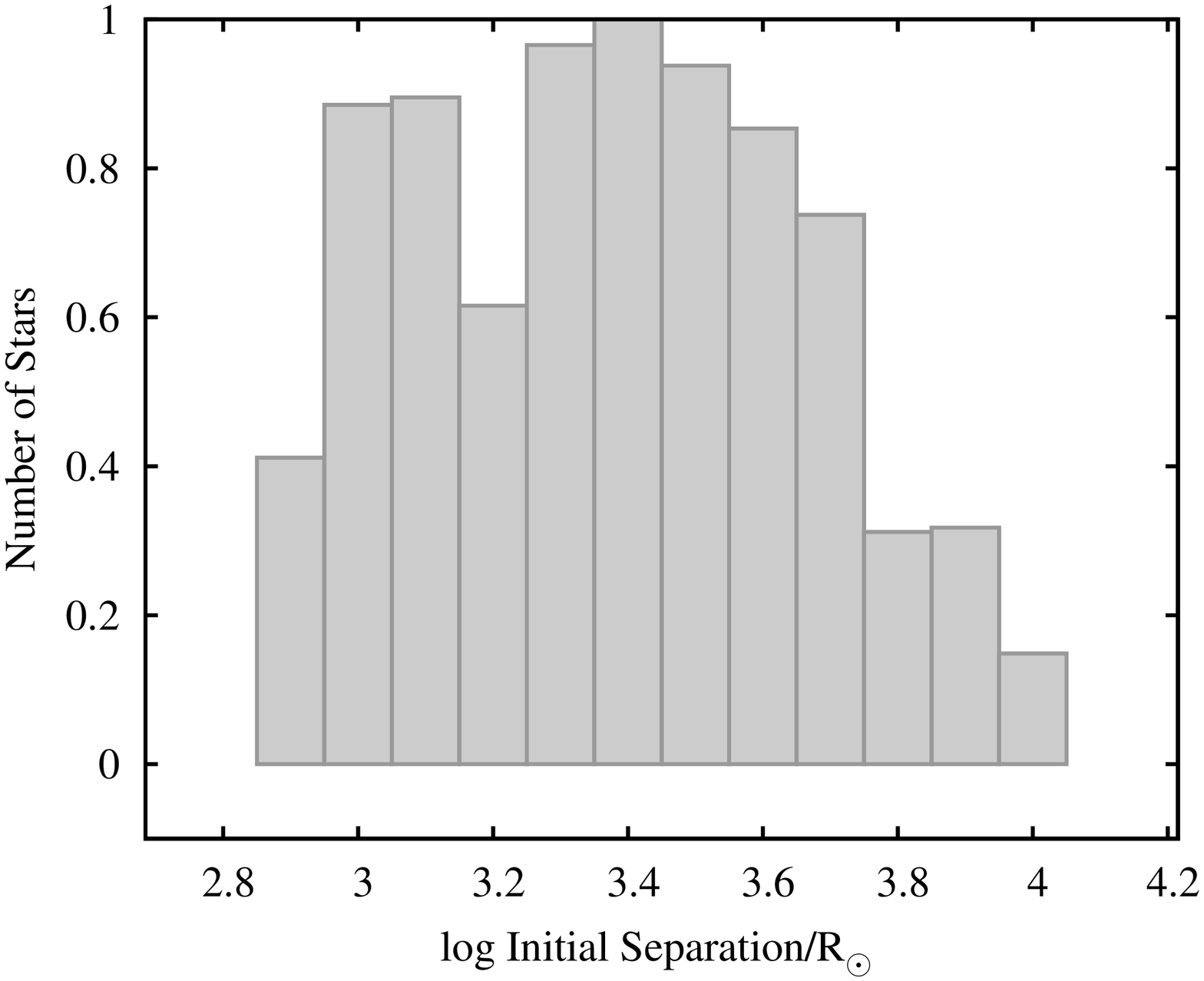} & \put(5,-3){\LabelG{d}}\includegraphics[bb=120bp 50bp 554bp 770bp,scale=0.34,angle=270]{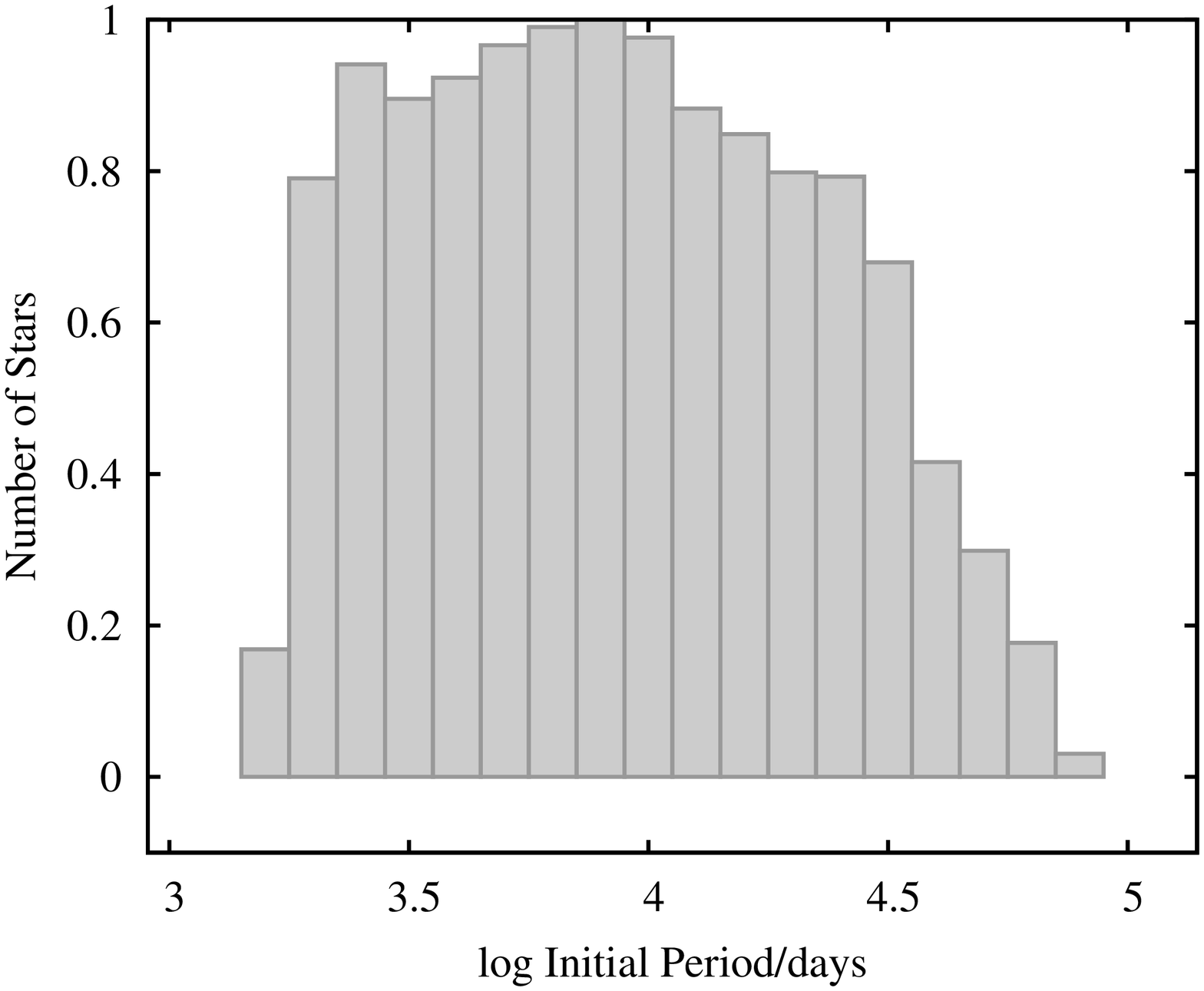}\tabularnewline
\end{tabular}

\caption{\label{fig:default-Initial-distributions}Time-weighted distributions
of initial masses, $M_{1}$, $M_{2}$, initial separation $a$ and
initial period $P$ in binaries which lead to the formation of a CEMP
star in our default model set \protect \modelset{1}. In this model
set almost all our CEMP stars are formed by accretion from primaries
with $M\gtrsim1.25\mathrm{\, M_{\odot}}$. The vertical scale, which
counts the number of stars per bin, is linear and arbitrarily normalized
to peak at one.}

\end{figure*}

\item A few sets come close to reproducing the observed CEMP to EMP ratio.
These are the sets with some combination of the following: (enhanced)
third dredge up in low-mass stars, no thermohaline mixing and accretion
in the common envelope phase. They represent a corner of the parameter
space which may be considered rather extreme, though not unfeasible.
We describe sets \modelset{49} and \modelset{54} in detail below.
\end{enumerate}

\subsection{Model set \protect \modelset{1}: Default physics}

\label{sub:default-physics-population}Our default model set -- set
\protect \modelset{1} -- represents a choice of physical parameters
which could be described as conservative. The parameters are not controversial
or extreme. As such they are a good starting point for our analysis
of the CEMP problem. 

First of all, the CEMP to EMP ratio in our default model set is $2.3\%$.
This is clearly at odds with the observed $9-25\%$ ratio, especially
if one factors in a binary fraction smaller than unity. However, the
number of NEMP stars in this simulation is small ($0.3\%$ of EMP
stars), which \emph{does} agree with the observations. In this section
we examine various properties of our default population with a view
to later sections which improve the match between our modelled CEMP
to EMP ratio and the observations.

\subsubsection{CEMP initial parameter space}

\label{sub:default-initial-parameter-space}The regions of the initial
$M_{1}-M_{2}-a\,\mathrm{or}\, P$ parameter space which form CEMP
stars are shown in Figure~\ref{fig:default-Initial-distributions}.
The distributions are time-weighted, as in Eq.~\ref{eq:popsyn-n-sums}.
The vast majority of our CEMP stars form via the wind-accretion channel
with a typical $M_{1}\sim1.2-1.5\mathrm{\, M_{\odot}}$, $M_{2}\sim0.8\mathrm{\, M_{\odot}}$,
separation around $10^{3}-10^{4}\mathrm{\, R_{\odot}}$ or, equivalently,
an orbital period of $10^{3}-10^{5}\,\mathrm{days}$. The initial
mass $M_{1}$ is limited at the low end by the minimum mass for third
dredge up ($\sim1.2\mathrm{\, M_{\odot}}$) and the distribution peters
out at the high end mainly as a result of the IMF. Stars with $M_{1}\gtrsim2.7\mathrm{\, M_{\odot}}$
undergo hot-bottom burning which results in $[\mathrm{C}/\mathrm{N}]<-1$.
These also classify as (C)NEMP stars. \Change{As mentioned in Section~\ref{sub:Physics}
some uncertainty exists regarding the mass of HBB onset which may
be greater than $2.7\mathrm{\, M_{\odot}}$. Because of the rapid
drop in the IMF with increasing mass, as shown in Figure~\ref{fig:default-Initial-distributions}a,
the number of CEMP stars affected by this uncertainty is small. An
increase in the HBB-onset mass reduces the number of NEMP stars.}

The secondary mass, $M_{2}\sim0.8\mathrm{\, M_{\odot}}$, is the mass
expected for a star that is approximately $10\,\mathrm{Gyr}$ old.
The shortest period (or separation) binary which forms a CEMP is limited
by the Roche limit. Closer binaries pass through a common-envelope
stage with little accretion on the secondary. The efficiency of wind
mass transfer drops as the initial separation increases. Beyond about
$10^{4}\mathrm{\, R_{\odot}}$ the secondary accretes too little carbon
to become a CEMP.

\subsubsection{The distribution of $\log g$ and potential selection effects}

\label{sub:default-logg-vs-obs}%
\begin{figure}
\includegraphics[scale=0.34,angle=270]{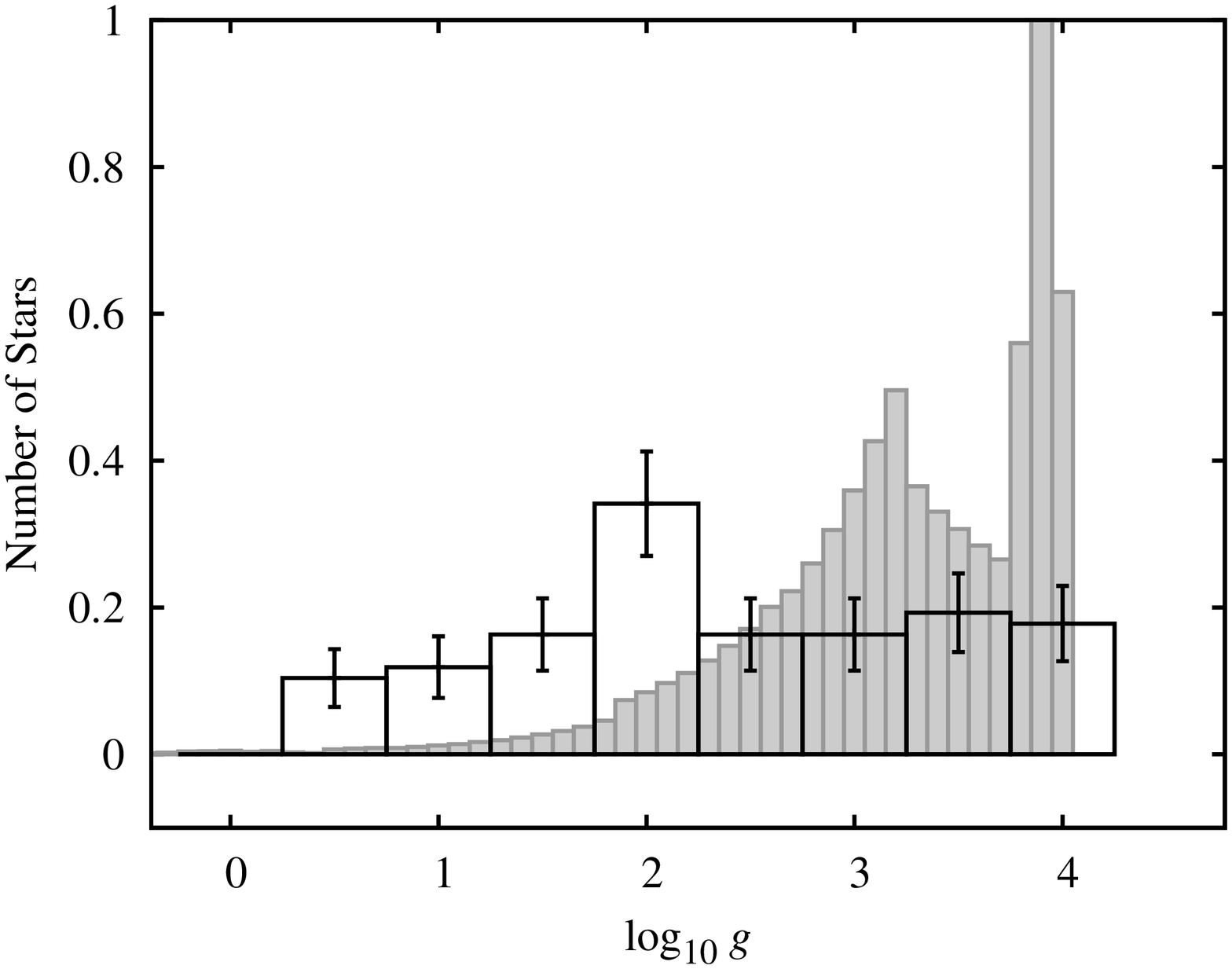}

\caption{\label{fig:logg-default-models-vs-obs}The distribution of $\log g$
in CEMP stars taken from our default stellar population model~\protect \modelset{1}
(filled histogram) vs our CEMP selection from the SAGA database (open
histogram with Poisson error bars).\protect \\
Note that in this and the similar plots that follow the area under
the graph (which is the total number of stars) is normalized such
that it is the same for both the observations and our model stars. }

\end{figure}

In Fig.~\ref{fig:logg-default-models-vs-obs} we compare the distribution
of $\log g$ in model set \protect \modelset{1} CEMP stars to our
selection from the SAGA database (see Fig.~\ref{fig:CEMP-frac-as-f-logg-and-FeH}b).
It is hard to understand why the distributions differ without invoking
selection effects such that low-gravity CEMP giants are preferred
(see Section~\ref{sec:Observational-database}). We have made no
attempt to model such selection effects but we do not expect this
to strongly affect our model predictions. The $\log g$ distribution
of non-CEMP stars in model set \protect \modelset{1} has a nearly
identical shape to the CEMP stars, so that the CEMP/EMP ratio for
our default population is independent of $\log g$ and hence the overall
CEMP fraction presented in Table~\ref{tab:percentages} is quite
robust. We also find that, at least in our default model as well as
other models that include thermohaline mixing of accreted material,
there is little surface abundance evolution in our CEMP stars after
accretion (with the exception of nitrogen, which increases somewhat
at first dredge up) so that the abundance distributions we present
in the following subsections are also hardly affected by this selection
bias.

Finally, we note that the peak in the model distribution at $\log g\sim3.2$
is due to horizontal branch stars, most of which have effective temperatures
close to $10,\!000\,\mathrm{K}$ in our models. The SAGA database
does not contain horizontal branch stars hotter than $\sim7,\!000\,\mathrm{K}$
probably because hotter stars are selected against. If we select only
red horizontal branch stars cooler than $7,\!000\,\mathrm{K}$ a small
peak at $\log g\sim2$ remains. This may correspond to the tentative
peak seen in the observed distribution.

\subsubsection{Carbon, nitrogen and fluorine}

\label{sub:default-C-and-F}
\begin{figure}
\includegraphics[scale=0.34,angle=270]{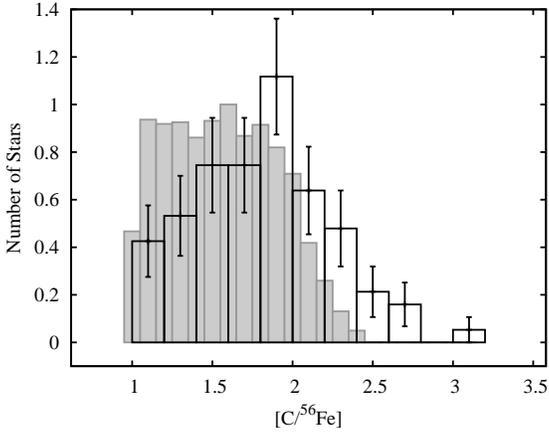}

\caption{\label{fig:C-Fe-default-population}The distribution of $[\mathrm{C}/\mathrm{Fe}]$
in our default CEMP population (model set \protect \modelset{1},
filled histogram) compared to observations (open histogram with Poisson
error bars).}

\end{figure}

The distribution of carbon in the CEMP
population of model set \protect \modelset{1} matches reasonably
well the observed distribution, as shown in Fig. \ref{fig:C-Fe-default-population}.%
 The observed distribution shows somewhat higher carbon enhancements
on average, while our models predict more stars in the range $1\lesssim[\mathrm{C}/\mathrm{Fe}]\lesssim1.5$
and none above $[\mathrm{C}/\mathrm{Fe}]=2.5$, compared to about
ten observed (11 per cent of the sample). This may be partly explained
by our assumption of complete thermohaline mixing: we assume the entire
star mixes, when in reality only a fraction mixes. 

\begin{figure}
\includegraphics[scale=0.34,angle=270]{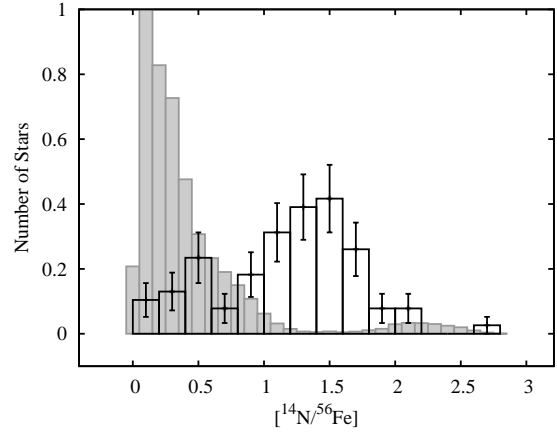}

\caption{\label{fig:N-Fe-default-population}The distribution of $[\mathrm{N}/\mathrm{Fe}]$
in our default CEMP population \protect \modelset{1} (filled histogram)
compared to observations (open histogram with Poisson error bars).}

\end{figure}
\begin{figure}
\includegraphics[bb=111bp 85bp 525bp 758bp,clip,angle=270,scale=0.43]{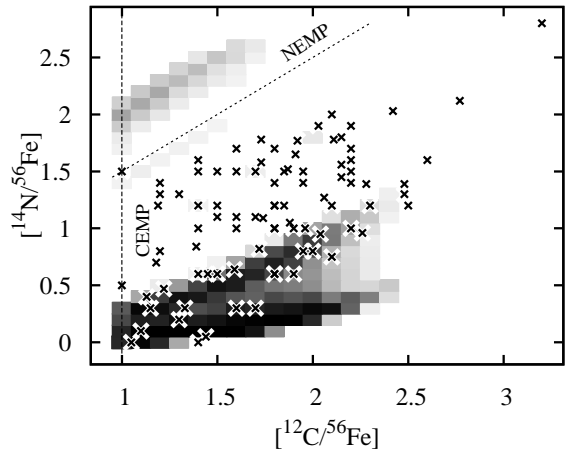}

\caption{\label{fig:C-N-default-population}The distribution of $[\mathrm{N}/\mathrm{Fe}]$
versus $[\mathrm{C}/\mathrm{Fe}]$ in our default CEMP population,
model set \protect \modelset{1} (darker grey indicates a larger density
of stars). The vertical dashed line indicates our CEMP selection criterion
($[\mathrm{C}/\mathrm{Fe}]\geq1$) and the diagonal dashed line shows
our NEMP selection criteria ($[\mathrm{N}/\mathrm{Fe}]\geq1$ and
$[\mathrm{N}/\mathrm{C}]>0.5$). Observed CEMP stars are indicated
by crosses. }

\end{figure}

The picture becomes much less favourable when we compare the distribution
of nitrogen, as shown in Figs.~\ref{fig:N-Fe-default-population}
and~\ref{fig:C-N-default-population}. Our default population contains
few (C)NEMPs (above the dashed line in Fig.~\ref{fig:C-N-default-population};
note that NEMPs with $[\mathrm{C}/\mathrm{Fe}]<1.0$ are not shown
in this figure), which agrees with both our observational sample and
that of \citet{2007ApJ...658.1203J}. However, most observed CEMP
stars are enhanced in nitrogen by $1-1.5\,\mathrm{dex}$, which in
fact coincides with a dearth of model CEMP stars. Our model includes
three mechanisms for increasing the nitrogen abundance: first dredge
up, hot-bottom burning and third dredge up of the hydrogen burning
shell. The latter is only a small effect and while first dredge up
enhances $[\mathrm{N}/\mathrm{Fe}]$ by typically $0.5\,\mathrm{dex}$,
this cannot reproduce the $1\,\mathrm{dex}$ nitrogen enhancements
seen in our observational sample. On the other hand, hot-bottom burning
converts most of the dredged-up carbon into nitrogen and thus results
in much larger nitrogen enhancements. If HBB were more effective than
we assume it would only raise the number of (C)NEMP stars, in contradiction
with the observations. We must therefore assume that either some kind
of extra-mixing mechanism or a dual core/shell flash is responsible.
This is beyond the scope of our present model.

Another indication of a missing ingredient in our models comes from
the carbon isotopic ratio. The $^{12}\mathrm{C}/\mathrm{^{13}}\mathrm{C}$
ratio in our models is always large ($10^{2}-10^{4}$), whereas the
observed ratio is generally less than the solar ratio (around 90),
from the equilibrium value of four up to about $50$ \citep{2005ApJ...635..349R}.
Again, this may result from extra mixing or a dual shell/core flash. 

\begin{figure}
\includegraphics[bb=111bp 85bp 525bp 758bp,clip,angle=270,scale=0.43]{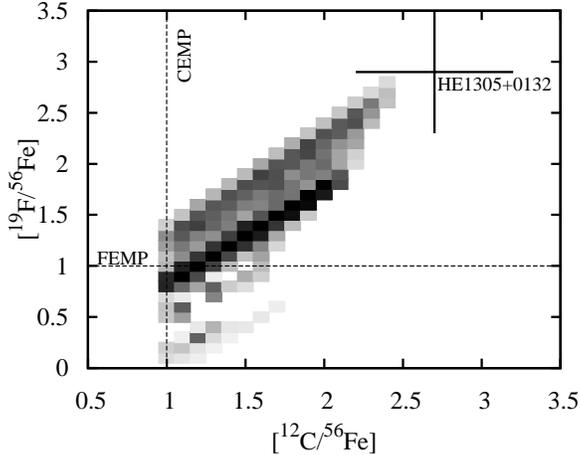}

\caption{\label{fig:C-F-default-population}The distribution of $[\mathrm{F}/\mathrm{Fe}]$
versus $[\mathrm{C}/\mathrm{Fe}]$ in our model set \protect \modelset{1}.
The horizontal dashed line indicates our FEMP selection criterion
($[\mathrm{F}/\mathrm{Fe}]\geq1$) and the vertical dashed line indicates
our CEMP selection criterion ($[\mathrm{C}/\mathrm{Fe}]\geq1$). The
position of the fluorine-rich CEMP star HE\,1305+0132 is also plotted.}

\end{figure}
Fluorine was recently measured in \Change{the CEMP star HE 1305+0132
\citep{2007ApJ...667L..81S} and the halo planetary nebula BoBn~1
\citep{2008ApJ...682L.105O}. \citet{2008A&A...484L..27L} pointed
out that fluorine is made in the progenitors of CEMP stars and therefore
that most CEMP stars \Change{should be} FEMP stars (EMP stars with
$[\mathrm{F}/\mathrm{Fe}]\geq1$). Fig.~\ref{fig:C-F-default-population}
confirms this for model set \protect \modelset{1}. Our model struggles
to reproduce the $[\mathrm{F}/\mathrm{Fe}]$ ratio of HE 1305+0132
although this star may represent the tail of the fluorine distribution.
Future observations of fluorine in CEMP stars should reveal this population.
Also, the abundance may be overestimated \citep{2009ApJ...694..971A}.}

\begin{figure}
\includegraphics[scale=0.34,angle=270]{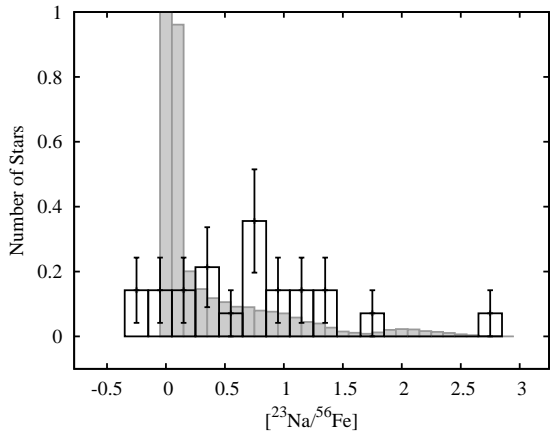}

\caption{\label{fig:Na-distribution-default}Sodium distribution in our default
model set \protect \modelset{1} (filled histogram) compared to observations
(open histogram with Poisson error bars).}

\end{figure}

\subsubsection{Sodium and other light elements\label{sub:default-sodium-and-light-elements}}

Sodium is enhanced at the surface of TPAGB stars through both third
dredge up and hot-bottom burning. In our model set \protect \modelset{1}
most stars dredge up little sodium, leading to the peak at $[\mathrm{Na}/\mathrm{Fe}]\sim+0$
in Fig. \ref{fig:Na-distribution-default}. A few of our stars, with
relatively massive AGB primaries, dredge up enough sodium that the
secondary reaches up to $[\mathrm{Na}/\mathrm{Fe}]\sim1.5$. The most
massive AGB primaries undergo HBB and give rise to the small peak
seen at $[\mathrm{Na}/\mathrm{Fe}]\sim2-2.5$: these stars are CNEMP
stars. Our model at least matches the range of observed stars, although
sodium-enriched stars are greatly underrepresented. \Change{Note
that there is great uncertainty in the yield of sodium from massive
AGB stars, and hence in our model CNEMP sodium abundances, because
of reaction rate uncertainties \citep{2007A&A...466..641I}.}

Observations of CEMP stars show an overabundance of oxygen and other
$\alpha$-elements typical of the halo, i.e. about $+0.4\,\mathrm{dex}$.
\Change{Our models include no explicit $\alpha$-enhancement but
do show enhancements in oxygen and magnesium because of third dredge
up (up to $+0.5\,\mathrm{dex}$ and $+0.6\,\mathrm{dex}$ respectively).}
The few observations of oxygen in CEMP stars range from $0$ to $+2\,\mathrm{dex}$
while magnesium is enhanced by up to $+1.5\,\mathrm{dex}$. It is
clear that our models struggle to reproduce these stars, especially
because they are often giants which should be well mixed.

There have been several recent lithium abundance measurements in CEMP
stars (see e.g. \citealp{2008ApJ...677..556T} and \citealp{2008ApJ...679.1549R}).
\Change{Modelling lithium nucleosynthesis is too complicated for
our synthetic models. Detailed discussions can be found in \citet{2008ApJ...679.1549R}
and \citet{2009MNRAS.394.1051S}. }

\subsubsection{The heavy elements\label{sub:default-heavy-elements}}

\begin{figure*}
\begin{centering}
\begin{tabular}{ccc}
\includegraphics[bb=50bp 145bp 554bp 540bp,angle=270,scale=0.33,angle=0]{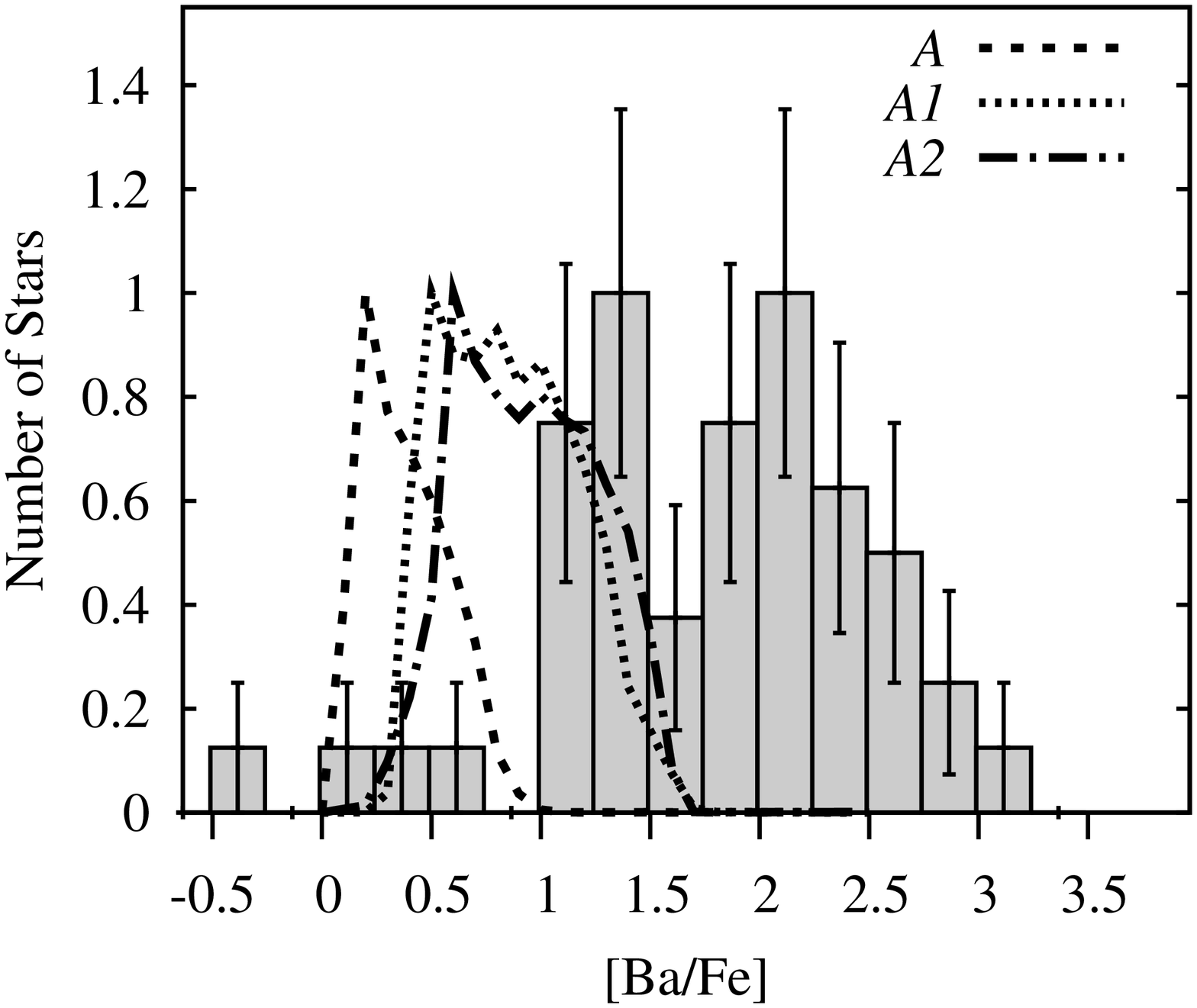} & \includegraphics[bb=50bp 67.5bp 554bp 500bp,angle=270,scale=0.33,angle=0]{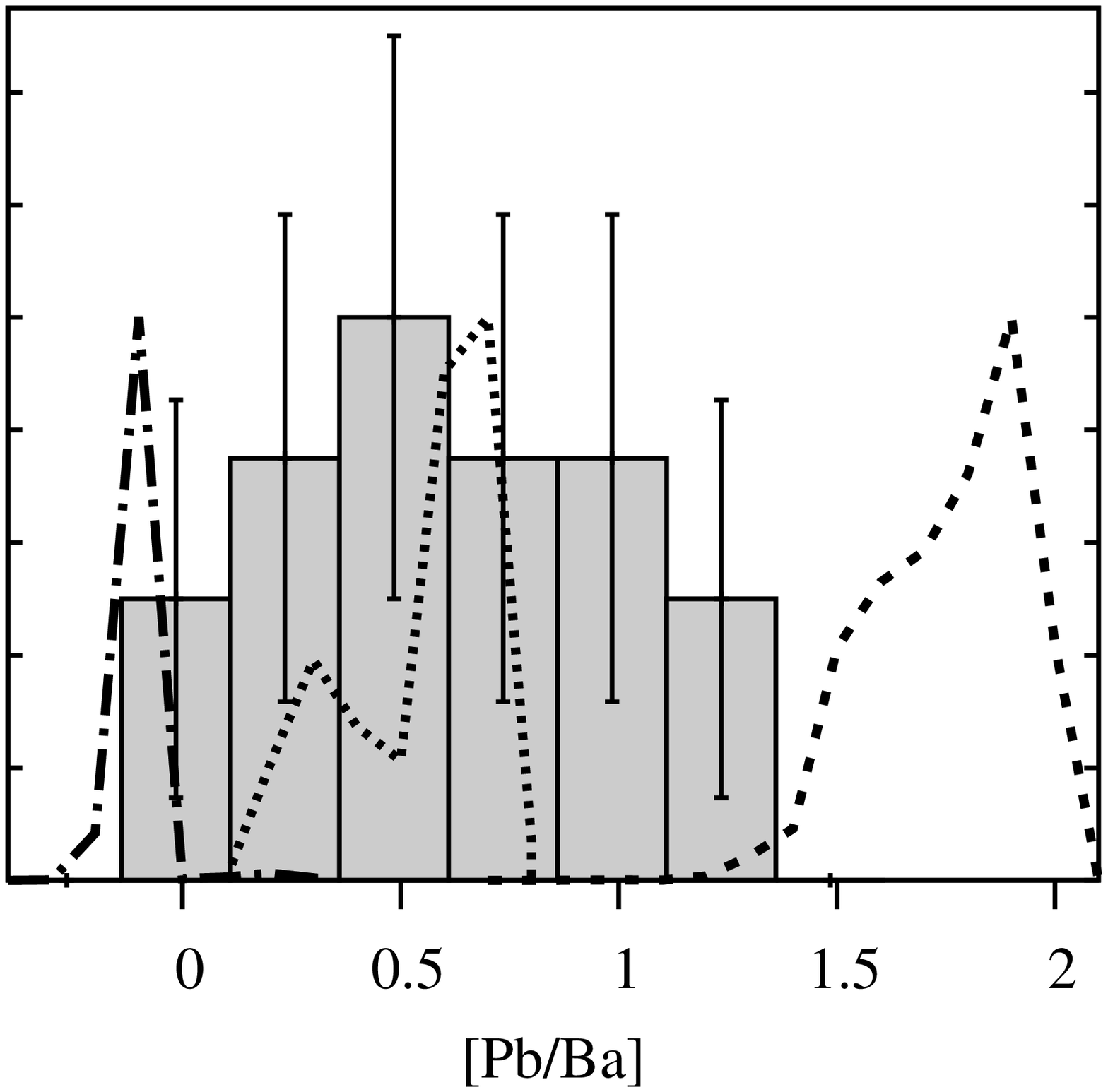} & \includegraphics[bb=50bp 59bp 554bp 770bp,angle=270,scale=0.33,angle=0]{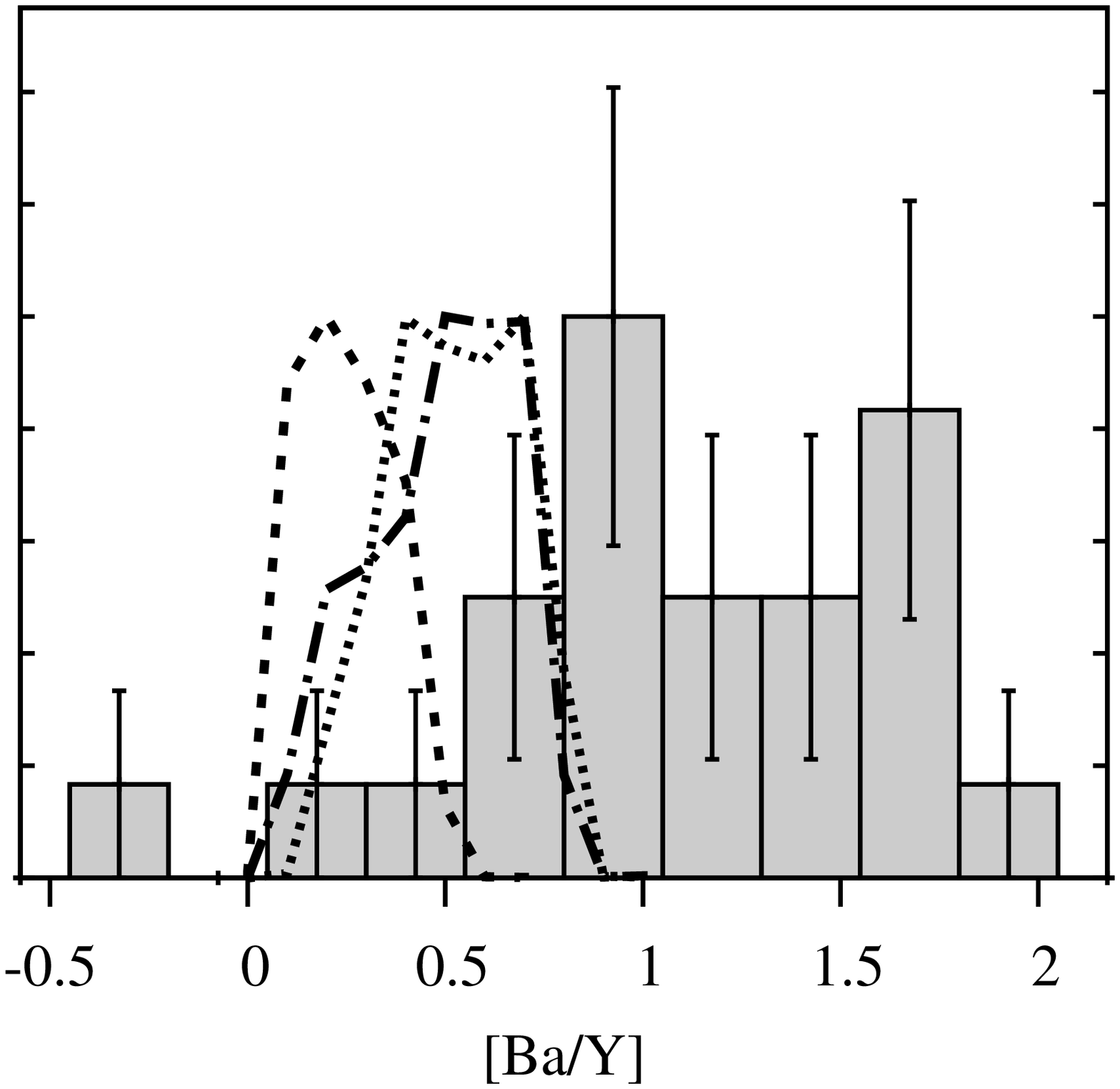}\tabularnewline
\end{tabular}
\par\end{centering}

\caption{\label{fig:Heavy-elements-default}\textcolor{red}{ }Heavy elements
in CEMP stars from model sets \protect \modelset{1}, \protect \modelset{17}
and \protect \modelset{18} with \Change{the $^{13}\mathrm{C}$ efficiency
parameter} $\xi_{13}=1$, $0.1$ and $0.01$ respectively. Our observational
sample is shown by the filled histograms (with Poisson error bars)
while the model sets \protect \modelset{1}, \protect \modelset{17}
and \protect \modelset{18} are plotted with long-dashed, short-dashed
and dot-dashed lines respectively. The distributions are all normalized
to peak at one. We choose Y, Ba and Pb to represent the $s$-process
peaks. }

\end{figure*}
Model \modelset{1} yields a CEMP-$s$/CEMP ratio of only 28 per cent,
smaller than is observed: of the 47 CEMP stars in our SAGA database
selection with both carbon and barium measured%
\footnote{We could assume that stars \emph{without} measured barium have little
barium, in which case the number of CEMP stars is 96 and the $s$-rich
fraction is $44/96=46\%$. %
} 44 have $[\mathrm{Ba}/\mathrm{Fe}]\geq0.5$ (94\%), while \citet{2007ApJ...655..492A}
report an $s$-rich fraction of 80\%. All our model CEMP stars are\emph{
}enriched in $s$-process elements, but most stars in this model
have $0<[\mathrm{Ba}/\mathrm{Fe}]<0.5$ and are thus not classified
as CEMP-$s$. \Change{This is because} the assumed $^{13}\mathrm{C}$
pocket efficiency $\xi_{13}=1$ gives such a high neutron exposure
that the $s$-process distribution is pushed to the lead peak \citep{1998ApJ...497..388G}.
\Change{A decreased $\xi_{13}$} (model sets \modelset{17} and
\modelset{18}) \Change{give}s larger barium abundances and hence
a larger CEMP-$s$ fraction (see Table~\ref{tab:percentages}).

\Change{The need to decrease the $^{13}\mathrm{C}$ efficiency at
low metallicity, to $\xi_{13}\sim0.1$ for $[\mathrm{Fe}/\mathrm{H}]<-1$,
was shown by \citet{2007A&A...469.1013B} on the basis of $[\mathrm{Pb}/\mathrm{hs}]$
ratios in lead stars. A comparison between model set \protect \modelset{1}
and the observed heavy element abundance distribution is shown in
Fig. \ref{fig:Heavy-elements-default}. We find a best match to the
$[\mathrm{Pb}/\mathrm{Ba}]$ ratio for model~\modelset{17} with
$\xi_{13}=0.1$. Although it is unsatisfactory that both the CEMP-$s$
to CEMP ratio and the $s$-abundance ratio distributions depend quite
sensitively on a free parameter in the model, we find a reasonable
match to all these constraints for a single value of $\xi_{13}$.}

\subsubsection{Orbital periods}

\label{sub:default-orbital-periods}
\begin{figure}
\includegraphics[bb=50bp 125bp 554bp 770bp,angle=270,scale=0.43]{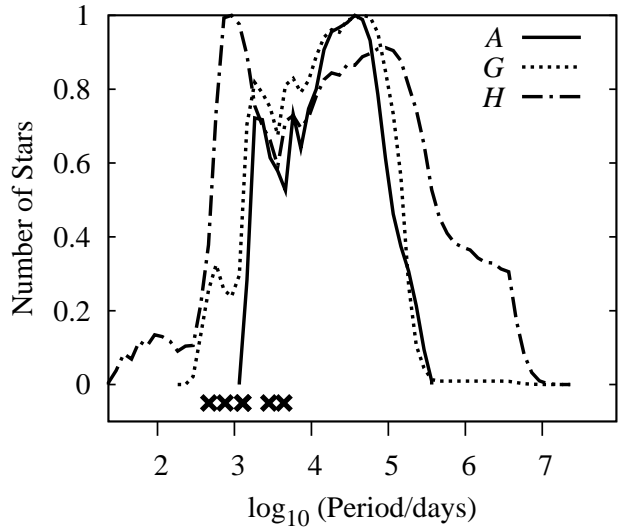}

\caption{\label{fig:default-period-distribution}Period distribution in our
CEMP model sets \protect \modelset{1}, \protect \modelset{49} and
\protect \modelset{54} (solid lines, dashes and dot-dashes respectively)
vs CEMP period measurements from our selection of the SAGA database
(crosses). }

\end{figure}

Comparison of observed CEMP orbital
periods with our models is difficult for two reasons. First, the number
of stars with known periods is small despite the fact that many are
binaries -- from our selection there are only six stars with orbital
solutions (see Table~\ref{tab:CEMP-binaries}). Second, long periods
are difficult to measure. Any period longer than about ten years,
which is approximately the time for which surveys have been ongoing,
is likely to remain unmeasured for some time to come. With this in
mind, Figure~\ref{fig:default-period-distribution} shows the distribution
of periods from model set \protect \modelset{1} compared to the six
observed stars. %
\begin{table}
\begin{tabular}{|c|c|c|c|}
\hline 
Object & Period & $e$ & Reference\tabularnewline
\hline
\hline 
CS22948-027 & 505d & 0.3 & {\scriptsize \citet{2001AJ....122.1545P}}\tabularnewline
\hline 
CS22948-027 & 426.5d & 0.02  & {\scriptsize \citet{2005AA...429.1031B}}\tabularnewline
\hline 
HD5223 & 755.2d & 0 & {\scriptsize \citet{1990mcclure_woodsworth}}\tabularnewline
\hline 
HD224959 & 1273d & 0.179 & {\scriptsize \citet{1990mcclure_woodsworth}}\tabularnewline
\hline 
CS22956-028 & 1290d & 0.22 & {\scriptsize \citet{2003ApJ...592..504S}}\tabularnewline
\hline 
CS22942-019 & 2800d & 0.19 & {\scriptsize \citet{2001AJ....122.1545P}}\tabularnewline
\hline
LP 625-44 & 12y &  & {\scriptsize \citet{2000ApJ...536L..97A}}\tabularnewline
\hline
\end{tabular}

\caption{\label{tab:CEMP-binaries}Periods and eccentricities of CEMP stars
in binaries from our selection ($[\mathrm{Fe}/\mathrm{H}]=-2.3\pm0.5$,
$\log_{10}g\leq4$). Note that two solutions are available for CS22948-027.}

\end{table}

The lower limit to the model period distribution is set by systems
that are just wide enough for the AGB primary to avoid filling its
Roche lobe when the star has its maximum radius. The shortest-period
CEMP binaries, CS22948-027 and HD5223, are not compatible with our
model set \protect \modelset{1}. Their short periods and small eccentricities
(if we adopt the \citealp{2005AA...429.1031B} result regarding CS22948-027)
suggest they underwent tidal circularisation, and perhaps RLOF and
common-envelope evolution. This does not explain their carbon enhancement.
The problem is reminiscent of the barium stars, which have too large
eccentricities and too short periods compared to models \citep{2003ASPC..303..290}.
Various explanations have been proposed to deal with this problem
which build on the fact that AGB stars have very extended, outflowing
atmospheres such that the canonical distinction between RLOF (when
the stellar surface fills the Roche lobe) and wind accretion (when
it does not) becomes blurred (e.g. see \citealp{2007BaltA..16..104F},
\citealp{2008A&A...480..797B}). Our understanding of binary evolution
in this transition region is still poor. An alternative solution may
be accretion during the common-envelope phase \citep{2008ApJ...672L..41R},
which we explore in model~\modelset{54} (see Section~\ref{sub:best-case-modelsets}).

The remaining stars, HD224959, CS22956-028, CS22942-019 and LP 625-44,
all lie at the short-period end of the range produced by our model
set \protect \modelset{1}, as is to be expected from the above-mentioned
selection effects.

\subsection{Model sets \protect \modelset{49} and \protect \modelset{54}: best
comparison to observations}

\label{sub:best-case-modelsets}The discussion of the previous section
suggests that to better match our models with the observations we
should consider an increase in the amount of third dredge up in low
mass stars and the effect of switching off thermohaline mixing. To
this end we consider model sets \modelset{49} and \modelset{54}.

\subsubsection{Model set \protect \modelset{49}: extra third dredge up}

\label{sub:Model-set-35}Model sets \modelset{9}, \modelset{26}
and \modelset{49} are the same as our default set (set \protect \modelset{1})
except that third dredge up is increased in efficiency in low mass
stars. In model~\modelset{9} we have set $\Delta M_{\mathrm{c},\mathrm{min}}=-0.07\mathrm{\, M_{\odot}}$
and $\lambda_{\mathrm{min}}=0.8$, which is the set of values found
by \citet{Izzard_et_al_2003b_AGBs} to be required to match the carbon-star
luminosity functions of the Magellanic Clouds. This results in only
a modest increase of the number of CEMP stars, see Table~\ref{tab:percentages}.
A larger effect is obtained by setting $\Delta M_{\mathrm{env},\mathrm{min}}=0.0$
in model~\modelset{26}, which allows for efficient dredge-up in
AGB stars of much smaller initial masses and increases the CEMP/EMP
ratio to 6.5 per cent. Model~\modelset{49} is a combination of these
parameter choices with $\Delta M_{\mathrm{c},\mathrm{min}}=-0.1\mathrm{\, M_{\odot}}$.
With this parameter combination \emph{all primary TPAGB stars} down
to initial masses of $0.8\mathrm{\, M_{\odot}}$ undergo efficient
third dredge up. The IMF peaks at low mass, so the number of stars
affected is large and the CEMP/EMP ratio increases to almost $10\%$
-- a factor of four increase compared to our default models. The NEMP/EMP
ratio remains small at $0.3\%$.

\begin{figure*}
\begin{tabular}{cc}
\put(10,-15){\LabelG{a}}\includegraphics[bb=125bp 100bp 525bp 758bp,angle=270,scale=0.5]{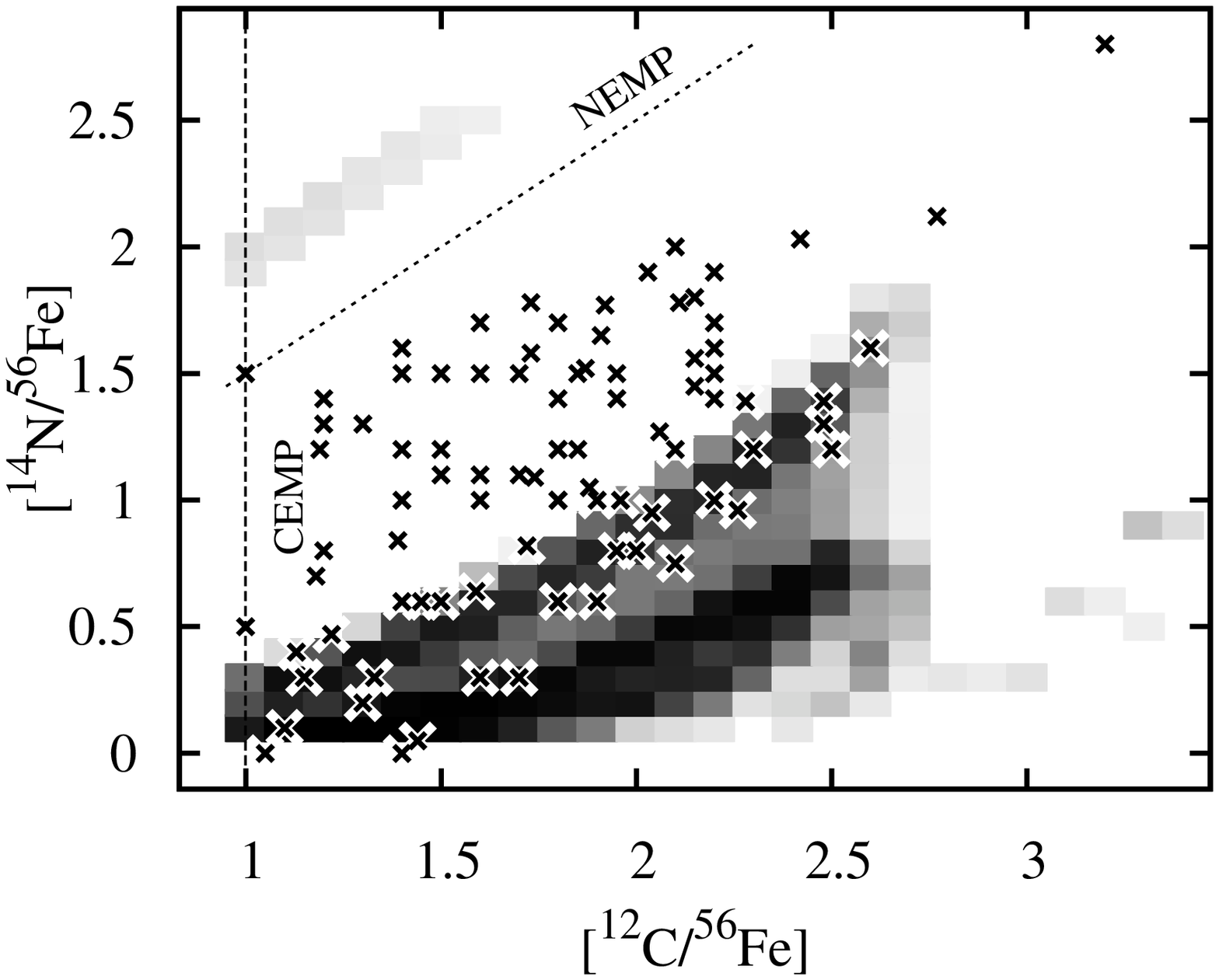} & \put(-75,-15){\LabelG{b}}\includegraphics[bb=125bp 250bp 525bp 758bp,angle=270,scale=0.5]{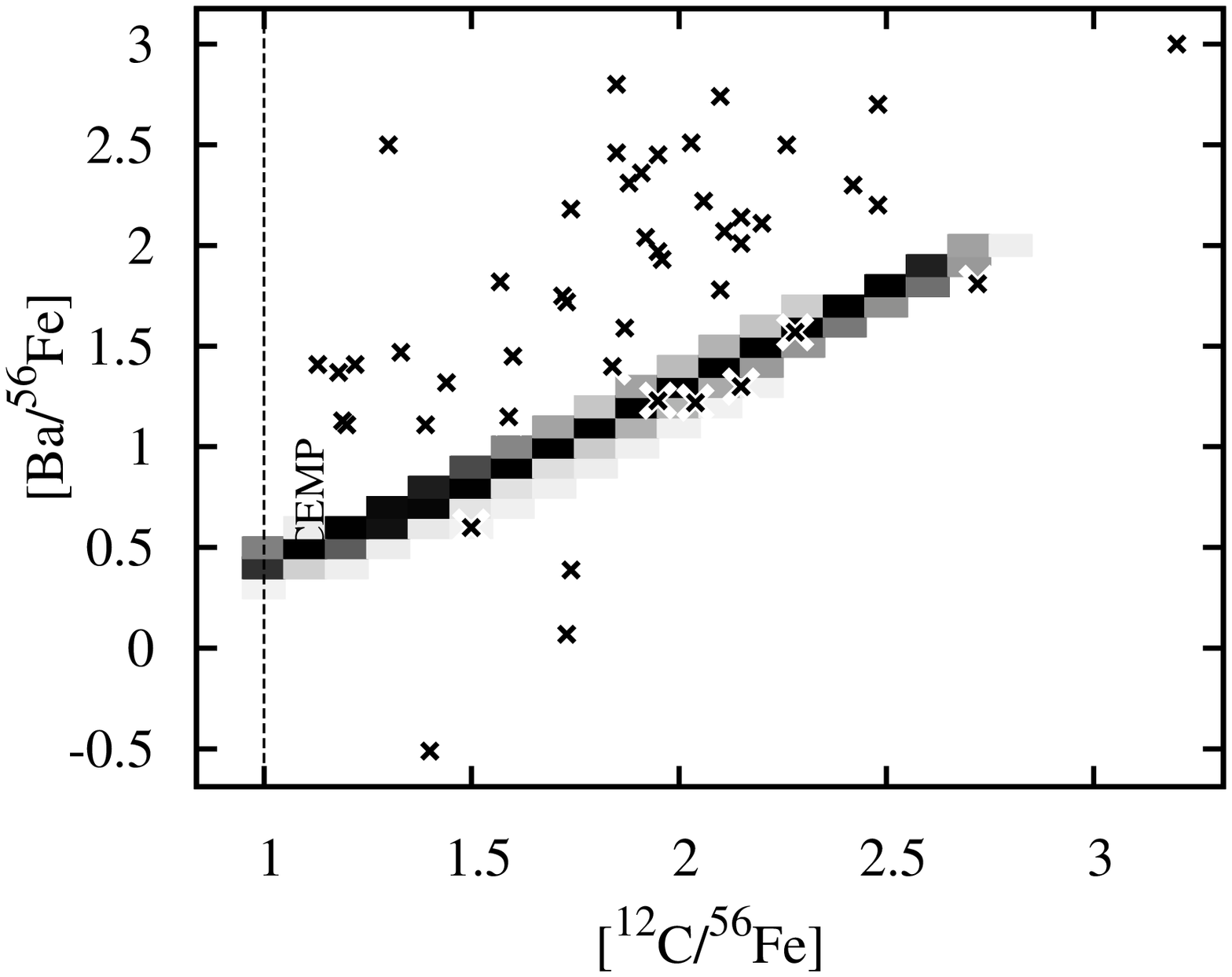}\tabularnewline
\end{tabular}

\caption{\label{fig:Model-set-35}Model set \protect \modelset{49}: \textbf{a}
$[\mathrm{N}/\mathrm{Fe}]$ vs $[\mathrm{C}/\mathrm{Fe}]$ and \textbf{b}
$[\mathrm{Pb}/\mathrm{Fe}]$ vs $[\mathrm{Ba}/\mathrm{Fe}]$. Crosses
show our observational sample. This model set is the same as our default
set (model set \protect \modelset{1}) but with efficient third dredge
up in stars down to $0.8\mathrm{\, M_{\odot}}$ and a reduced $^{13}\mathrm{C}$
efficiency.\textcolor{red}{ }}

\end{figure*}
The distribution of $[\mathrm{C}/\mathrm{Fe}]$ reaches up to $+2.7$
and is in good agreement with the observations (see Fig.~\ref{fig:Model-set-35}a).
The distributions of nitrogen (Fig.~\ref{fig:Model-set-35}a) and
sodium suffer the same problems as in the default model set. Although
the total amount of $\mathrm{C}+\mathrm{N}$ (and hence the total
amount of third dredge up) as well as the observed trend of $[\mathrm{N}/\mathrm{Fe}]$
vs $[\mathrm{C}/\mathrm{Fe}]$ roughly match the observations, there
is clearly a need for some additional CN cycling to convert carbon
into nitrogen.

Model set \modelset{49} has a reduced $^{13}\mathrm{C}$ efficiency,
$\xi_{13}=0.1$ as in model~\modelset{17}. This yields a reasonably
good fit to the observations for the light-$s$ elements strontium,
yttrium and zirconium, but the stars most strongly enriched in heavy-$s$
elements such as barium (with $[\mathrm{Ba}/\mathrm{Fe}]>+2.7$) are
not well reproduced. Fig.~\ref{fig:Model-set-35}b reveals that,
firstly, our models predict a strong correlation between barium and
carbon (as well as between other $s$-process elements) which is not
seen. The spread in the observations is only partially explained by
measurement errors. Secondly, the models trace out roughly the lowest
observed barium abundances for any value of $[\mathrm{C}/\mathrm{Fe}]$.
More barium can be produced in our models by further reducing the
$^{13}\mathrm{C}$ efficiency, as discussed in Section~\ref{sub:default-heavy-elements},
but only at the expense of lead. With the adopted value of $\xi_{13}$
the average observed barium to lead ratio is well matched, although
the most lead-rich stars (with $[\mathrm{Pb}/\mathrm{Fe}]>+2.7$)
are not reproduced. The CEMP-$s$/CEMP ratio for this model set is
$94\%$, in agreement with observations. 

Interestingly, some CEMP binaries are made in model set \modelset{49}
which have periods of one to a few years, similar to the observed
short-period giant CEMP stars (see Fig.~\ref{fig:default-period-distribution}).
These arise from binaries with low-mass AGB primaries which have smaller
maximum radii and can thus avoid filling their Roche lobes in tighter
orbits. Because the initial mass ratio is close to unity and the primary
may even be \emph{less} massive than the secondary due to mass loss
prior to the AGB, these binaries may also undergo stable RLOF without
a common envelope phase.

\subsubsection{Model set \protect \modelset{54}: extra third dredge up, no thermohaline
mixing, common envelope accretion}

\label{sub:Model-set-43}Model set \modelset{54} is similar to \modelset{49},
as described in the previous section, but is tuned to maximise the
CEMP/EMP ratio. Thermohaline mixing is turned off so that accreted
material remains in the surface convection zone. Before the star ascends
the giant branch and its convection zone deepens, its surface composition
is therefore essentially the same as that of the primary TPAGB star
ejecta. Furthermore, during the common envelope phase accretion of
$0.05\mathrm{\, M_{\odot}}$ of material is allowed, so some stars
that undergo RLOF can become CEMP stars. These changes mean that CEMP
stars can form out to longer periods and more turn-off stars and subgiants,
with $\log g\gtrsim3.5$, become CEMP stars because there is no dilution
until first dredge up. While the individual effects of these changes
on the number of CEMP stars are modest (see models~\modelset{16}
and \modelset{19} in Table~\ref{tab:percentages}), in combination
with extra third dredge-up they yield a CEMP/EMP ratio of almost $16\%$.

\begin{figure*}
\begin{tabular}{cc}
\put(10,-15){\LabelG{a}}\includegraphics[bb=125bp 100bp 525bp 758bp,angle=270,scale=0.5]{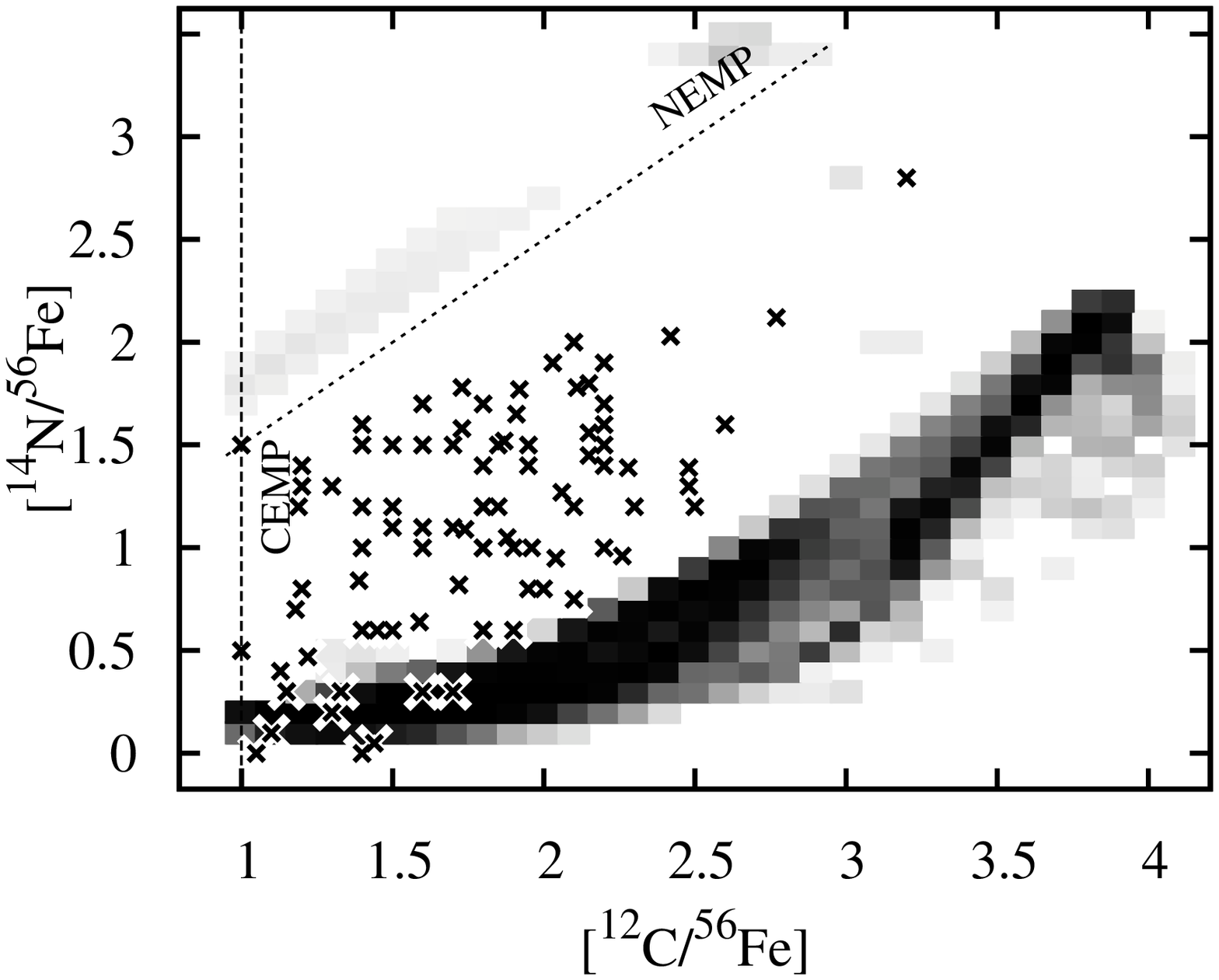} & \put(-70,-15){\LabelG{b}}\includegraphics[bb=125bp 250bp 525bp 758bp,angle=270,scale=0.5]{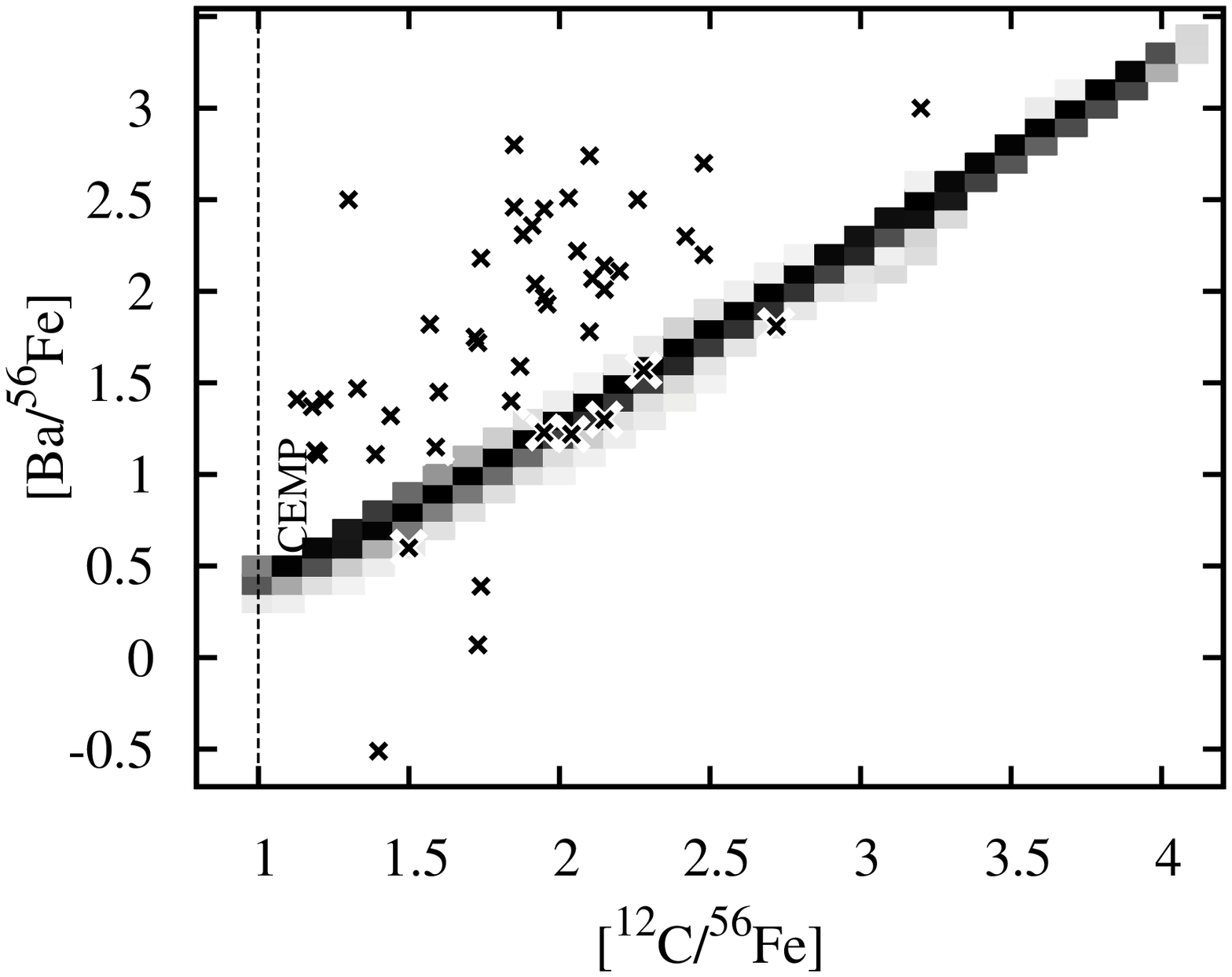}\tabularnewline
\end{tabular}

\caption{\label{fig:Model-set-43}As Fig. \ref{fig:Model-set-35} for model
set \protect \modelset{54}.\textcolor{red}{ }}

\end{figure*}

The effect on our model abundance distributions is significant: the
sub-giant and turn-off CEMP stars form a group of stars with undiluted
abundances. In the case of $[\mathrm{C}/\mathrm{Fe}]$ these are between
$+3$ and $+4\,\mathrm{dex}$ (see Fig.~\ref{fig:Model-set-43}a),
which does not match the observations -- only one star in our observed
selection has $[\mathrm{C}/\mathrm{Fe}]=+3$ and none are more enhanced
than this. It may be argued that this excess carbon is converted to
nitrogen by extra mixing processes, but then $[\mathrm{N}/\mathrm{Fe}]$
should rise to approximately $+3\,\mathrm{dex}$ in some stars, which
is not observed either. Our model $[\mathrm{N}/\mathrm{Fe}]$ distribution
ranges up to $+2\,\mathrm{dex}$, as do most of the observations,
but while in our model these correspond to the most C-enriched, undiluted
stars, the observed nitrogen-rich stars have more modest carbon enhancements.

The undiluted turn-off stars in model~\modelset{54} result in a
peak in the sodium distribution at around $0.8\,\mathrm{dex}$, in
better agreement with the (broad) peak in the observations than our
models with thermohaline mixing. Similarly, a broad distribution of
oxygen abundances is obtained, out to about $+1.7\,\mathrm{dex}$,
also in better agreement with the observations. 

The heavy element distributions are reasonably well matched by this
model, although the maximum abundances predicted (for undiluted stars)
are sometimes in excess of what is observed. This is especially true
for the light-$s$ elements, for which this model set makes too many
stars with large enhancements (up to $+2.5\,\mathrm{dex}$ for $[\mathrm{Zr}/\mathrm{Fe}]$
and$+2.2\,\mathrm{dex}$ for $[\mathrm{Y},\,\mathrm{Sr}/\mathrm{Fe}]$)
which are not seen in the observations. Lanthanum and barium are enriched
up to $+3.2\,\mathrm{dex}$ in this model which is in agreement with
observations, although the most barium-rich objects have too much
carbon (see Fig.~\ref{fig:Model-set-43}b). The models cover the
full range of lead observations, up to $\sim3.5\,\mathrm{dex}$.

We must keep in mind that the undiluted stars in our models, which
have the largest overabundances for all the elements discussed above,
are also high-gravity stars against which there is an apparent observational
bias (see Sections~\ref{sec:Observational-database} and \ref{sub:default-logg-vs-obs}).
They may therefore be overrepresented in our model distributions compared
to the observational samples. We also note that in our observational
selection, no correlation of the abundance distributions against $\log g$
is apparent for any of the elements discussed above. However, \citet{2008ApJ...679.1541D}
and \citet{2008ApJ...678.1351A} do find evidence for significantly
higher average $[\mathrm{C}/\mathrm{Fe}]$ values in turn-off CEMP
stars compared to bright giants.

This model set also includes accretion of up to $0.05\mathrm{\, M_{\odot}}$
of material during the common envelope phase. This allows CEMP formation
in the narrow range of initial separation corresponding to primary
TPAGB stars which undergo a few pulses before overflowing their Roche
lobe, and leads to a peak in the orbital period distribution around
one year and a tail extending down to $100\,\mathrm{days}$ (see Fig.~\ref{fig:default-period-distribution}).
This is just the range which includes the stars described in Section~\ref{sub:default-orbital-periods}
and may explain their origin.

\subsection{Model set conclusions}

It appears from the previous sections that our models struggle to
reproduce both the observed frequency of CEMP stars and the full range
of their abundance patterns. Unless we choose a rather extreme combination
of model parameters our model CEMP/EMP ratio falls short of the range
of values deduced from observations and does not exceed $16\%$ even
in our most favourable model.

In order to come close to reproducing the observations, we must assume
that TPAGB stars with masses as low as $0.8\mathrm{\, M_{\odot}}$
experience efficient third dredge up. This yields a good fit of the
distribution of carbon enhancements, but still falls short of reproducing
the largest observed $s$-process abundances. If we switch off thermohaline
mixing we find, on the other hand, that the $s$-process elements
are quite well reproduced, but our model CEMP stars have too much
carbon (and perhaps also lead). We note that both the $s$-process
abundance ratios and the CEMP-$s$/CEMP ratio depend on the $^{13}\mathrm{C}$
pocket efficiency. We can find a reasonable match to these constraints
by choosing this efficiency to be about 0.1 times its default value.

The orbital periods of CEMP binaries in our observational sample agree
well with the shortest period CEMP stars in both these model sets,
so it seems that -- at least for the few systems with measured periods
-- the binary mass-transfer scenario is compatible with the observations.

\section{Discussion}

\label{sec:Discussion}We have modelled the observed properties --
$\log g$, chemistry and orbital period -- of CEMP stars and have
attempted to match our models to observed stars. Our models are successful
in reproducing several key observed properties of CEMP stars, but
struggle to match the full range of the observations, most notably
the high CEMP fraction, even with rather extreme choices of model
parameters. 

The first criticism that can be levelled at our work is probably our
choice of stars from, and subsequent processing of, the SAGA observational
database. We chose a group of stars limited by metallicity, $[\mathrm{Fe}/\mathrm{H}]=2.3\pm0.5$,
because our stellar models have $Z=10^{-4}$. As shown in Section~\ref{sec:Observational-database},
the statistics (number of stars, CEMP/EMP ratios etc.) of our selection
vary remarkably little as a function of $[\mathrm{Fe}/\mathrm{H}]$
so the selection is justified in this respect. However, there may
be a selection effect inherent in the SAGA database because papers
are selected for inclusion in it if they contain stars with $[\mathrm{Fe}/\mathrm{H}]\leq-2.5$.
Stars with higher metallicities, of which there are many in the database,
are included because they just happen to be in those papers.

We should then consider the selection criterion $\log g\leq4$. While
we select identically in the models it is reasonable to ask which
model constraints are weakened by this choice. The answer is few:
most stars in the SAGA database are giants. This is also the result
of a bias against faint, high-gravity stars that is inherent in the
SAGA database. We have made no attempt to model such a selection bias
in any detail and our simple gravity criterion is not entirely successful
in selecting against turn-off stars which may be affected by the uncertain
strength of thermohaline mixing. On the other hand, giants are well
mixed as their surface convection zones deepen on the giant branch
and so whether we assume efficient thermohaline mixing, or do not,
is of minor importance provided we look only at giants and avoid elements
which may be processed in the envelope (see e.g. \citealp{2008MNRAS.389.1828S}
and \citealp{2009MNRAS.394.1051S}).

The exception to this rule lies with carbon and nitrogen. The amount
of nitrogen dredged up as the convection zone deepens depends on whether
accreted carbon has mixed deep enough to be burned by the CN cycle.
In the case of efficient thermohaline mixing this is certainly the
case, although to some extent the \emph{dilution} associated with
such deep mixing reduces the effect of first dredge up. On the other
hand, if accreted carbon sits at the stellar surface it does not burn
to nitrogen. Almost all the nitrogen seen in the CEMP star must then
come from the primary star, posing the questions of whether extra
mixing, dual-shell flashes and dual-core flashes are important. Interesting
progress is being made regarding these questions (e.g. \citealp{2009arXiv0904.2393S}).

We have tried to push the physics of the \emph{canonical} third dredge
up as far as possible, inducing it in stars down to $0.8\mathrm{\, M_{\odot}}$
with large efficiency. Previous models suggest that stars with envelope
masses less than $0.5\mathrm{\, M_{\odot}}$ do not undergo third
dredge up, although many of these models are for higher metallicity
than we are considering here.

\citet{2008MNRAS.389.1828S} have made a model of a $0.9\mathrm{\, M_{\odot}}$
star with $Z=10^{-4}$ which \emph{does} undergo third dredge up.
The efficiency is not high, $\lambda=0.16$, but carbon is dredged
to the surface. At the time that third dredge up occurs the star has
an envelope mass of just $0.19\mathrm{\, M_{\odot}}$ and the dredge
up is sufficient to increase the surface carbon abundance dramatically,
from just $6\times10^{-6}$ up to $3.7\times10^{-3}$ by mass fraction.

If third dredge up is as efficient as assumed in our models that provide
the best match to observations we would expect to see a population
of (single) CEMP stars that are currently in the TPAGB phase with
$\log g<0.5$. \citet{2006A&A...455.1059M} have suggested that CS
30322-023 is just such a star. They construct a $0.8\mathrm{\, M_{\odot}}$,
$[\mathrm{Fe}/\mathrm{H}]=-3.3$ detailed model sequence with the
\textsc{starevol} code \citep{2006A&A...448..717S} for comparison
with observations of CS 30322-023. They find no third dredge up, which
is not surprising given that models constructed with \textsc{starevol}
do not show third dredge up unless some kind of neutral-convective-boundary
method or overshooting is invoked. However the authors make clear
our lack of quantitative understanding of mixing processes in these
stars. Other mechanisms such as dual shell and/or core flashes, proton
ingestion and so-called {}``canonical'' extra mixing may, to some
extent, mimic the nucleosynthetic signature of third dredge up. At
the very least, if CS 30322-023 is a (single) TPAGB star, as \citet{2006A&A...455.1059M}
suggest, it certainly has undergone some nucleosynthetic processing
in order to reach $[\mathrm{C}/\mathrm{Fe}]\sim+0.6$ and $[\mathrm{N}/\mathrm{Fe}]\sim+2.8$.
Convective overshooting may also play a role in enhancing third dredge
up in low-mass stars: prescriptions such as that of \citet{2000A&A...360..952H}
contain free parameters which have the same effect as our $\Delta M_{\mathrm{c},\mathrm{min}}$
and $\lambda_{\mathrm{min}}$.

Even with enhanced third dredge up efficiency our models barely make
enough CEMP stars to match the observed fraction. There are more exotic
methods for increasing the number of CEMP stars. One is to accrete
material from a TPAGB star on to a main-sequence star during the common-envelope
phase which should occur for most TPAGB primaries that overflow their
Roche lobe. We assumed $0.05\mathrm{\, M_{\odot}}$ is accreted on
to the main-sequence star, which is probably too much \citep{2008ApJ...672L..41R},
so gives an upper limit. In any case, if there is no thermohaline
mixing in the main-sequence star even a tiny amount of accreted carbon-rich
material should turn it into a CEMP star. Still, even with $0.05\,\mathrm{M}_{\odot}$
of accreted material, only a few extra CEMP stars are made. This is
because the period range in which AGB stars both have enough pulses
to become carbon rich \emph{and then} undergo RLOF is rather narrow.
At a slightly smaller period RLOF occurs too early, i.e. after too
few (or no) third dredge up episodes and little or no carbon enhancement.

We also considered a reduction in the common-envelope parameter such
that $\alpha_{\mathrm{CE}}=0.1$. This does \emph{not} increase the
number of CEMP stars -- instead it reduces the number of \emph{non-CEMP
stars} by forcing many short-period systems to merge. After the merger
the stellar mass is so large that the star evolves quickly and by
the present age of the Galaxy it is a white dwarf. We do not pretend
that this mechanism is realistic, but it may help to improve the CEMP/EMP
ratio match with observations.

A non-canonical dredge up event may lead to the formation of CEMP
(and possibly NEMP) stars. Previous works, such as \citet{2007ApJ...658..367K},
have suggested that below a certain threshold metallicity, in their
work $[\mathrm{Fe}/\mathrm{H}]=-2.5$, dual shell flashes make all
the required nitrogen and some of the carbon and $s$-process elements
seen in CEMP stars. The recent works of \citet{2007ApJ...667..489C}
and \citet{2008A&A...490..769C} indeed show some agreement with the
observed carbon and nitrogen abundances. However, at the metallicity
under consideration here ($[\mathrm{Fe}/\mathrm{H}]\sim-2.3$) these
proton-ingestion events are not expected to occur -- or at least only
in the lowest mass stars -- and the number of CEMP stars should drop
as the metallicity increases. Extra mixing processes, and a dependence
on stellar properties other than the metallicity, may well blur the
apparently sharp boundary between stars that undergo dual-shell flashes
and those that do not. Incorporation of the results of \citet{2007ApJ...667..489C}
and \citet{2008A&A...490..769C} into our models is planned for future
work.

A deliberate omission in the above is discussion of the binary fraction
in low-metallicity stars. In order to obtain anywhere near the observed
CEMP fraction we must assume a $100\%$ binary fraction. This should
be compared to about $60\%$ in solar-neighourhood G dwarfs \citep{1991A&A...248..485D}
and probably higher among stars more massive than the Sun%
\footnote{We completely ignore triples and systems of high multiplicity -- these
are likely to be hierarchical in nature.%
}. 

We are left with the situation that in order for our models to come
close to reproducing the observed CEMP/EMP fraction several physical
parameters must be pushed to the ends of their reasonable range of
values. This is not a very satisfactory solution, and probably indicates
that other phenomena require our consideration, such as a shift in
the initial mass function \citep{2007ApJ...658..367K,2006ApJ...652L..37L},
massive-star pollution, primordial supernovae, accretion from the
interstellar medium etc. All of these solutions to reproducing the
CEMP/EMP fraction have their own problems. Most likely, some combination
of our physical models with, say, a \emph{slightly}-shifted IMF (or
other initial distribution), will better reproduce the CEMP/EMP fraction.
Our investigation into this is ongoing and will comprise future work.

Finally, we note that both the observational statistics and the observed
abundances of CEMP stars are still uncertain. Interesting results
regarding three-dimensional stellar atmosphere models \citep{2007A&A...469..687C}
may be of relevance to CEMP studies. They conclude that the abundances
of C, N and O in red giants with metallicity $[\mathrm{Fe}/\mathrm{H}]=-3$
may be overestimated by up to $1\,\mathrm{dex}$ in traditional one-dimensional
LTE model atmosphere analyses. If we apply these corrections to the
Suda database, with the crude assumption of a linear scaling as a
function of metallicity%
\footnote{The shift in each abundance is then $\Delta X\,\mathrm{dex}=f\times\delta X$
where $\delta X$ is the shift given by \citet{2007A&A...469..687C}
for their $[\mathrm{Fe}/\mathrm{H}]=-3$ models and $f=\max\left(0,\min\left[1,\left\{ [\mathrm{Fe}/\mathrm{H}]/-3\right\} \right]\right)$.%
}, the resulting CEMP/EMP ratio drops to about $14\%$ which well fits
our enhanced dredge up models. We are not suggesting this is the answer
to the CEMP/EMP ratio problem, but it highlights the fact that observed
carbon and nitrogen abundances, and hence CEMP number counts, are
still quite uncertain.

\section{Conclusions}

In an attempt to reproduce the observed CEMP to EMP number ratio we
have simulated populations of low-metallicity ($[\mathrm{Fe}/\mathrm{H}]=-2.3$)
binary stars with a variety of input physics. Our model sets with
efficient third dredge up in low-mass (down to $0.8\mathrm{\, M_{\odot}}$)
stars have CEMP to EMP ratios of up to $15\%$, comparable with the
observed $\sim20\%$. They also have low NEMP to EMP number ratios,
in agreement with the observations. Other parameters in our simulations,
such as the efficiency of wind accretion, the common envelope parameter
etc., have only relatively minor effects on our results.

\label{sec:Conclusions}
\begin{acknowledgements}
We thank Sara Bisterzo, Simon Campbell, Roberto Gallino, Laura Husti,
Amanda Karakas, Maria Lugaro and the Utrecht stellar evolution group
for useful criticism and discussion. We thank very much the authors
of the SAGA database for their willingness to share their database
before its publication. RGI thanks the NWO for his fellowship in Utrecht
and is the recipient of a Marie Curie-Intra European Fellowship at
ULB. EG acknowledges support from the NWO under grant 614.000.303
and NSERC. RJS is funded by the Australian Research Council's Discovery
Projects scheme under grant DP0879472. He is grateful to Churchill
College for his Junior Research Fellowship, under which this work
commenced. \ChangeA{We would also like to thank the referee, Achim
Weiss, for many useful suggestions.}
\end{acknowledgements}

\bibliography{/home/izzard/svn/tex/references}

\appendix

\section{Dredge Up Prescriptions}

\subsection{First dredge up}

\label{sub:First-DUP-appendix}The change in surface abundance of
isotopes $j$ at first dredge up, $\Delta X_{j}$, is interpolated
from a table of detailed models \Change{with} $Z=10^{-4}$ in mass
range $0.5\leq M/\mathrm{M_{\odot}}\leq12$. A correction factor $f_{\mathrm{CNO}}=X_{\mathrm{CNO}}(\mathrm{TMS})/X_{\mathrm{CNO}}(\mathrm{ZAMS})$,
the ratio of CNO mass fraction at the terminal-age main sequence (TMS)
and zero-age main sequence (ZAMS), is then applied to CNO elements
to take into account accretion during the main sequence.

In the \citet{2006A&A...460..565I} model first dredge up is considered
as an instantaneous event. In terms of \emph{time evolution} this
is a reasonable assumption because giant-branch evolution is fast,
but in terms of \emph{luminosity or gravity} this approximation is
not good and it proves difficult to compare to e.g. the $[\mathrm{C}/\mathrm{Fe}]$
vs $\log\left(L/\mathrm{L_{\odot}}\right)$ data of \citet{2006ApJ...652L..37L}.
To resolve this problem the changes in abundances are modulated by
a factor $f_{\mathrm{p}}=\min\left[\left(M_{\mathrm{c}}-M_{\mathrm{c,BAGB}}\right)/\left(M_{\mathrm{c,1DUPMAX}}-M_{\mathrm{c,BAGB}}\right),1\right]$
where $M_{\mathrm{c}}$ is the core mass, $M_{\mathrm{C,1DUPMAX}}$
is the core mass at which first dredge up reaches its maximum depth
and $M_{\mathrm{c,BAGB}}$ is the core mass at the base of the giant
branch, before first dredge up starts. $M_{\mathrm{c,BAGB}}$ is known
from the stellar evolution prescription and $M_{\mathrm{C,1DUPMAX},}$
is interpolated from a grid of models constructed with the TWIN stellar
evolution code \citep{2002ApJ...575..461E}.

In summary, the surface abundances changes at first dredge up are
given by $f_{\mathrm{CNO}}f_{\mathrm{p}}\Delta X_{j}$. They agree
well with the detailed models, as a function of $M_{\mathrm{c}}$,
$\log L$ and time.

\subsection{Third dredge up}

\label{sub:Third-DUP-appendix}Abundance changes at third dredge up
are treated in a similar way to the prescription of \citet{Izzard_et_al_2003b_AGBs}
and \citet{2006A&A...460..565I}. Intershell abundances are interpolated
from tables based on the \citet{Parameterising_3DUP_Karakas_Lattanzio_Pols}
detailed models the metallicities of which extend down to $Z=10^{-4}$. 

In low-metallicity TPAGB stars dredge up of the hydrogen-burning shell
enhances the surface abundance of $^{13}\mathrm{C}$ and $^{14}\mathrm{N}$
(at higher metallicity the effect is negligible because the initial
abundance of $^{13}\mathrm{C}$ and $^{14}\mathrm{N}$ is relatively
large). This is modelled by dredging up $\delta M$ of hydrogen-burnt
material during each third dredge up, where the abundance mixture
in this material is enhanced in $^{13}\mathrm{C}$ and $^{14}\mathrm{N}$
according to \begin{eqnarray}
X_{\mathrm{C}13} & = & 0.006\times X_{\mathrm{C}12}\,\mathrm{and}\\
X_{\mathrm{N14}} & = & 0.28\times X_{\mathrm{C}12}\,,\label{eq:c13-n14-X}\end{eqnarray}
where \begin{eqnarray}
\delta M & = & \left(\frac{0.01}{1+0.1^{2.2-M(t)}}\right)\times\min\left(1,\left[\frac{N_{\mathrm{TP}}}{10}\right]^{2}\right)\times\nonumber \\
 &  & \left(\frac{1}{1+\epsilon^{M_{\mathrm{env}}(t)-0.5}}\right)\label{eq:c13-n14-deltam}\end{eqnarray}
and $M(t)$ is the instantaneous stellar mass, $M_{\mathrm{env}}(t)$
is the instantaneous envelope mass, $N_{\mathrm{TP}}$ is the thermal
pulse number, $X_{12}$ is the envelope abundance of $^{12}\mathrm{C}$
and $\epsilon=10^{-20}$. The first term gives the amount of H-burnt
material dredged up, the second term is a turn-on effect as the star
reaches the asymptotic regime and the third term is a turn-off effect
for small envelopes.

\section{Mass-loss prescriptions}

\label{sec:Mass-loss-prescriptions}We consider three mass-loss prescriptions
for TPAGB stars.
\begin{description}
\item [{VW93}] The formalism of \citet[VW93]{1993ApJ...413..641V} relates
the mass-loss rate to the Mira pulsation period of the star, given
by\begin{eqnarray}
P_{0} & = & 10^{-2.07-0.90L/\mathrm{L_{\odot}}+1.94\log_{10}\left(R/\mathrm{R_{\odot}}\right)}\,\mathrm{days}.\end{eqnarray}
The mass loss rate is then given by, as in \citealp{Parameterising_3DUP_Karakas_Lattanzio_Pols},
i.e. without the $M/\mathrm{M_{\odot}}-2.5$ term of the original
VW93 prescription, \begin{eqnarray}
\dot{M}=\dot{M}_{\mathrm{VW0}} & = & f_{\mathrm{VW}}\left(-11.4+0.0125P_{0}\right)\mathrm{\, M_{\odot}}\,\mathrm{yr}^{-1}\end{eqnarray}
unless $P_{0}>\left(500-\Delta P_{\mathrm{VW}}\right)\,\mathrm{days}$
in which case a superwind is applied \begin{eqnarray}
\dot{M} & = & \max\left(\dot{M}_{\mathrm{VW0}},f_{\mathrm{VW}}\frac{Lc}{v_{\mathrm{w}}}\right)\end{eqnarray}
where \begin{eqnarray}
v_{\mathrm{w}} & = & 10^{5}\left(-13.5+0.056P_{0}\right)\,\mathrm{cm}\,\mathrm{s}^{-1}\,.\end{eqnarray}
The free parameters $f_{\mathrm{VW}}$ and $\Delta P_{\mathrm{VW}}$
subtly affect the mass-loss rate. The factor $f_{\mathrm{VW}}$ is
a simple multiplier, which is $1$ by default (see model set \modelset{27}).
The period shift $\Delta P_{\mathrm{VW}}$ allows the onset of the
superwind to be delayed, e.g. $\Delta P_{\mathrm{VW}}=-100\,\mathrm{days}$
in model set \modelset{33} -- it is zero by default.
\item [{Reimers}] The Reimers mass-loss rate is given by \begin{eqnarray}
\dot{M} & = & 4\times10^{-13}\eta\frac{RL}{M}\mathrm{\, M_{\odot}}\,\mathrm{yr}^{-1}\,,\end{eqnarray}
where $\eta$ is a parameter of order unity \citep{1975psae.book..229R}
which we vary in model sets \modelset{10}, \modelset{11} and \modelset{12}.
\item [{van~Loon}] In model set \modelset{13} we use the split form of
\citet{2005A&A...438..273V} appropriate to oxygen-rich red giants,\begin{align}
\log_{10}\left[\dot{M}/(\mathrm{M_{\odot}}\mathrm{yr}^{-1})\right] & =\nonumber \\
 & \mbox{}{\hspace{-25mm}\left\{ \begin{array}{cc}
-5.6+1.10l_{4}-5.2\mathrm{t_{35}} & l_{4}<0.9\,\mathrm{and\,}\\
-5.3+0.82l_{4}-10.8t_{35} & l_{4}\geq0.9\,,\end{array}\right.}\end{align}
where $l_{4}=\log_{10}\left(L/10^{4}\mathrm{L_{\odot}}\right)$ and
$t_{\mathrm{35}}=\log_{10}\left(T_{\mathrm{eff}}/3500\mathrm{K}\right)$.
Note, if $T_{\mathrm{eff}}>4000\,\mathrm{K}$ we enforce a minimum
mass-loss rate of $10^{-4}\mathrm{\, M_{\odot}}\,\mathrm{year}^{-1}$
because the above formula can approach zero as the temperature rises
(and the envelope mass becomes small) as a star approaches the white-dwarf
cooling track.
\end{description}

\section{Binary distributions}

\label{sec:Binary-distributions}Our default binary-star distribution
is the combination of 
\begin{enumerate}
\item The initial mass function (IMF) of \citet[KTG93]{KTG1993MNRAS-262-545K}
for the initial primary mass $M_{1}$\begin{equation}
\psi(M_{1})=\left\{ \begin{array}{ll}
0 & M_{1}/\mathrm{\, M_{\odot}}\leq m_{0}\\
a_{1}(M_{1}/\mathrm{M_{\odot})}^{p_{1}} & m_{0}<M_{1}/\mathrm{M_{\odot}}\leq m_{1}\\
a_{2}(M_{1}/\mathrm{M_{\odot})}^{p_{2}} & m_{1}<M_{1}/\mathrm{M_{\odot}}\leq m_{2}\\
a_{3}(M_{1}/\mathrm{M_{\odot}})^{p_{3}} & m_{2}<M_{1}/\mathrm{M_{\odot}}\leq m_{\textrm{max}}\\
0 & m>m_{\textrm{max}}\end{array}\right.\label{eq:KTG93-IMF}\end{equation}
where $p_{1}=-1.3$, $p_{2}=-2.2$, $p_{3}=-2.7$, $m_{0}=0.1$, $m_{1}=0.5$,
$m_{2}=1.0$ and $m_{\textrm{max}}=80.0$. Continuity requirements
and $\int\psi(M)dM=1$ give the constants $a_{1}$, $a_{2}$ and $a_{3}$.
\item A distribution flat in $q=M_{2}/M_{1}$ for the initial secondary
mass $M_{2}$, where $M_{2}\leq M_{1}$
\item A distribution flat in $\ln a$ (i.e. probability $\sim1/a$) for
the separation $a$ where $3\leq a\leq10^{5}$.
\item Initially circular binaries (except for model set \modelset{7}).
\end{enumerate}

\section{All Model Sets and Results}

\label{sec:all-models}Table~\ref{tab:The-full-list-of-population-parameters}
shows the full set of models we considered, of which Table~\ref{tab:modelsets}
is a subset.

Table~\ref{tab:percentages-full} shows the full set of CEMP, CNEMP
and NEMP to EMP ratios for all our model sets of which Table~\ref{tab:percentages}
is a subset.

\begin{table*}
\begin{tabular}{|c|c|}
\hline 
Model Set & Physical Parameters (differences from model set \protect \modelset{1})\tabularnewline
\hline
\modelset{1} & -\tabularnewline
\modelset{4}, \modelset{5} & $\alpha_{\mathrm{CE}}=0.1$ and $3$ respectively\tabularnewline
\modelset{6} & Bondi-Hoyle $\alpha_{\mathrm{BH}}=5$\tabularnewline
\modelset{7} & Initial eccentricity $e=0.5$\tabularnewline
\modelset{8} & CRAP parameter $B=10^{3}$\tabularnewline
\modelset{9} & Third DUP calibration: $\Delta M_{\mathrm{c},\mathrm{min}}=-0.07\mathrm{\, M_{\odot}}$,
$\lambda_{\mathrm{min}}=0.8$\tabularnewline
\modelset{10}, \modelset{11}, \modelset{12} & Reimers AGB wind: $\eta=0.1$, $1$ and $5$ respectively\tabularnewline
\modelset{13} & AGB wind of Van Loon\tabularnewline
\modelset{14}, \modelset{15}, \modelset{16} & Common envelope accretion $0.01$, $0.02$ and $0.05\mathrm{\, M_{\odot}}$
respectively\tabularnewline
\modelset{17}, \modelset{18} & $^{13}\mathrm{C}$ efficiency parameter $0.1$ and $0.01$ respectively\tabularnewline
\modelset{19} & Thermohaline mixing disabled\tabularnewline
\modelset{21}, \modelset{22} & Minimum CEMP age $12$ and $8\,\mathrm{Gyr}$ respectively\tabularnewline
\modelset{26} & Third DUP $M_{\mathrm{env},\mathrm{min}}=0\mathrm{\, M_{\odot}}$\tabularnewline
\modelset{27} & $M_{\mathrm{env},\mathrm{min}}=0\mathrm{\, M_{\odot}}$, $\xi_{\mathrm{13}}=0.01$,
$\Delta M_{\mathrm{c,min}}=-0.1\mathrm{\, M_{\odot}}$, $\lambda_{\mathrm{min}}=0.5$,
$f_{\mathrm{VW}}=0.1$\tabularnewline
\modelset{28} & $M_{\mathrm{env},\mathrm{min}}=0\mathrm{\, M_{\odot}}$, $\xi_{\mathrm{13}}=0.01$,
$\Delta M_{\mathrm{c,min}}=-0.1\mathrm{\, M_{\odot}}$, $\lambda_{\mathrm{min}}=0.5$\tabularnewline
\modelset{29}, \modelset{30} & $\left[\mathrm{C}/\mathrm{Fe}\right]_{\mathrm{min}}=0.5$ and $0.7$
respectively\tabularnewline
\modelset{31}, \modelset{32} & As \modelset{27} with $\left[\mathrm{C}/\mathrm{Fe}\right]_{\mathrm{min}}=0.5$
and $0.7$ respectively\tabularnewline
\modelset{33} & $\Delta P_{\mathrm{VW}}=-100\,\mathrm{days}$\tabularnewline
\modelset{34}, \modelset{35} & $M_{\mathrm{env},\mathrm{min}}=0\mathrm{\, M_{\odot}}$, $\xi_{\mathrm{13}}=0.01$,
$\Delta M_{\mathrm{c,min}}=-0.1\mathrm{\, M_{\odot}}$ with $\lambda_{\mathrm{min}}=0.1$
and $0.8$ respectively\tabularnewline
\modelset{36} & $M_{\mathrm{env},\mathrm{min}}=0.25\mathrm{\, M_{\odot}}$, $\xi_{\mathrm{13}}=0.01$,
$\Delta M_{\mathrm{c,min}}=-0.1\mathrm{\, M_{\odot}}$, $\lambda_{\mathrm{min}}=0.8$\tabularnewline
\modelset{37}, \modelset{38}, \modelset{39}, \modelset{40} & Nelemans common-envelope prescription, $\gamma=0.5$, $1$, $1.5$
and $2$ respectively\tabularnewline
\modelset{41} & As \modelset{39} but Nelemans prescription for $q>0.2$ only (otherwise
the default prescription)\tabularnewline
\modelset{42} & As \modelset{41} but Nelemans prescription only for the first common-envelope
phase\tabularnewline
\modelset{43} & $\alpha_{\mathrm{CE}}=0.1$, $M_{\mathrm{env},\mathrm{min}}=0\mathrm{\, M_{\odot}}$,
$\Delta M_{\mathrm{c,min}}=-0.1\mathrm{\, M_{\odot}}$, $\lambda_{\mathrm{min}}=0.5$
, Comenv accretion $0.05$, No thermohaline\tabularnewline
\modelset{44} & \modelset{38} and \modelset{43} combined\tabularnewline
\modelset{45} & As \modelset{35} with $\xi_{\mathrm{13}}=0.1$\tabularnewline
\modelset{48} & $M_{\mathrm{env},\mathrm{min}}=0\mathrm{\, M_{\odot}}$, $\Delta M_{\mathrm{c,min}}=-0.1\mathrm{\, M_{\odot}}$,
$\lambda_{\mathrm{min}}=0.5$, $\xi_{\mathrm{13}}=0.01$\tabularnewline
\modelset{49} & $M_{\mathrm{env},\mathrm{min}}=0\mathrm{\, M_{\odot}}$, $\Delta M_{\mathrm{c,min}}=-0.1\mathrm{\, M_{\odot}}$,
$\lambda_{\mathrm{min}}=0.8$, $\xi_{\mathrm{13}}=0.1$\tabularnewline
\modelset{50} & $M_{\mathrm{env},\mathrm{min}}=0\mathrm{\, M_{\odot}}$, $\Delta M_{\mathrm{c,min}}=-0.1\mathrm{\, M_{\odot}}$,
$\lambda_{\mathrm{min}}=0.5$, $\xi_{\mathrm{13}}=0.001$\tabularnewline
\modelset{51}, \modelset{52}, \modelset{53} & %
\begin{minipage}[t]{0.8\columnwidth}%
\begin{center}
Comenv accretion $0.05\mathrm{\, M_{\odot}}$, $M_{\mathrm{env},\mathrm{min}}=0\mathrm{\, M_{\odot}}$,
$\Delta M_{\mathrm{c,min}}=-0.1\mathrm{\, M_{\odot}}$, $\lambda_{\mathrm{min}}=0.5$,\\
 $\xi_{\mathrm{13}}=0.1,\,0.01\,\mathrm{and}\,0.001$ respectively
\par\end{center}%
\end{minipage}\tabularnewline
\modelset{54}, \modelset{55}, \modelset{56} & %
\begin{minipage}[t]{0.8\columnwidth}%
\begin{center}
Comenv accretion $0.05\mathrm{\, M_{\odot}}$, $M_{\mathrm{env},\mathrm{min}}=0\mathrm{\, M_{\odot}}$,
$\Delta M_{\mathrm{c,min}}=-0.1\mathrm{\, M_{\odot}}$, $\lambda_{\mathrm{min}}=0.8$,\\
 $\xi_{\mathrm{13}}=0.1,\,0.01\,\mathrm{and}\,0.001$ respectively
\par\end{center}%
\end{minipage}\tabularnewline
\modelset{57}, \modelset{58}, \modelset{59} & %
\begin{minipage}[t][1.1\totalheight]{0.8\columnwidth}%
\begin{center}
$M_{\mathrm{env},\mathrm{min}}=0\mathrm{\, M_{\odot}}$, $\Delta M_{\mathrm{c,min}}=-0.1\mathrm{\, M_{\odot}}$,
$\lambda_{\mathrm{min}}=0.5$,\\
 $\xi_{\mathrm{13}}=0.1,\,0.01\,\mathrm{and}\,0.001$ respectively
\par\end{center}%
\end{minipage}\tabularnewline
\hline
\end{tabular}

\caption{\label{tab:The-full-list-of-population-parameters}The full list of
our binary population models (a subset is shown in Table~\ref{tab:modelsets}).
The meanings of the symbols are given in Section~\ref{sec:Model}.}

\end{table*}
\begin{table*}
\begin{centering}
\begin{tabular}{|c|c|c|c|c|}
\hline
Model Set & CEMP/EMP $\%$ & CNEMP/EMP $\%$ & NEMP/EMP $\%$ \tabularnewline
\hline
$\modelset{1}, \modelset{17}, \modelset{18}$ & $2.300 \pm 0.034$ & $0.098 \pm 0.002$ & $0.267 \pm 0.007$ \tabularnewline
$\modelset{4}$ & $2.370 \pm 0.029$ & $0.100 \pm 0.002$ & $0.273 \pm 0.007$ \tabularnewline
$\modelset{5}$ & $2.300 \pm 0.033$ & $0.097 \pm 0.002$ & $0.267 \pm 0.006$ \tabularnewline
$\modelset{6}$ & $3.050 \pm 0.038$ & $0.157 \pm 0.003$ & $0.323 \pm 0.006$ \tabularnewline
$\modelset{7}$ & $2.150 \pm 0.033$ & $0.086 \pm 0.002$ & $0.249 \pm 0.007$ \tabularnewline
$\modelset{8}$ & $2.460 \pm 0.034$ & $0.101 \pm 0.002$ & $0.268 \pm 0.006$ \tabularnewline
$\modelset{9}$ & $2.900 \pm 0.040$ & $0.098 \pm 0.002$ & $0.267 \pm 0.006$ \tabularnewline
$\modelset{10}$ & $2.640 \pm 0.032$ & $0.208 \pm 0.003$ & $0.310 \pm 0.005$ \tabularnewline
$\modelset{11}$ & $1.490 \pm 0.021$ & $0.053 \pm 0.001$ & $0.091 \pm 0.002$ \tabularnewline
$\modelset{12}$ & $0.489 \pm 0.009$ & $0.005 \pm 0.000$ & $0.015 \pm 0.001$ \tabularnewline
$\modelset{13}$ & $0.101 \pm 0.004$ & $0.000$ & $0.000$ \tabularnewline
$\modelset{14}$ & $2.460 \pm 0.035$ & $0.099 \pm 0.002$ & $0.291 \pm 0.007$ \tabularnewline
$\modelset{15}$ & $2.640 \pm 0.036$ & $0.100 \pm 0.002$ & $0.303 \pm 0.007$ \tabularnewline
$\modelset{16}$ & $2.940 \pm 0.038$ & $0.103 \pm 0.002$ & $0.311 \pm 0.007$ \tabularnewline
$\modelset{19}$ & $4.210 \pm 0.039$ & $0.290 \pm 0.003$ & $0.409 \pm 0.004$ \tabularnewline
$\modelset{21}$ & $2.230 \pm 0.047$ & $0.093 \pm 0.003$ & $0.258 \pm 0.009$ \tabularnewline
$\modelset{22}$ & $2.480 \pm 0.028$ & $0.110 \pm 0.002$ & $0.278 \pm 0.005$ \tabularnewline
$\modelset{26}$ & $6.470 \pm 0.030$ & $0.103 \pm 0.002$ & $0.267 \pm 0.006$ \tabularnewline
$\modelset{27}$ & $8.840 \pm 0.041$ & $0.171 \pm 0.003$ & $0.273 \pm 0.004$ \tabularnewline
$\modelset{28}$ & $8.320 \pm 0.039$ & $0.103 \pm 0.002$ & $0.266 \pm 0.006$ \tabularnewline
$\modelset{29}$ & $3.460 \pm 0.042$ & $0.203 \pm 0.003$ & $0.267 \pm 0.004$ \tabularnewline
$\modelset{30}$ & $2.990 \pm 0.039$ & $0.158 \pm 0.003$ & $0.267 \pm 0.005$ \tabularnewline
$\modelset{31}$ & $11.600 \pm 0.053$ & $0.256 \pm 0.003$ & $0.273 \pm 0.004$ \tabularnewline
$\modelset{32}$ & $10.400 \pm 0.048$ & $0.221 \pm 0.003$ & $0.273 \pm 0.004$ \tabularnewline
$\modelset{33}$ & $2.490 \pm 0.035$ & $0.116 \pm 0.003$ & $0.340 \pm 0.008$ \tabularnewline
$\modelset{34}$ & $6.520 \pm 0.030$ & $0.103 \pm 0.002$ & $0.267 \pm 0.006$ \tabularnewline
$\modelset{35}$ & $9.430 \pm 0.044$ & $0.103 \pm 0.002$ & $0.266 \pm 0.006$ \tabularnewline
$\modelset{36}$ & $5.640 \pm 0.060$ & $0.102 \pm 0.002$ & $0.267 \pm 0.006$ \tabularnewline
$\modelset{37}$ & $2.260 \pm 0.036$ & $0.096 \pm 0.002$ & $0.262 \pm 0.006$ \tabularnewline
$\modelset{38}$ & $2.270 \pm 0.036$ & $0.096 \pm 0.002$ & $0.264 \pm 0.006$ \tabularnewline
$\modelset{39}$ & $2.280 \pm 0.035$ & $0.097 \pm 0.002$ & $0.265 \pm 0.007$ \tabularnewline
$\modelset{40}$ & $2.300 \pm 0.036$ & $0.098 \pm 0.002$ & $0.265 \pm 0.007$ \tabularnewline
$\modelset{41}, \modelset{42}$ & $2.280 \pm 0.035$ & $0.097 \pm 0.002$ & $0.265 \pm 0.006$ \tabularnewline
$\modelset{43}$ & $14.900 \pm 0.063$ & $0.303 \pm 0.003$ & $0.425 \pm 0.004$ \tabularnewline
$\modelset{44}$ & $14.600 \pm 0.066$ & $0.286 \pm 0.003$ & $0.389 \pm 0.004$ \tabularnewline
$\modelset{45}, \modelset{49}$ & $9.430 \pm 0.044$ & $0.103 \pm 0.002$ & $0.266 \pm 0.006$ \tabularnewline
$\modelset{47}$ & $2.060 \pm 0.028$ & $0.000$ & $0.002 \pm 0.000$ \tabularnewline
$\modelset{48}, \modelset{50}$ & $8.320 \pm 0.039$ & $0.103 \pm 0.002$ & $0.266 \pm 0.006$ \tabularnewline
$\modelset{51}, \modelset{52}, \modelset{53}$ & $14.700 \pm 0.064$ & $0.298 \pm 0.003$ & $0.426 \pm 0.004$ \tabularnewline
$\modelset{54}, \modelset{55}, \modelset{56}$ & $15.500 \pm 0.068$ & $0.298 \pm 0.003$ & $0.426 \pm 0.004$ \tabularnewline
$\modelset{57}, \modelset{58}, \modelset{59}$ & $12.900 \pm 0.057$ & $0.278 \pm 0.003$ & $0.387 \pm 0.004$ \tabularnewline
\hline
\end{tabular}

\par\end{centering}

\caption{\label{tab:percentages-full}Percentage of CEMP, CNEMP and NEMP (sub-)giants
relative to total EMP giants in all our model binary populations (see
Section~\ref{sub:model-Selection-criteria} for selection criteria).
The errors convey Poisson statistics only. The final column gives
the number ratio of CEMP-$s$ to CEMP stars.}

\end{table*}

\section{Observation database}

\label{sec:appendix-Observation-Database}Our observational selection
is taken from the SAGA database as compiled by \citet{2008PASJ...60.1159S}
combined with data \citet{2006ApJ...652L..37L}. 

When data exists for the same star from more than one source, we \Change{take
the arithmetic mean of} the values and add errors in quadrature.
In the case of log-values, e.g. $[\mathrm{Fe}/\mathrm{H}]$ or $\log g$,
we simply average the log-values rather than attempt a more sophisticated
approach. This makes little difference to our final results. In the
case of data limits (e.g. $x<4$) we ignore the data -- few data are
of this type and the general result is not affected.

We ignore error bars in the sense that, e.g. a star with $[\mathrm{Fe}/\mathrm{H}]=-2.9\pm0.2$
is not included in our selection, even though it may well have --
in reality -- $[\mathrm{Fe}/\mathrm{H}]=-2.7$ and hence qualify.
This is the price we pay for a simple selection procedure and in the
large number limit (the database has about 1300 stars) it is not a
problem.

\end{document}